\documentclass[seceq,supplement]{ptptex}

\usepackage{graphicx}
\usepackage{wrapft}
\usepackage{bm}
\usepackage{multirow}

\def\la{\mathrel{\mathpalette\fun <}}
\def\ga{\mathrel{\mathpalette\fun >}}
\def\fun#1#2{\lower3.6pt\vbox{\baselineskip0pt\lineskip.9pt
\ialign{$\mathsurround=0pt#1\hfil##\hfil$\crcr#2\crcr\sim\crcr}}}

\def\ve{\varepsilon}

\newcommand{\beq}{\begin{equation}}
\newcommand{\eeq}{\end{equation}}
\newcommand{\bea}{\begin{eqnarray}}
\newcommand{\eea}{\end{eqnarray}}

\newcommand{\bfi}[1]{\mbox{\boldmath $#1$}}

\newcommand{\vk}{{\bfi k}}
\newcommand{\vK}{{\bfi K}}
\newcommand{\vp}{{\bfi p}}
\newcommand{\vq}{{\bfi q}}
\newcommand{\vs}{{\bfi s}}

\newcommand{\vrr}{{\bfi r}}
\newcommand{\vR}{{\bfi R}}

\newcommand{\vy}{{\bfi y}}

\newcommand{\br}{{\bf r}}

\def\dfrac#1#2{{\displaystyle\frac{#1}{#2}}}

\def\beq{\begin{equation}}
\def\eeq{\end{equation}}
\def\bea{\begin{eqnarray}}
\def\eea{\end{eqnarray}}

\def\la{\mathrel{\mathpalette\fun <}}
\def\ga{\mathrel{\mathpalette\fun >}}
\def\fun#1#2{\lower3.6pt\vbox{\baselineskip0pt\lineskip.9pt
  \ialign{$\mathsurround=0pt#1\hfil##\hfil$\crcr#2\crcr\sim\crcr}}}
\title{
The continuum discretized coupled-channels method \\
and its applications
}

\author{Masanobu \textsc{Yahiro}$^{1}$,
Kazuyuki \textsc{Ogata}$^{2}$, Takuma \textsc{Matsumoto}$^{1}$
and Kosho \textsc{Minomo}$^{1}$
}

\inst{
$^{1}$Department of Physics, Kyushu University, Fukuoka 812-8581, Japan \\
$^{2}$Research Center of Nuclear Physics (RCNP), Osaka University,
Ibaraki 567-0047, Japan}

\recdate{\today}%            %Editorial Office will fill in this.

\abst{This is a review on recent developments of
the continuum discretized coupled-channels method (CDCC) and
its applications to nuclear physics, cosmology and astrophysics,
and nuclear engineering.
The theoretical foundation of CDCC is shown, and
a microscopic reaction theory for nucleus-nucleus scattering is
constructed as an underlying theory of CDCC.
CDCC is then extended to treat Coulomb breakup and four-body breakup.
We also propose a new theory that makes CDCC applicable
to inclusive reactions.
}

\begin{document}

\maketitle

\tableofcontents

\section{Introduction}
\label{sec1}

Nuclear reaction is one of fundamental reactions in Nature.
Understanding of nuclear reaction is important not only in
nuclear physics but also in cosmology and astrophysics.
The study of nuclear reaction is necessary to understand the mechanism of
nucleosynthesis in the early universe, stellar evolution,
and supernova explosions, and
to eventually know the origin of elements in the universe.
Application of the study to nuclear engineering is another important direction.
The Li$(d,n)$ reaction, for example, is regarded as one of the most promising
reactions to produce intense neutron beams at the international
fusion material irradiation facility (IFMIF).
The study of the Li$(d,n)$ reaction is indispensable
for the engineering design of accelerator-driven neutron sources.
Theoretical calculations play an important role there, since
experimental data are not available systematically.

One of the most important current subjects in nuclear physics is
to elucidate unstable nuclei.
This subject is interesting also from the viewpoint
of the origin of elements through nucleosynthesis.
Unstable nuclei are considered to have exotic properties such as
the halo structure~\cite{Tan92,Jen04,Jon04} and
the loss of magicity in the
``island of inversion''~\cite{War90},
i.e., the region of unstable nuclei from $^{30}$Ne to $^{34}$Mg.
In the island of inversion,
the first-excited states have low excitation energies
and large $B({\rm E}2)$ values~\cite{Mot95,Cau98,Uts99,Iwa01,Yan03}.
This indicates that the $N=20$ neutron magic number is not valid anymore.
These novel quantum properties have inspired
extensive experimental and theoretical studies.

Important experimental tools of investigating properties of
unstable nuclei are the interaction cross section $\sigma_{\rm I}$
and the nucleon-removal cross section
$\sigma_{-N}$~\cite{Tan92,Jen04,Jon04,Gad08}.
The experimental exploration of halo nuclei
is moving from lighter nuclei such
as He and C isotopes to relatively heavier nuclei such as Ne isotopes.
Very lately $\sigma_{\rm I}$ was measured
by Takechi and collaborators~\cite{Tak10} for $^{28\mbox{--}32}$Ne located near or
in the island of inversion.
Furthermore, a halo structure of $^{31}$Ne was reported by
Nakamura and his group~\cite{Nak09} with the experiment on $\sigma_{-N}$.
The $^{31}$Ne nucleus resides in the island of inversion and is
the heaviest halo nucleus suggested experimentally in the present stage.

The elucidation of unstable nuclei
can be accomplished with high-accuracy measurements
of the scattering of unstable nuclei and accurate analyses
of the measurements with reliable reaction theories.
The scattering of unstable nuclei have two features.
First, the projectile is fragile and thereby the projectile breakup becomes
important.
Second, measurements of the elastic scattering are not easy
because of weak intensity of the secondary beam,
and consequently, there is no reliable phenomenological optical potential.
It is then important to construct
a microscopic reaction theory that can treat the projectile breakup and
does not need phenomenological optical potentials.
This construction is precisely a goal of the nuclear reaction theory.

A pioneering work on the microscopic description of nucleon-nucleus
scattering was done by Watson~\cite{Wat53}, which was reformulated
by Kerman, McManus, and Thaler~\cite{Ker59}
as series expansion in terms of an underlying nucleon-nucleon (NN)
$t$~matrix.
Another important microscopic model particularly to treat inclusive reactions
is the Glauber model~\cite{Gla59}.
The Glauber model is based on the eikonal and the adiabatic approximation.
It is well known that the adiabatic approximation makes
the breakup cross section diverge when
the Coulomb breakup is included.
For this reason the Glauber model has been applied mainly for
lighter targets in which the role of the Coulomb interaction is
negligible; see, e.g.,
Refs.~\citen{Gad08,HM85,Hen96,Yab92,AT96,BM92,BH04}.
Recently, a way of making Coulomb corrections to the calculations
has been proposed,~\cite{AS04,Cap08} which works when Coulomb breakup is
not so strong.

The continuum discretized coupled-channels method
(CDCC)~\cite{Kam86,Aus87}
is an accurate method of treating the projectile breakup
in exclusive reactions such as elastic scattering,
elastic-breakup reactions, and transfer reactions.
CDCC has succeeded in reproducing experimental data on the scattering of
both stable and unstable nuclei;
see, e.g., Refs.~\citen{Hir91,Rus00,HS92,Tor03,Gom05,Bec07,Rus04,%
Rus05,Mor07,Far10,Tos01,Dav01,Mor02,MN06,Hus06,Oga09b,DT02,Tak03,%
Can09,Oga03,Oga06,Mat03,Mat04,Ega04,Mat06,Ega09,Mat09,Mat10,Mor01,%
Rod05,Mor06,Rod08,Mor09,Rod09}
and references therein.
Another reliable method of
treating the projectile breakup in exclusive reactions
is the dynamical eikonal approximation (DEA)~\cite{Bay05,Gol06}.
The DEA is accurate at intermediate and high incident energies
where the eikonal approximation is reliable.

CDCC was first proposed as a method
of treating the three-body system composed
of a target and two fragments
of a projectile.
The original version of CDCC is called three-body CDCC.
In the review articles \citen{Kam86} and \citen{Aus87} on
three-body CDCC, the method
treated nuclear breakup but not Coulomb breakup,
and the phenomenological optical potentials were mainly used
as the interactions between the target and the two fragments
of the projectile.
For the scattering of $^{6,7}$Li and $^{12}$C,
there was an attempt to obtain the interactions microscopically.
However, some phenomenological treatments of the imaginary parts
of the interactions were necessary in that period.
Three-body CDCC thus had some limitations.
After the review articles, CDCC has been developed
in several aspects, and applied to various studies in wide
research fields.
The main purpose of the present article is to review
these developments and applications of CDCC.

One of the most important developments is the construction of
the microscopic reaction theory for nucleus-nucleus
scattering~\cite{Yah08} based on the multiple scattering theory.
This gives a foundation of
CDCC and other reaction calculations that adopt microscopic
nucleus-nucleus interactions based on the NN $t$ or $g$~matrix.
The microscopic reaction theory is reduced to the double-folding model
when both nuclear and Coulomb breakup are weak.
Then it allows us to practically obtain a microscopic optical potential.
As mentioned above,
CDCC is designed to accurately treat the breakup of
a projectile into some fragments. The double-folding model
can be used the input potentials of CDCC, i.e.,
the microscopic optical potential between
the target and each constituent of the projectile.

In \S\ref{sec2},
we recapitulate the original version of CDCC, i.e., three-body CDCC, and
show its theoretical foundation, with
clarifying the relation between the Faddeev method and
CDCC.
In \S\ref{sec3}, we present
the microscopic reaction theory for nucleus-nucleus scattering.
We also discuss the validity of the Glauber model and
examine the localization prescription by Brieva and Rook~\cite{BR77}
for the nonlocal microscopic optical potentials.
The double-folding model is then applied to the scattering of stable nuclei
and unstable Ne isotopes at intermediate energies, in which
breakup effects are expected to be small.
We propose in \S\ref{sec4} eikonal-CDCC (E-CDCC) that treats Coulomb-dominated
breakup processes very accurately and efficiently. Inclusion of
dynamical relativistic effects in breakup reactions is also
accomplished. In \S\ref{sec5} we extend three-body CDCC to treat
breakup processes of a projectile that has a three-body structure,
i.e., four-body CDCC. We describe also how to smooth the
discrete breakup cross sections obtained by CDCC.
The eikonal reaction theory (ERT) is proposed in \S\ref{sec6},
which makes CDCC applicable to inclusive breakup reactions.
CDCC and the extended theories are applied to
the scattering of unstable nuclei in \S\ref{sec7},
to reactions essential in cosmology and astrophysics in \S\ref{sec8},
and to reactions concerning nuclear engineering in \S\ref{sec9}.
Finally, we give summary in \S\ref{sec10}.

\section{Theoretical foundation of CDCC}
\label{sec2}

In this section, we recapitulate
the original version of CDCC~\cite{Kam86,Aus87},
i.e., three-body CDCC, and show the theoretical
foundation~\cite{Aus89,Aus96,Del07}
in order to understand what CDCC is.

\subsection{Three-body CDCC}
\label{sec2-1}
\begin{figure}[htbp]
\begin{center}
 \includegraphics[width=0.5\textwidth,clip]{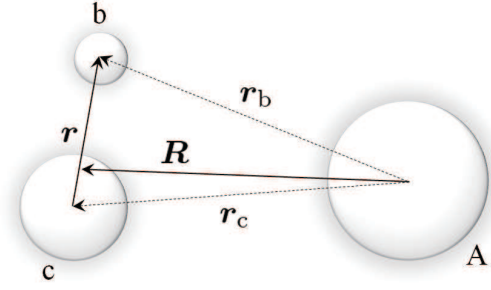}
 \caption{Illustration of the A+b+c three-body system.
 Vector $\vrr_{\gamma}$ ($\gamma =$b or c)
is the relative coordinate between $\gamma$ and A, whereas
$\vR$ and $\vrr$ form a set of Jacobi coordinates.
}
\label{Figure-3body-system-s2}
\end{center}
\end{figure}

We consider the projectile that is composed of two particles b and c
exactly or approximately. In this case, the scattering of the projectile
on a target (A) can be described by the A+b+c three-body system.
Figure~\ref{Figure-3body-system-s2}
is an illustration of the three-body system and
coordinates among the three particles.
The three-body scattering is governed
by the Schr\"odinger equation
\bea
  [H -E] \Psi =0
\label{three-body-equation-s2}
\eea
with the Hamiltonian
\bea
 H = K_{r} + K_{R} + v(\vrr) + U_{\rm b}(\vrr_{\rm b}) + U_{\rm c}(\vrr_{\rm c}),
\label{three-body-H-s2}
\eea
where $K_{r}$ and $K_{R}$ are the kinetic energy operators
associated with $\vrr$ and $\vR$, respectively, and $v(\vrr)$ is
the interaction between b and c, while
$U_{\rm b}$ ($U_{\rm c}$) is an optical potential of the
scattering of b (c) on A.

%%%%%%%%%%%%%%%%%%%%%%%
%%%  Figure 1
%%%%%%%%%%%%%%%%%%%%%%%
In CDCC, the total wave function $\Psi$ is expanded
by the complete set of
eigenfunctions of the Hamiltonian $h=K_{r}+v(\vrr)$ of the b+c subsystem.
The eigenfunctions are composed of bound and continuum
states. The continuum states are characterized by
the orbital angular momentum $\ell$ and
the linear momentum $k$ of the b+c subsystem,
and they are truncated as
\bea
k \le k_{\rm max}, \quad \ell \le \ell_{\rm max}.
\label{Truncation-s2}
\eea
The $\ell$- and $k$-truncations are the primary approximation in CDCC.
After making the truncations, we further discretize the $k$-continuum.
The model space ${\cal P'}$ thus constructed is described by
\bea
{\cal P'}=\sum_{i=0}^{N}| \phi_i \rangle \langle \phi_i| ,
\eea
where the $\phi_i$ represent the bound and discretized-continuum states of
$h$, and $N$ is the number of the $\phi_i$.
The total wave function $\Psi$ is hence approximated into
\bea
\Psi \approx \Psi_{\rm CDCC} \equiv {\cal P'}\Psi =
\sum_{i=0}^{N} \phi_i(\vrr)\chi_i(\vR),
\eea
where the coefficients $\chi_i(\vR)$ of the expansion describe motions
between the projectile in the states $\phi_i$ and A.
The approximate total wave function $\Psi_{\rm CDCC}$ is obtained by solving
the three-body Schr\"odinger equation \eqref{three-body-equation-s2}
in the model space ${\cal P'}$
\bea
  {\cal P'}[H -E]{\cal P'} \Psi_{\rm CDCC} =0.
\eea
This leads to a set of coupled-channel equations for $\chi_i(\vR)$,
\bea
   [E - K_{R} - \ve_i] \chi_i(\vrr)= \sum_{j}^{N}
   \langle \phi_i|U| \phi_j \rangle \chi_j(\vR),
\label{CC-eq-s2}
\eea
where
$U=U_{\rm b}(\br_{\rm b}) + U_{\rm c}(\br_{\rm c})$
and
the $\ve_i$ are energies of the b+c subsystem in the $\phi_i$.
Solving the CDCC equations \eqref{CC-eq-s2}
with the ordinary boundary condition,
one can get the $S$-matrix elements of the elastic scattering and the
projectile-breakup reactions~\cite{Kam86,Aus87}.

As for the discretization prescription, three methods have been
proposed so far: i) the average (Av) method,~\cite{Kam86,Aus87,Rod09}
ii) the midpoint (Mid) method,~\cite{Aus87,Piy89,Piy99} and
iii) the pseudo-state (PS) method.~\cite{Kam86,Mor01,Ras89,Mat03,Ega04,Mat04,Mat06,Rod08}
In the Av and Mid methods, the $k$-continuum is
divided into a finite number of bins.
In the Av method, the continuum states within each bin are
averaged into a single state, whereas in the Mid method they are
represented by a single state at a midpoint of the bin.
The Av and Mid methods are valid,
since the two methods yield the same converged $S$-matrix elements
as the width $\Delta$ of bins decreases~\cite{Piy89,Piy99}.
The Av method is more practical than the Mid one,
because the former requires less numerical tasks than the latter.
Since then the Av method has widely been used as a standard way
of the discretization.

In the PS method, on the other hand, $h$ is diagonalized in a space spanned
by a finite number of $L^2$-type basis functions.
The resulting eigenstates can be regarded as bound states
when the eigenenergies are negative and
as discretized-continuum states when the eigenenergies are positive.
The discretized-continuum states are thus obtained
by a superposition of analytic basis functions in the PS method.
This makes numerical tasks much easier.
We could not derive, however, a smooth breakup cross section from
the discrete breakup cross section calculated with CDCC.
For this reason, the PS method was applied to virtual breakup processes
in the intermediate state of elastic scattering and transfer reactions,
but not to breakup reactions themselves.
This problem has recently been solved by the smoothing method proposed in
Refs.~\citen{Mat03,Ega04,Mat04,Mat06}.
This smoothing method is presented in \S\ref{sec5}.

The $S$-matrix elements calculated with CDCC depend on the parameters,
$\ell_{\rm max}$, $k_{\rm max}$, $\Delta$, i.e., the size of the model space
$P'$. This artifact should be removed by confirming
that the calculated $S$-matrix elements converge
as the model space is expanded.
Indeed, the convergence was shown
in Refs.~\citen{Kam86,Aus87,Piy89,Piy99}.
The next question to be addressed is
whether the converged $S$-matrix elements are exact.
This point is discussed in the next subsection.

\subsection{Relation between CDCC and Faddeev solutions}
\label{sec2-2}

As mentioned in \S\ref{sec2-1},
CDCC is based on three approximations, the $\ell$-truncation,
the $k$-truncation, and the discretization of the $k$-continuum.
The $\ell$-truncation is most essential among these approximations
as shown below.
Now we introduce the projection operator ${\cal P}$ that only selects
$\ell$ up to $\ell_{\rm max}$.
Obviously, ${\cal P'}$ tends to ${\cal P}$ in the limit of
large $k_{\rm max}$ and small $\Delta$.
The component ${\cal P}\Psi$ has no asymptotic amplitudes in
the rearrangement channels.
For example, let us consider a simple case of $\ell_{\rm max}=0$.
Then,
${\cal P}U{\cal P}$
is
the average of $U$ over the angle of vector $\vrr$. After the angle average,
the potential ${\cal P}U{\cal P}$ becomes a function of $r$ and $R$.
Thus, ${\cal P}U{\cal P}$ is a
three-body potential that vanishes at large $R$ and/or large $r$,
so that it does not generate any rearrangement channel.

The insertion of three-body distorting potentials does not change
the mathematical properties of the Faddeev equations~\cite{BR82}.
Now we consider ${\cal P}U{\cal P}$ as such a distorting potential
in order to obtain the distorted Faddeev equations,
\bea
&&(E-K_{r} - K_{R}-v-{\cal P}U{\cal P})\psi_{\rm A}
=v(\psi_{\rm b}+\psi_{\rm c}) ,
\label{distorted-Faddeev-eq-1-s2} \\
&&(E-K_{r} - K_{R}-U_{\rm b})\psi_{\rm b}
=(U_{\rm b}-{\cal P}U_{\rm b}{\cal P})\psi_{\rm A}+U_{\rm b}\psi_{\rm c} ,
\label{distorted-Faddeev-eq-2-s2} \\
&&(E-K_{r} - K_{R}-U_{\rm c})\psi_{\rm c}
=(U_{\rm c}-{\cal P}U_{\rm c}{\cal P})\psi_{\rm A}+U_{\rm c}\psi_{\rm b} ,
\label{distorted-Faddeev-eq-3-s2}
\eea
where $\psi_{\rm A}$, $\psi_{\rm b}$, and $\psi_{\rm c}$ satisfy the relation
$\Psi=\psi_{\rm A}+\psi_{\rm b}+\psi_{\rm c}$.
If Eqs.~\eqref{distorted-Faddeev-eq-1-s2}--\eqref{distorted-Faddeev-eq-3-s2}
are added, the distorting potential is canceled and the
original three-body Schr\"odinger equation
\eqref{three-body-equation-s2} is recovered.
In an iterative approach to
Eqs.~\eqref{distorted-Faddeev-eq-1-s2}--\eqref{distorted-Faddeev-eq-3-s2},
the zeroth-order solution for $\psi_{\rm A}$ is obtained by
setting the right-hand side of Eq.~\eqref{distorted-Faddeev-eq-1-s2} to zero.
The zeroth-order solution is nothing but $\Psi_{\rm CDCC}$.
When $\Psi_{\rm CDCC}$ is inserted in
Eqs.~\eqref{distorted-Faddeev-eq-2-s2} and \eqref{distorted-Faddeev-eq-3-s2},
the equations do not generate any disconnected diagram,
since $\Psi_{\rm CDCC}$ has no rearrangement channel in the asymptotic region.
Furthermore, the subtractions, $U_{\rm b}-{\cal P}U_{\rm b}{\cal P}$ and
$U_{\rm c}-{\cal P}U_{\rm c}{\cal P}$, sizably weaken couplings of
$\Psi_{\rm CDCC}$ with $\psi_{\rm b}$ and $\psi_{\rm c}$.
Thus, $\Psi_{\rm CDCC}$ is a good solution to
the three-body Schr\"odinger equation \eqref{three-body-equation-s2},
when $\ell_{\rm max}$ is large enough.
Very lately, the CDCC solution has directly been compared
with the Faddeev solution through numerical calculations and it has been shown
that the two solutions agree very well with each other~\cite{Del07}.

The discussion mentioned above is made
without using the $k$-truncation and the discretization of $k$-continuum.
The $\ell$-truncation is thus most essential in CDCC.

\section{Microscopic reaction theory for nucleus-nucleus scattering}
\label{sec3}

In this section, we present a microscopic
reaction theory for nucleus-nucleus scattering,
following Ref.~\citen{Yah08}.
The validity of the Glauber model is discussed with
the microscopic reaction theory.
When the projectile breakup is weak, the theory is reduced
to the double-folding model with the effective
nucleon-nucleon (NN) interaction.
The microscopic optical potential constructed with the double-folding model
is applied to measured reaction cross sections
for the scattering of stable nuclei and
neutron-rich Ne isotopes around 240~MeV/nucleon.
Through the analyses, it is confirmed that
the microscopic optical potential is reliable and hence can be used in
CDCC as the potentials between a target and fragments of a projectile.
This microscopic version of CDCC can be applied to many
nucleus-nucleus scattering.

\subsection{Microscopic reaction theory}
\label{sec3-1}

The most fundamental equation to describe nucleus-nucleus scattering is
the many-body Schr\"odinger equation with the realistic
NN interaction $v_{ij}$
\bea
(K+h_{\rm P}+h_{\rm A}+ \sum_{i \in {\rm P}, j \in {\rm A}} v_{ij}-E)
\Psi^{(+)}=0,
\label{schrodinger-vij-s3}
\eea
where $E$ is the energy of the total system,
$K$ is the kinetic-energy operator for the relative motion
between a projectile (P) and a target (A), and
$h_{\rm P}$ ($h_{\rm A}$) is the internal Hamiltonian of P (A).
The scattering of P from A can be described with a series of multiple
scattering in terms of $v_{ij}$.
In the series, one can first take a summation of
ladder diagrams between the same NN pair.
The summation can be described by an effective NN
interaction $\tau_{ij}$ in nuclear medium.
Taking a resummation of the series in terms of $\tau_{ij}$,
one can get the many-body
Schr\"odinger equation with $\tau_{ij}$~\cite{Yah08}
\bea
(K+h_{\rm P}+h_{\rm A}+ \sum_{i \in {\rm P}, j \in {\rm A}} \tau_{ij}-E)
{\hat \Psi}^{(+)}=0  \; ,
\label{schrodinger-effective-s3}
\eea
where it has been assumed that
the number of pairs $(i,j)$ is much larger than unity.
We further assumed that
the antisymmetrization between incident nucleons in P and
target nucleons in A can be approximated by using $\tau_{ij}$ that is
properly antisymmetrical with respect to the exchange
of the colliding nucleons.
The first assumption is satisfied well for nucleus-nucleus scattering,
and the second one is known to be accurate
at intermediate and high incident energies~\cite{TW55,PT81}.
This is an extension of the Kerman-McManus-Thaler formalism~\cite{Ker59}
for nucleon-nucleus scattering to nucleus-nucleus scattering.

As mentioned above, $\tau_{ij}$ describes nucleon-nucleon scattering
in nuclear medium. A possible simplification of $\tau_{ij}$ is
to replace $\tau_{ij}$ with the Br\"uckner $g$-matrix interaction,
which has been done in many applications; see, e.g.,
Refs.~\citen{BR77,Ber77,Jeu76,Sat79,Sat83,Yam83,Rik84,Amo00,Fur08}.
The $g$-matrix interaction, however, does not include effects induced
by finite nucleus,
e.g., effects of projectile breakup and target collective excitations,
because the interaction is evaluated in infinite nuclear matter.
In other words, $\tau_{ij}$ is much more complicated than
the $g$-matrix interaction.
Therefore, in general, it is not easy to solve
the many-body Schr\"odinger equation \eqref{schrodinger-effective-s3}.
However, it becomes feasible at least in the following cases.

First, if we consider a) nucleus-nucleus scattering
at high incident energies,
or, b) scattering of lighter projectiles from lighter targets
at intermediate incident energies, effects induced by finite nucleus
such as projectile breakup and target collective excitations
are small. Then, the double-folding model becomes reliable, with which
we can analyze the elastic-scattering cross section or the
total reaction cross section $\sigma_{\rm R}$.
We discuss the validity of the double-folding model for the scattering
of type b) in \S\ref{sec3-4} and \S\ref{sec3-5}.
It should be noted that in this case the Glauber model also becomes
reliable, since Coulomb breakup is negligible. The observables to be
analyzed with the Glauber model are $\sigma_{\rm R}$ or
inclusive cross sections such as the one-neutron removal cross section.
The validity of the Glauber model is discussed in \S\ref{sec3-2}.

Second, let us consider a projectile that is a weakly bound system of
two nucleus b and c. Then the many-body Schr\"odinger
equation~\eqref{schrodinger-effective-s3} is approximated well into
the three-body Schr\"odinger equation \eqref{three-body-equation-s2},
in which
the potential $U_{\gamma}({\bm r}_{\gamma})$ ($\gamma=$b or c)
is constructed to reproduce the scattering of $\gamma$ on A.
When $\gamma$ is not a weakly bound system,
the potential can be constructed microscopically
with the double-folding model, i.e., by folding
the effective NN interaction with the densities of A and $\gamma$.
Thus, we can solve with high accuracy Eq.~\eqref{schrodinger-effective-s3}
by three-body CDCC with all the input optical potentials
obtained microscopically. Note that the microscopic potential
is nonlocal in general, which makes it very difficult to solve
Eq.~\eqref{three-body-equation-s2}.
It is, however, possible to derive a local potential equivalent to
the nonlocal potential by using the method proposed by
Brieva and Rook~\cite{BR77}.
The validity of their localization method is discussed in \S\ref{sec3-3}.

\subsection{Validity of the Glauber model}
\label{sec3-2}

\begin{figure}[htb]
\begin{center}
\includegraphics[width=0.8\textwidth,clip]{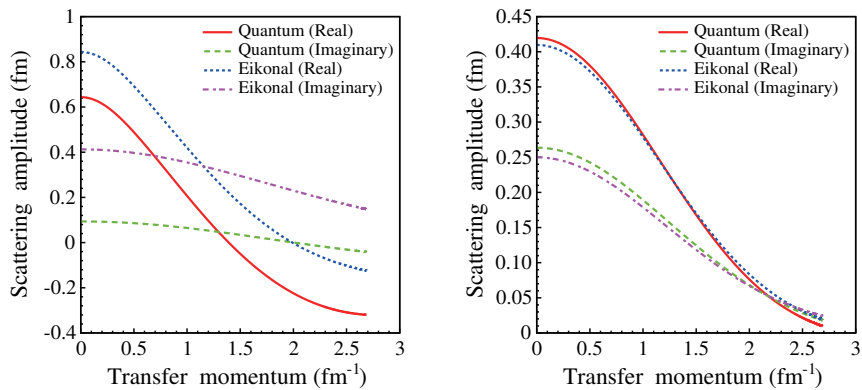}
\caption{
The on-shell NN scattering amplitude $f_{\rm {NN}}(\vq)$
at the laboratory energy $E_{\rm {NN}}=150$~MeV
calculated with the bare NN potential AV18 in the left panel and
with the JLM $g$~matrix \cite{Jeu76} in the right panel.
The solid (dashed) and dotted (dash-dotted) lines show, respectively,
the real and imaginary parts of $f_{\rm NN}(\vq)$ of the exact (eikonal)
calculation.
}
\label{fig:f-NN-s3}
\end{center}
\end{figure}

The Glauber model is based on the eikonal approximation
for NN scattering and
the eikonal and adiabatic approximations for nucleus-nucleus scattering.
The condition for the eikonal approximation
to be good for NN collisions in both free space and nucleus-nucleus scattering
is that
\beq
v({\vrr})/e \ll 1, \quad  ka \gg 1,
\label{condition-s3}
\eeq
where $e$ ($k$) is the kinetic energy (wave number) of
the NN collision, ${\vrr}$ is the relative coordinate between two nucleons
and $a$ is the range of the realistic NN interaction $v$.
Obviously, this condition is not well satisfied, because $v$
has a strong short-range repulsive core;
for example, $v \sim 2,000$~MeV at $r=0$ for the AV18 force~\cite{Wir95}.
In fact, the eikonal approximation is not good
for NN scattering at intermediate energies,
as shown in the left panel of Fig.~\ref{fig:f-NN-s3}.
To avoid this problem, a slowly-varying function such as the Gaussian form
has been used as a profile function in the Glauber model~\cite{GM70}.
%
%---figure 1 and 2-----------------------------------------------------
%----------------------------------------------------------------

The use of a slowly-varying profile function and hence of
a slowly-varying NN interaction can be justified
by using the many-body Schr\"odinger equation
\eqref{schrodinger-effective-s3}.
Applying the adiabatic and eikonal approximations to
Eq.~\eqref{schrodinger-effective-s3}, one can obtain the $S$~matrix
of nucleus-nucleus scattering as
\bea
S=\exp\Big[
-\frac{i}{\hbar v_{\rm rel}}\sum_{ij}\int_{-\infty}^{\infty}dz_{ij} \tau_{ij}
\Big],
\label{S-effective-NN-s3}
\eea
where $v_{\rm rel}$ stands for a velocity of P relative to A and $z_{ij}$
is the $z$-component of the relative coordinate $\vrr_{ij}$
between two nucleons.
In general, $\tau_{ij}$ has much milder $r$ dependence than
the bare NN potential $v_{ij}$.
At high incident energies,
for instance,
$\tau_{ij}$ is reduced to the $t$~matrix of NN scattering
that is a product of $v_{ij}$
and the wave operator of the NN scattering.
When $v_{ij}$ has a strong repulsive core at small $r$,
the wave operator provides a large suppression there. This leads to the
fact that the $t$~matrix is a slowly-varying function of
$r$~\cite{Yah08}.
This is also the case with the $g$~matrix.
The $t$- or $g$-matrix is thus more suitable than $v_{ij}$
as an input of the Glauber model.
In fact, as shown in the right panel of Fig.~\ref{fig:f-NN-s3},
the eikonal approximation is quite good
for NN scattering at intermediate energies, say, 150~MeV,
when the $g$~matrix proposed by Jeukenne, Lejeune, and Mahaux
(JLM)~\cite{Jeu76} is adopted.
The use of the $g$-matrix interaction has another merit in the sense that
the effective interaction includes, at least in part, nuclear medium effects.

\subsection{Localization of nonlocal microscopic optical potential}
\label{sec3-3}

In this subsection, we focus our discussion on nucleon-nucleus
scattering for simplicity.
The multiple scattering theory proposed by Watson~\cite{Wat53}
was reformulated
by Kerman, McManus, and Thaler~\cite{Ker59}
as series expansion in terms of an underlying NN
$t$~matrix.
The expansion was developed as the spectator expansion~\cite{Ern77}, and
the corresponding first-order optical potential was
successful in reproducing
experimental data in a wide range of incident energies $E_{\rm in}$
from 65~MeV to 400~MeV~\cite{Chi95}.
The microscopic optical potential thus produced is complex and nonlocal.

%
%%------------------------------------------------------------------%%
%% Figure
%%------------------------------------------------------------------%%
%\begin{figure}[htb]
\begin{figure}[b]
\begin{center}
 \includegraphics[clip,width=0.85\textwidth]{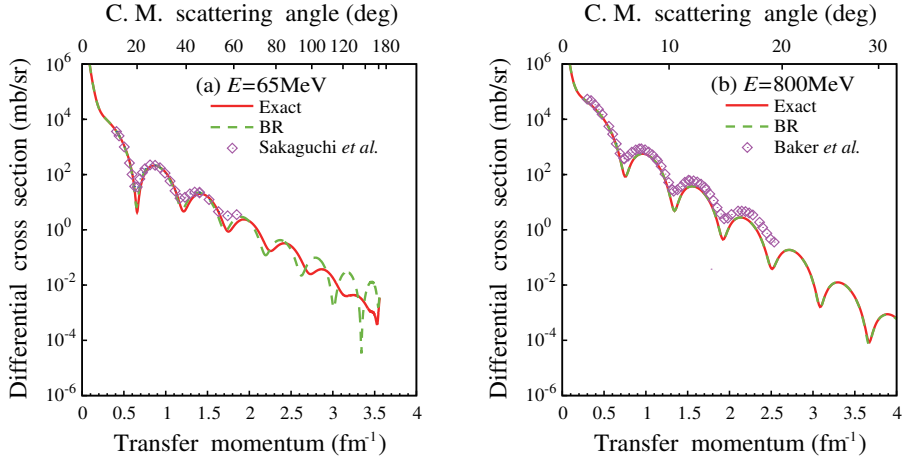}
 \caption{
 The differential cross sections of the proton elastic
 scattering from $^{90}$Zr
 at (a) 65~MeV and (b) 800~MeV.
The lower (upper) horizontal scale shows
the transferred wave number $q$ (the scattering angle ${\theta_{\rm cm}}$).
The solid curves represent the results of the exact calculation, whereas
the dashed lines correspond to those of the equivalent local potential.
Experimental data are taken from Refs.~\citen{Sak82} and \citen{Bak83}.
}
 \label{fig:elastic-Zr-s3}
\end{center}
\end{figure}%
%%------------------------------------------------------------------%%
%
Another way of obtaining the microscopic optical potential is
the single-folding model with the $g$-matrix
interaction~\cite{BR77,Ber77,Jeu76,Sat79,Sat83,Yam83,Rik84,Amo00,Fur08}.
The $g$-matrix interaction is evaluated
in infinite nuclear matter, and the potential between an incident nucleon and
a target is constructed by folding the $g$-matrix interaction with
the target density by using the local-density approximation.
This approach also is highly successful in reproducing
nucleon-nucleus elastic scattering data~\cite{Amo00,DA00}
for $40$~MeV $< E_{\rm in} < 800$~MeV from light to heavy targets.
The optical potential constructed in this approach is, again,
complex and nonlocal.

Thus, there have been several works on microscopic optical potentials.
The use of a nonlocal optical potential is, however, not practical
in many applications. This is true also in CDCC calculations,
since the nonlocality makes it difficult to
solve the CDCC equation \eqref{CC-eq-s2}.
In Ref.~\citen{BR77}, Brieva and Rook proposed
an approximate form of the equivalent
local potential. This Brieva-Rook (BR) localization has then been
commonly used in many studies~\cite{Rik84,Yam83,Fur08}, the
accuracy of which has not yet been examined numerically.

In Ref.~\citen{Min10}, we tested
the validity of the BR localization for the proton
elastic scattering from $^{90}$Zr
in a wide range of incident energies from 65 to 800~MeV.
As shown in Fig.~\ref{fig:elastic-Zr-s3},
the BR localization works very well for $q \la 2~{\rm fm}^{-1}$ at
65~MeV and for $q \la 4~{\rm fm}^{-1}$ at 800~MeV.
The BR localization is thus accurate in a wide range of
$65 \la E_{\rm in} \la 800$~MeV, unless $q$ is large.

%
%%------------------------------------------------------------------%%
%% Figure
%%------------------------------------------------------------------%%
\begin{figure}[htb]
\begin{center}
 \includegraphics[clip,width=0.85\textwidth]{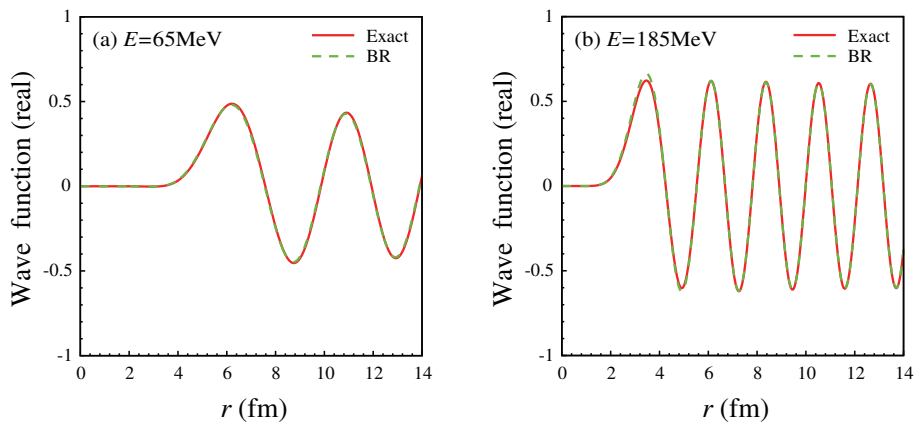}
 \caption{
 The real part of the exact and approximate wave functions
 for the p+$^{90}$Zr scattering at the grazing angular momentum.
 See the text for details.
 }
 \label{fig:WF-s3}
\end{center}
\end{figure}%
%%------------------------------------------------------------------%%
In Fig.~\ref{fig:WF-s3}, the real part of the exact wave function
(solid line) of
the Schr\"odinger equation with the nonlocal microscopic potential
is compared with that of the approximate wave function (dashed line) calculated
with the equivalent local potential
for the $p$+$^{90}$Zr scattering at the grazing angular momentum.
For both $E_{\rm in}=65$~MeV (left panel) and $185$~MeV (right panel),
the two results agree very well with each other.
The exact wave function is thus not
suppressed at small $R$ by the nonlocality of the optical potential, i.e.,
the Perrey suppression factor~\cite{PB62}
is not necessary in the BR-type localization.

\subsection{Application of the double-folding model
to the scattering of stable nuclei}
\label{sec3-4}

In this subsection, we consider the scattering of stable nuclei
from lighter targets at intermediate energies.
Then, projectile breakup and collective target excitations
are considered to be small, and consequently,
the double-folding model becomes reliable.
In the double-folding model, the potential $U$ is obtained as a sum of
the direct part $U^{\rm DR}$ and
the exchange part $U^{\rm EX}$~\cite{Sin75,Fur10}
\bea
\label{eq:UD-s3}
U^{\rm DR}(\vR) &=&
\sum_{\mu,\nu}\int \rho^{\mu}_{\rm P}(\vrr_{\rm P})
            \rho^{\nu}_{\rm A}(\vrr_{\rm A})
            g^{\rm DR}_{\mu\nu}(s;\rho_{\mu\nu}) d \vrr_{\rm P} d \vrr_{\rm A}, \\
\label{eq:UEX-s3}
U^{\rm EX}(\vR) &=& \sum_{\mu,\nu}
\int \rho^{\mu}_{\rm P}(\vrr_{\rm P},\vrr_{\rm P}-\vs)
\rho^{\nu}_{\rm A}(\vrr_{\rm A},\vrr_{\rm T}+\vs) \nonumber \\
            &&~~\times g^{\rm EX}_{\mu\nu}(s;\rho_{\mu\nu})
\exp{[-i\vK(\vR) \cdot \vs/M]}
            d \vrr_{\rm P} d \vrr_{\rm A}
            \label{U-EX-s3}
\eea
with $\vs=\vrr_{\rm P}-\vrr_{\rm A}+\vR$.
The relative coordinate between P and A is denoted by $\vR$,
and $\vrr_{\rm P}$ ($\vrr_{\rm A}$) is the coordinate of a nucleon in P (A)
from the center-of-mass (c.m.) of P (A);
$\mu$ ($\nu$) stands for the $z$-component of the isospin of a nucleon in
P (A). The original form of $U^{\rm EX}$ is nonlocal,
but it has been localized in Eq.~\eqref{U-EX-s3}
with the BR localization, where $M=A_{\rm P} A_{\rm A}/(A_{\rm P} +A_{\rm A})$
with $A_{\rm P}$ ($A_{\rm A}$) the mass number of P (A).
The direct part $g^{\rm DR}_{\mu\nu}$ and the exchange part
$g^{\rm EX}_{\mu\nu}$ of the $g$~matrix are assumed to depend
on the local density
\bea
 \rho_{\mu\nu}=\rho^{\mu}_{\rm P}(\vrr_{\rm P}-\vs/2)
 +\rho^{\nu}_{\rm A}(\vrr_{\rm A}+\vs/2)
\label{local-density approximation-s3}
\eea
at the midpoint of the interacting nucleon pair.
As for the projectile density $\rho^{\mu}_{\rm P}$
and the target density $\rho^{\nu}_{\rm A}$,
we use the phenomenological proton-density~\cite{Vri87} deduced
from electron scattering, and assume that the neutron density has
the same geometry as the corresponding proton one,
since the root-mean-square radii of proton and neutron
agree with each other with more than 99\% accuracy
in the Hartree-Fock (HF) calculation.

%
%%%%%%%%%%%%%%%%%%%%%%%
%%%  Figure
%%%%%%%%%%%%%%%%%%%%%%%
\begin{figure}[htbp]
\begin{center}
 \includegraphics[width=0.43\textwidth,clip]{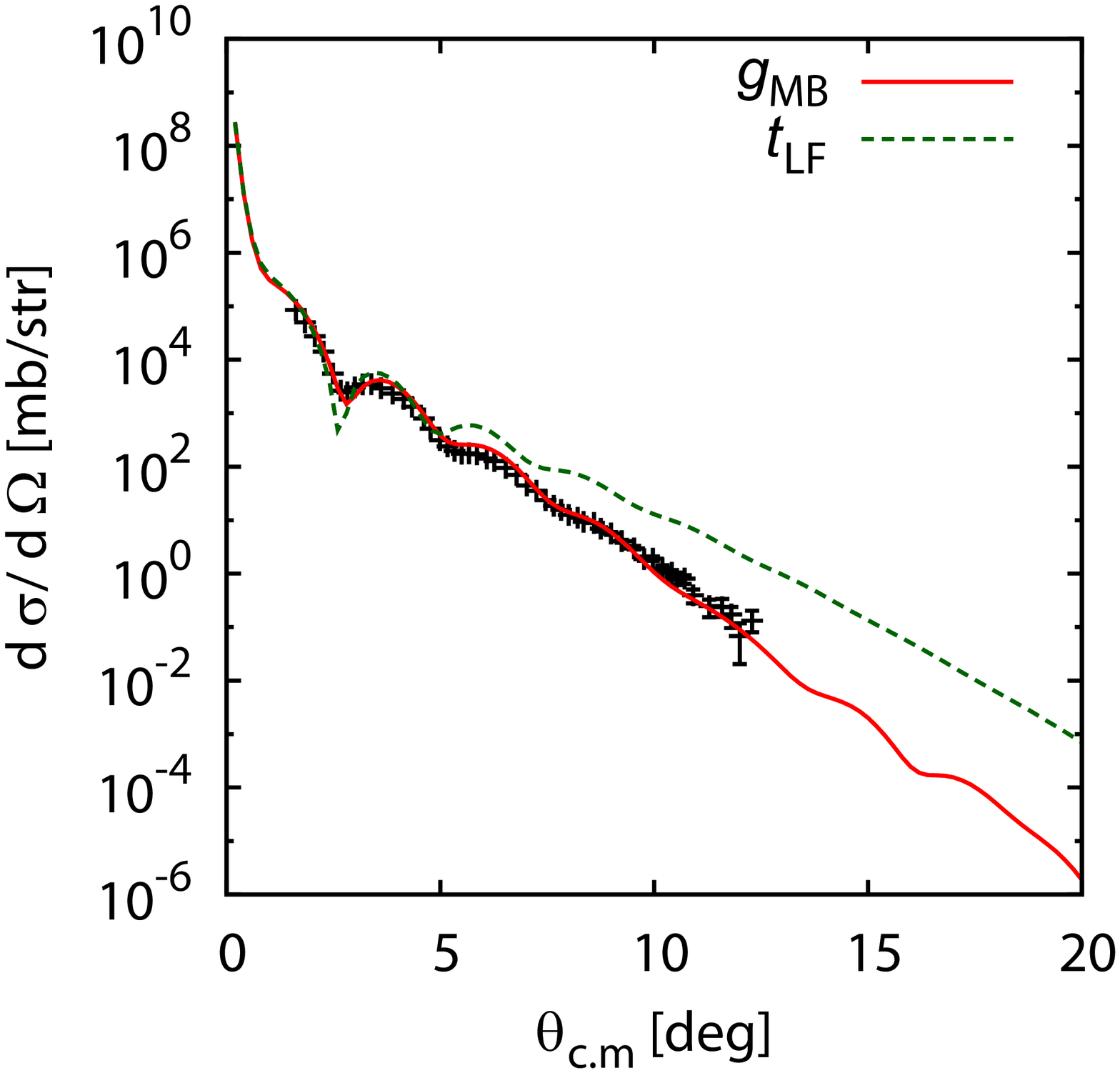}
 \includegraphics[width=0.43\textwidth,clip]{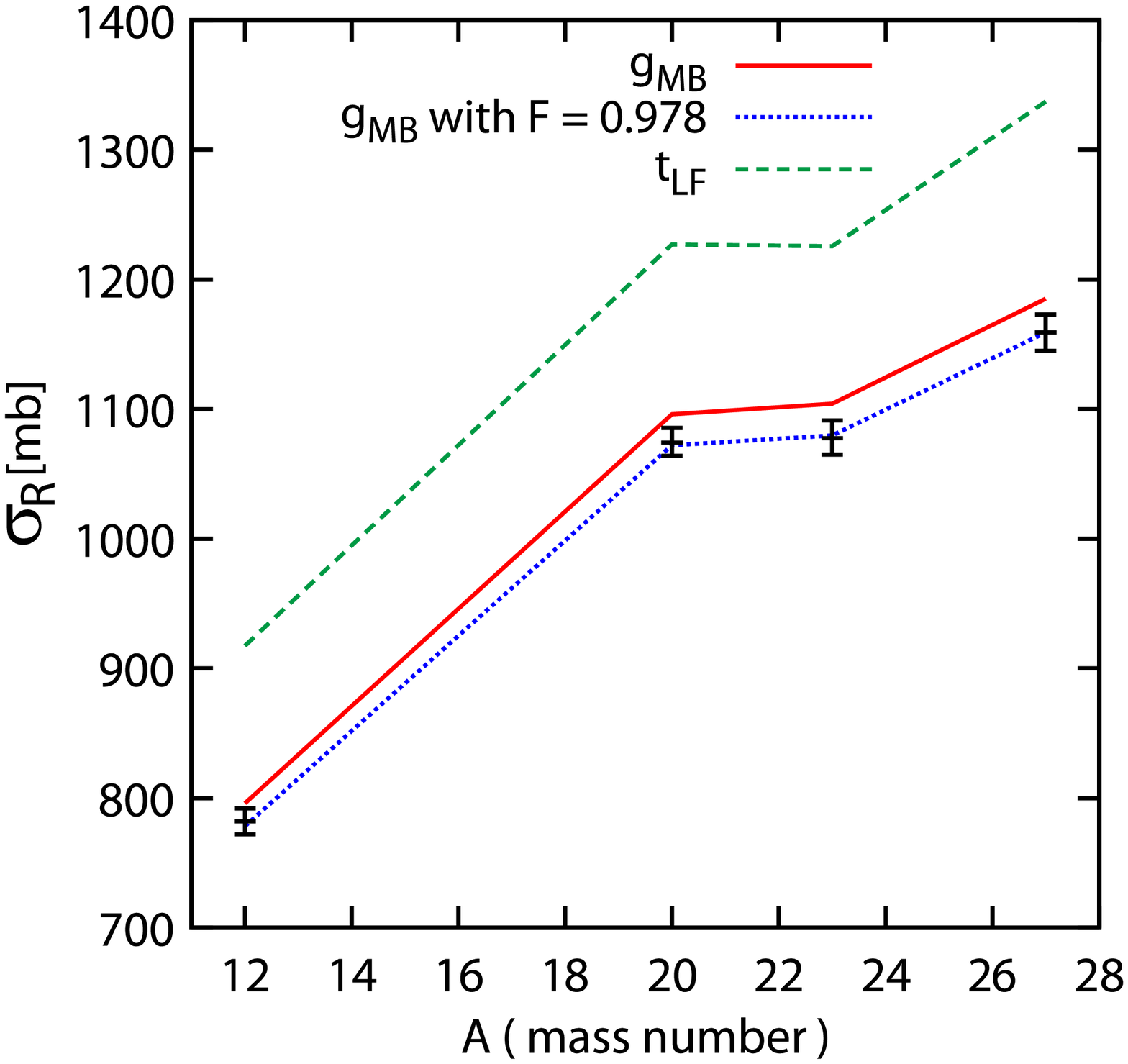}
 \caption{Angular distribution of the $^{12}$C+$^{12}$C
 elastic cross section at $135$~MeV/nucleon (left panel)
 and reaction cross sections
 for $^{12}$C scattering from $^{12}$C, $^{20}$Ne, $^{23}$Na, and $^{27}$Al
 (right panel).
 In each panel, the solid (dashed) line shows the result with
 $g_{\rm MB}$ ($t_{\rm LF}$). The dotted line in the right panel
 represents the result with $g_{\rm MB}$ multiplied by 0.978.
 The experimental data are taken from Ref.~\citen{Ich94} (left panel)
 and Refs.~\citen{Tak10,Suz95,Chu96,Tak09} (right panel).
 }
 \label{Fig-reaction-Xsec-stable-s3}
\end{center}
\end{figure}
We show in the left panel of Fig.~\ref{Fig-reaction-Xsec-stable-s3}
the angular distribution of the  $^{12}$C+$^{12}$C
elastic scattering at $135$~MeV/nucleon.
The folding model calculation with the Melbourne $g$~matrix ($g_{\rm MB}$)
reproduces well the data~\cite{Ich94}, whereas that with the
Love-Franey $t$-matrix interaction ($t_{\rm LF}$)~\cite{LF81} does not.
The medium effect is thus important, and the double-folding model
with $g_{\rm MB}$ is found to be quite reliable at intermediate incident energies.
The right panel shows reaction cross sections $\sigma_{\rm R}$
for $^{12}$C scattering from
$^{12}$C, $^{20}$Ne, $^{23}$Na, and $^{27}$Al targets at 250.8~MeV/nucleon.
The solid and dotted lines represent the results of folding model
calculations with $g_{\rm MB}$
before and after the normalization of $F=0.978$, respectively.
Before the normalization procedure, the results (solid line)
slightly overestimate the mean values of the experimental
data~\cite{Tak10,Suz95,Chu96,Tak09} for $A=20$--27.
After the normalization, the results (dotted line) agree
with the mean values for all the targets.
The dashed line corresponds to the results of double-folding model
calculations with $t_{\rm LF}$ and no normalization.
One sees clearly that the medium effect on $\sigma_{\rm R}$
is also significant at this energy.

\subsection{Application of the double-folding model
to reaction cross sections for Ne isotopes}
\label{sec3-5}

In this subsection, the double-folding model is applied to
the scattering of Ne isotopes from a $^{12}$C target at 240~MeV/nucleon.
The projectile densities are constructed by either (I)
antisymmetrized molecular dynamics (AMD)~\cite{Kim04}
with the Gogny D1S interaction~\cite{DG80,Ber91}
or (II) the deformed Woods-Saxon (DWS) model with
the deformation evaluated by AMD.
%
%%%%%%%%%%%%%%%%%%%%%%%
%%%  Figure
%%%%%%%%%%%%%%%%%%%%%%%
\begin{figure}[htbp]
\begin{center}
 \includegraphics[width=0.43\textwidth,clip]{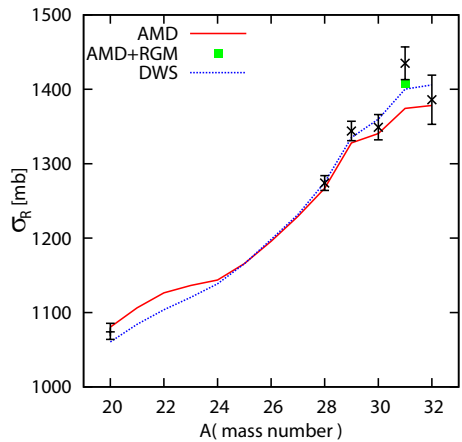}
 \caption{Reaction cross sections for the scattering of Ne
isotopes from a $^{12}$C target at 240~MeV/nucleon.
 The experimental data
 are taken from Refs.~\citen{Tak10} and \citen{Chu96}.
The dotted line represents the result of
the DWS model (method II),
whereas the solid line corresponds to the result of AMD calculations
(method I).
The closed square represents the result of the AMD-RGM calculation.
  }
 \label{Fig-AMD-HF-sigma-R-s3}
\end{center}
\end{figure}

Figure~\ref{Fig-AMD-HF-sigma-R-s3} represents the $\sigma_{\rm R}$
for the scattering of Ne isotopes
from a $^{12}$C target at 240~MeV/nucleon.
The result of the double-folding model with the AMD density (method I)
shown by the solid line reproduces well the data~\cite{Tak10}
except for $^{31}$Ne.
It turns out that deformation effects give a significant increase
in the $\sigma_{\rm R}$ and thus very important for the
reproduction of the data; even no bound state is obtained with
a spherical HF calculation for $A > 30$.

The underestimation of method I for $^{31}$Ne comes from
the inaccuracy of the AMD density in its tail region.
To remove this shortcoming, we describe the ground state
of $^{31}$Ne by a $^{30}$Ne+$n$ cluster model
including several configurations of the valence neutron
as well as excited states of $^{30}$Ne below 10~MeV.
We adopt the resonating group method (RGM) to include antisymmetrization
of nucleons explicitly.~\cite{Min12}
The result of this AMD-RGM calculation is shown by
the square symbol, which reproduces well the experimental
data~\cite{Tak10} with no adjustable parameter.
It is found that the tail correction gives
an increase in $\sigma_{\rm R}$ by 35~mb.
From these findings, we conclude that
$^{31}$Ne is a halo nucleus with large deformation.

Although the AMD-RGM calculation is highly successful,
it is quite time consuming. Thus, the DWS model (method II)
becomes an important alternative. As shown by the dotted line,
method II simulates well the results of method I for $A\neq 31$
and the result of AMD-RGM for $A=31$. An important advantage
of method II is that the tail correction mentioned above is
not necessary. This suggests that the DWS model is a handy and
promising way of simulating the AMD model
with the tail correction. The difference between method I and method II
for $^{28\mbox{--}32}$Ne may show the tail correction to method I.

Further investigations on
effects of breakup, dynamical deformation, and reorientation
were made in Refs.~\citen{Min11} and \citen{Sum12}.
The findings obtained on $\sigma_{\rm R}$ for
the $^{30,31}$Ne$+^{12}$C scattering
at 240~MeV/nucleon are summarized as follows.
\begin{enumerate}
\item[1)] The breakup effects of $^{31}$Ne to $^{30}$Ne$+n$ is 0.7\%.
\item[2)] Coupled-channel effects due to the deformation, i.e.,
couplings to the rotational states, of $^{30}$Ne is less than 1\%.
\item[3)] Difference between $\sigma_{\rm R}$ and the interaction
cross section $\sigma_{\rm I}$, i.e., the total inelastic cross section,
is 0.2\% for $^{30}$Ne.
\item[4)] The reorientation effect due to the intrinsic spin of
$^{31}$Ne$(3/2^{-})$ is 0.2\%.
\end{enumerate}

Thus, these so-called higher-order effects on $\sigma_{\rm R}$ are
found to be very small for the present reaction systems.
It should be noted, however, that
an elastic breakup or inelastic excitation to a rotational state
itself is an interesting subject of nuclear physics.
Then, CDCC or ordinary coupled-channel calculation for these
processes will be necessary.
Nevertheless, as mentioned in \S\ref{sec3-1}, the microscopic
reaction theory constructed is indispensable, because it gives
a theoretical foundation as well as input potentials
to these reaction calculations.

\section{Proposal of eikonal-CDCC (E-CDCC) as a method of treating
Coulomb breakup}
\label{sec4}

For some decades, Coulomb breakup (or Coulomb excitation)
processes have been regarded as one of the most effective tools to
investigate nuclear properties.~\cite{Ber09} The standard theory to
describe such processes is the virtual photon theory
(VPT),~\cite{Jac75,BB88}
or sometimes called equivalent photon method, which is based on the picture that
a nucleus moves along with its classical trajectory, and dissociates
by the electromagnetic field constructed by the target nucleus.
Although the VPT has been highly successful in many cases,
various higher-order processes that are not included in the VPT
can play some important roles, i.e.,
nuclear breakup, interference between nuclear and Coulomb
amplitudes, and multistep breakup processes mainly due to
strong continuum-continuum couplings, and so forth.
Quantitative evaluation of these contributions with more sophisticated
quantum mechanical models is very important
for genuine understanding of the {\lq\lq}Coulomb'' breakup processes.

The first application of CDCC to Coulomb breakup was done by
Hirabayashi and Sakuragi~\cite{HS92}
for $^{208}$Pb($^6$Li,$\alpha+d$) at
156~MeV. Later CDCC analyses of
$^{58}$Ni($^8$B,$^7$Be$+p$) at 26~MeV~\cite{Tos01} and
$^{208}$Pb($^8$B,$^7$Be$+p$) at 44 and 83~MeV/nucleon~\cite{Dav01}
were performed. The most essential characteristic of these CDCC
calculations is that the number of partial waves, $L_{\rm max}$,
between the projectile (P) and the target (A) is extremely large;
typically, $L_{\rm max}=15,000$ was taken
in Ref.~\citen{Dav01}. This is due to the
very long-range Coulomb coupling potentials, the dipole components
in particular.
These calculations definitely show the applicability of
CDCC to Coulomb breakup processes. However, inclusion of such
large numbers of partial waves is time-consuming and requires
very careful numerical treatment of the coupled-channel calculations.

In Ref.~\citen{Oga03} we proposed a new framework for accurately
and efficiently describing Coulomb breakup processes,
i.e., {\it eikonal-CDCC (E-CDCC)}.
Let us consider a reaction system of P and A, in which P consisting
of a core nucleus (c) and a valence particle (b).
E-CDCC uses the Coulomb-eikonal approximation to the scattering
waves between P and A. Consequently, the wave function of the
three-body system is given by
\beq
\Psi({\bm R},{\bm r})
=
\sum_{c}
\Phi_{c}({\bm r})
e^{-i(m-m_0)\phi_R}
\psi_{c}(b,z)
\phi_{{\bm K}_c}^{{\rm C}}(b,z),
\label{wf3b-s4}
\eeq
where $\Phi_{c}({\bm r})$ is the internal wave function of
P with $c$ the channel indices \{$i$, $\ell$, $m$\}; $i>0$
($i=0$) stands for the $i$th discretized-continuum (ground) state,
and $\ell$ and $m$\ are, respectively, the orbital angular momentum
between the constituents (c and b) of P and its
projection on the $z$-axis taken to be parallel to the incident
beam. $m_0$ is the value of $m$ in the incident channel.
We here neglect the internal spins of c and b for
simplicity; see Ref.~\citen{Oga06} for the inclusion of the intrinsic spins
of P in the E-CDCC formalism.
$b$ is the impact parameter (or transverse coordinate)
in the collision of P and A,
which is defined by $b=\sqrt{x^2+y^2}$ with ${\bm R} = (x,y,z)$,
the relative coordinate of P from A in the Cartesian representation.
Note that $\Psi({\bm R},{\bm r})$ properly takes into account the
dependence of the wave function on the azimuthal angle $\phi_R$
around the $z$-axis, which corresponds to the {\it coherent choice}
of the wave function in the terminology of the
dynamical eikonal approximation (DEA).~\cite{Bay05,Gol06}

The use of the Coulomb incident wave
$\phi_{{\bm K}_c}^{{\rm C}}(b,z)$
instead of the plane wave $\exp ({\bm K}_c \cdot {\bm R})$
in the eikonal approximation is one of the most important
features of E-CDCC;
${\bm K}_c$ is the asymptotic wave-number vector of P in channel $c$
from A.
In actual calculations we use the following approximate asymptotic form
\beq
\phi_{{\bm K}_c}^{{\rm C}}(b,z)
\approx
\dfrac{1}{(2\pi)^{3/2}}
e^{i(K_c z+\eta_c\ln{(K_c R-K_c z)})},
\label{Coulwf-s4}
\eeq
which is valid for large values of $b$, with
$\eta_c$ the Sommerfeld parameter for channel $c$.
We further assume
\beq
{\bm \nabla}_{\bm R}\phi_{{\bm K}_c}^{{\rm C}}(b,z)
\approx
iK_c(R) \phi_{{\bm K}_c}^{{\rm C}}(b,z)
\dfrac{{\bm z}}{z}
\eeq
with
\beq
K_c(R)
=
\sqrt{
K^2_c
-
\dfrac{2\mu_R}{\hbar^2}
\dfrac{Z_{\rm P}Z_{\rm A}e^2}{R}},
\label{klcl-s4}
\eeq
where $\mu_R$ is the reduced mass of P and A, and
$Z_{\rm P}e$ ($Z_{\rm A}e$) is the charge of P (A).

As in the usual eikonal approximation, the wave function
$\psi_{c}(b,z)$ is assumed to be a slowly varying function
and its second-order derivative $\Delta_{\bm R}\psi_{c}(b,z)$
is negligibly small. Then we obtain the following coupled-channel
equations
\begin{equation}
\dfrac{i\hbar^2}{\mu_R}K_c^{(b)}(z)
\dfrac{d}{d z}\psi_{c}^{(b)}(z) = \sum_{c'}
{\mathfrak{F}}^{(b)}_{cc'}(z) \; {\cal R}^{(b)}_{cc'}(z) \;
\psi_{c'}^{(b)}(z) \ e^{i\left(K_{c'}-K_c \right) z},
\label{cceq-s4}
\end{equation}
where
${\cal R}_{cc'}^{(b)}(z)=(K_{c'} R-K_{c'}
z)^{i\eta_{c'}}/ (K_c R-K_c z)^{i\eta_c}$.
The reduced coupling potential ${\mathfrak{F}}^{(b)}_{cc'}(z)$ is given by
\begin{equation}
{\mathfrak{F}}^{(b)}_{cc'}(z)
=
\left\langle
\Phi_{c}
|
U_{\rm c}+U_{\rm b}
|
\Phi_{c'}
\right\rangle_{\bm r} e^{i(m-m')\phi_R}
-\dfrac{Z_{\rm P}Z_{\rm A}e^2}{R}\delta_{cc'},
\label{FF-s4}
\end{equation}
where
$U_{\rm c}$ ($U_{\rm b}$) is the sum of the nuclear and Coulomb
interactions between c (b) and A.
The explicit form of the coupling potential
is given in Refs.~\citen{Oga03} and \citen{Oga06}.
Note that in Eq.~(\ref{cceq-s4})
$b$ is relegated to a superscript
since it is not a dynamical variable.

Solving Eq.~(\ref{cceq-s4}) under the boundary condition
\begin{equation}
\lim_{z \to -\infty}\psi_{c}^{(b)}(z)=\delta_{c0},
\end{equation}
where $0$ denotes the incident channel, one obtains the
following form of the eikonal scattering amplitude~\cite{Oga03}
\begin{equation}
f^{\rm E}_{c0}
=
f^{\rm Ruth}\delta_{c0}
+
\dfrac{2\pi}{i K_0}
\sum_L
f'^{\rm E}_{L;c0}\;
Y_{L \,m-m_0}(\hat{\bm K}'_c),
\label{fC7-s4}
\end{equation}
where $f^{\rm Ruth}$ is the Rutherford amplitude. The
partial scattering amplitude $f'^{\rm E}_{L;c0}$ is defined by
\begin{equation}
f'^{\rm E}_{L;c0}
=
\dfrac{K_0}{K_{c}}
{\cal H}^{(b_{c;L})}_c
\sqrt{\dfrac{2L+1}{4\pi}}
i^{(m-m_0)}
\left[
{\cal S}_{c0,L}
-
\delta_{c0}
\right],
\label{fC8-s4}
\end{equation}
where
\begin{equation}
{\cal S}_{c0,L} \equiv
\displaystyle
{\lim_{z \to \infty}}
\psi_{c}^{(b_{c;L})}(z)
\end{equation}
with $b_{c;L} \equiv (L + 1/2)/K_c$. The eikonal Coulomb-phase
factor ${\cal H}_{c,L}$ is defined by
\beq
{\cal H}_{c,L} = \exp[2i\eta_c\ln{(L+1/2)}].
\label{ECPF-s4}
\eeq

It should be noted that in the derivation of Eq.~\eqref{fC7-s4},
we have used
\beq
\int_0^{2\pi} e^{-i M \phi_R} e^{-iK_c b_{c;L} \theta \cos \phi_R} d\phi_R
=\dfrac{2\pi}{i^M} J_M((L+1/2)\theta)
\approx
2\pi i^M
\sqrt{\dfrac{4\pi}{2L+1}} Y_{LM} (\theta,0),
\eeq
with assuming $L \gg 1$ and small scattering angles $\theta$. Note also that
$\eta_c\ln{(L+1/2)}$ in the exponent of ${\cal H}_{c,L}$ is
the asymptotic form of the Coulomb phase shift $\sigma_L (\eta_c)$, i.e.,
\beq
\sigma_L (\eta_c) \to \eta_c\ln{(L+1/2)} \quad ({\rm for } \; L \gg \eta_c).
\eeq
Furthermore, the formal solution to Eq.~\eqref{cceq-s4} will be used
as the starting equation of the eikonal reaction theory (ERT)
described in \S\ref{sec6}.

The quantum mechanical (QM) correction in the scattering amplitude
can be performed if one replaces $f'^{\rm E}_{L;c0}$ for
small $L$, say, $L < L_{\rm C}$, with the QM partial amplitude
obtained by conventional QM CDCC,
\begin{eqnarray}
f'^{\rm Q}_{L;c0}
&\equiv&
\sum_{J=|L-\ell|}^{L+\ell}
\sum_{L_0=|J-\ell_0|}^{J+\ell_0}
\sqrt{\dfrac{2L_0+1}{4\pi}}
(\ell_0 m_0 L_0 0 | J m_0)
(\ell m L \;m_0\!\!-\!m| J m_0)
\nonumber \\
& &
\times
(S_{i L \ell,i_0 L_0 \ell_0}^{J}-\delta_{i i_0}
\delta_{L L_0}\delta_{\ell \ell_0})
e^{i(\sigma_L+\sigma_{L_0})}
(-)^{m-m_0},
\label{fCq-s4}
\end{eqnarray}
where $J$ is the total angular momentum of the three-body system
and $S$ is the scattering matrix.
This correction is valid if $f'^{\rm E}_{L;c0}\approx f'^{\rm Q}_{L;c0}$
with high precision
for $L\ge L_{\rm C}$; this is in fact the definition of $L_{\rm C}$.
We show how this is satisfied below (see Fig.~\ref{fig2-s4}).
Thus, we use the following {\it hybrid} scattering amplitude
\begin{equation}
f_{c0}
=
f^{\rm Ruth}\delta_{c0}
+
\dfrac{2\pi}{i K_0}
\sum_{L < L_{\rm C}}
f'^{\rm Q}_{L;c0}\;
Y_{L \,m-m_0}(\hat{\bm K}'_c)
+
\dfrac{2\pi}{i K_0}
\sum_{L \ge L_{\rm C}}
f'^{\rm E}_{L;c0}\;
Y_{L \,m-m_0}(\hat{\bm K}'_c),
\label{fh-s4}
\end{equation}
which enables one to perform coupled-channel calculations for
Coulomb and nuclear breakup processes with high computational speed,
freely from numerical difficulties due to extremely large values
of $L$, and with the same accuracy as of conventional QM CDCC.

%%%%%%%%%%%%%%%%%%%%%%%
%%%  Figure 1
%%%%%%%%%%%%%%%%%%%%%%%
\begin{figure}[htpb]
%\begin{figure}[t]
\centerline{
\includegraphics[width=1.0\textwidth,clip]{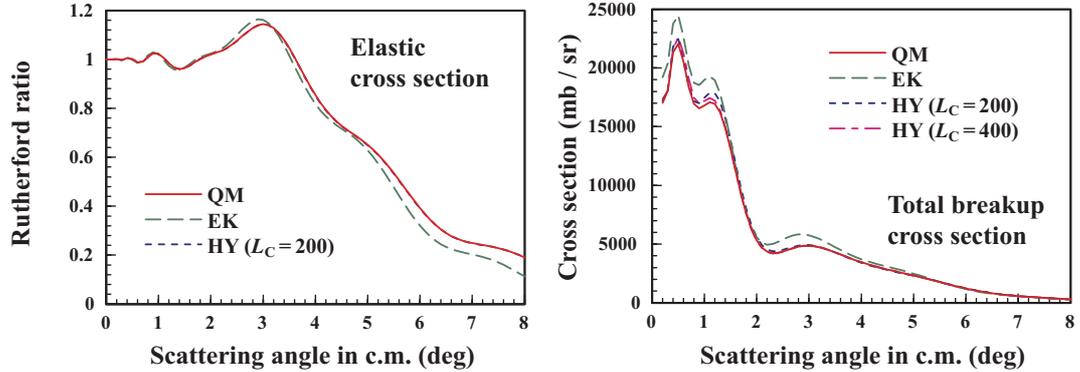}
}
\caption{\label{fig1-s4}
Angular distribution of
the elastic (left panel) and total breakup (right panel)
cross sections for
the $^8$B$+^{58}$Ni scattering at 240~MeV.
The solid and dashed lines
show, respectively, the results of the QM
and eikonal calculations.
The dotted (dash-dotted) line represents the hybrid result
with $L_{\rm C}=200$ (400).
}
\end{figure}
In the left and right panels of Fig.~\ref{fig1-s4}, respectively, we show
the elastic cross section (Rutherford ratio) and the total breakup cross
section for the $^8$B$+^{58}$Ni scattering at 240~MeV (30~MeV/nucleon),
as a function of the scattering angle $\theta$ in the center-of-mass (c.m.)
frame. Details of the numerical inputs are given in Ref.~\citen{Oga03};
we here remark only that $L_{\rm max}=4,000$ is taken in the calculation.
The solid, dashed, and dotted lines
represent the results of the QM,
eikonal, and hybrid calculations, where $L_{\rm C}$ is
taken to be 4,000, 0, and 200, respectively.
In the right panel the result of the hybrid calculation with $L_{\rm C}=400$
is also shown by the dash-dotted line.
The hybrid calculations with an appropriate
value of $L_{\rm C}$,
namely, 200 (400) for the elastic (breakup) cross
section, agree very well with the QM results;
the difference is only less than 1\%.
Another interesting finding is that the difference between the QM and eikonal
calculations is quite small up to even $\theta=8^\circ$.
This encourages one to use the eikonal approximation in analysis
of reactions around 30~MeV/nucleon. If very high accuracy is
required as in the study on determination of the astrophysical
factor $S_{17}$, however, inclusion of the QM correction
for lower partial waves is necessary (see \S\ref{sec8-1}).
In any case, use of the eikonal amplitude is very
effective to save computational time and eliminate numerical
difficulties.

\vspace{3mm}

It is well known that a theoretical description of breakup reactions
at intermediate energies requires a relativistic treatment of the
reaction dynamics, though only a relativistic modification on the
kinematics has been usually included.
In Refs.~\citen{OB09} and \citen{OB10},
we developed a full coupled-channel calculation
including a relativistic treatment of not only the kinematics but
also the dynamics, based on E-CDCC.
An essential ingredient of the relativistic coupled-channel calculation
is the proper treatment of nuclear and Coulomb coupling potentials
between P and A.
We adopt the form of the Coulomb dipole and quadrupole
interactions shown in Ref.~\citen{Ber05}, which is obtained from
a relativistic Li\'{e}nard-Wiechert potential with so-called
far-field approximation; the validity of this
approximation is confirmed in Ref.~\citen{OB10}.
As for the nuclear potential, the conjecture of Feshbach and Zabek
\cite{FZ77} is adopted.

It should be noted that E-CDCC can easily incorporate the
relativistic Coulomb and nuclear coupling potentials,
because it adopts the cylindrical coordinate representation.
On the other hand, inclusion of these potentials in conventional CDCC,
which is based on a partial wave decomposition in the spherical
coordinate representation, is very complicated.
Fortunately, however, the dynamical
relativistic corrections are found to be necessary only for large $L$,
where the scattering processes
are described well by the eikonal approximation.
%
%%%%%%%%%%%%%%%%%%%%%%%
%%%  Figure 2
%%%%%%%%%%%%%%%%%%%%%%%
\begin{figure}[htpb]
%\begin{figure}[t]
\centerline{
\includegraphics[width=70mm,keepaspectratio]{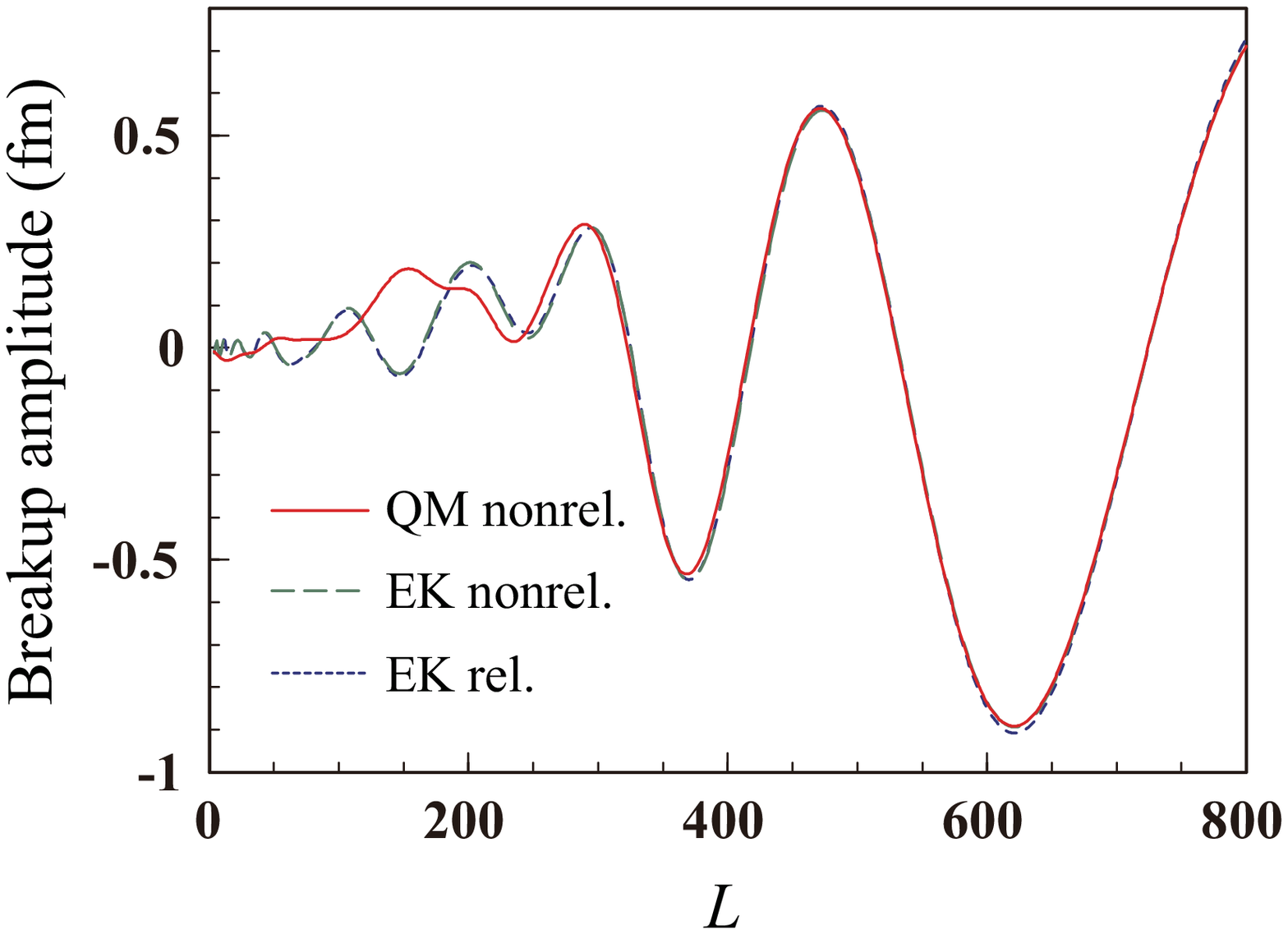}
\includegraphics[width=70mm,keepaspectratio]{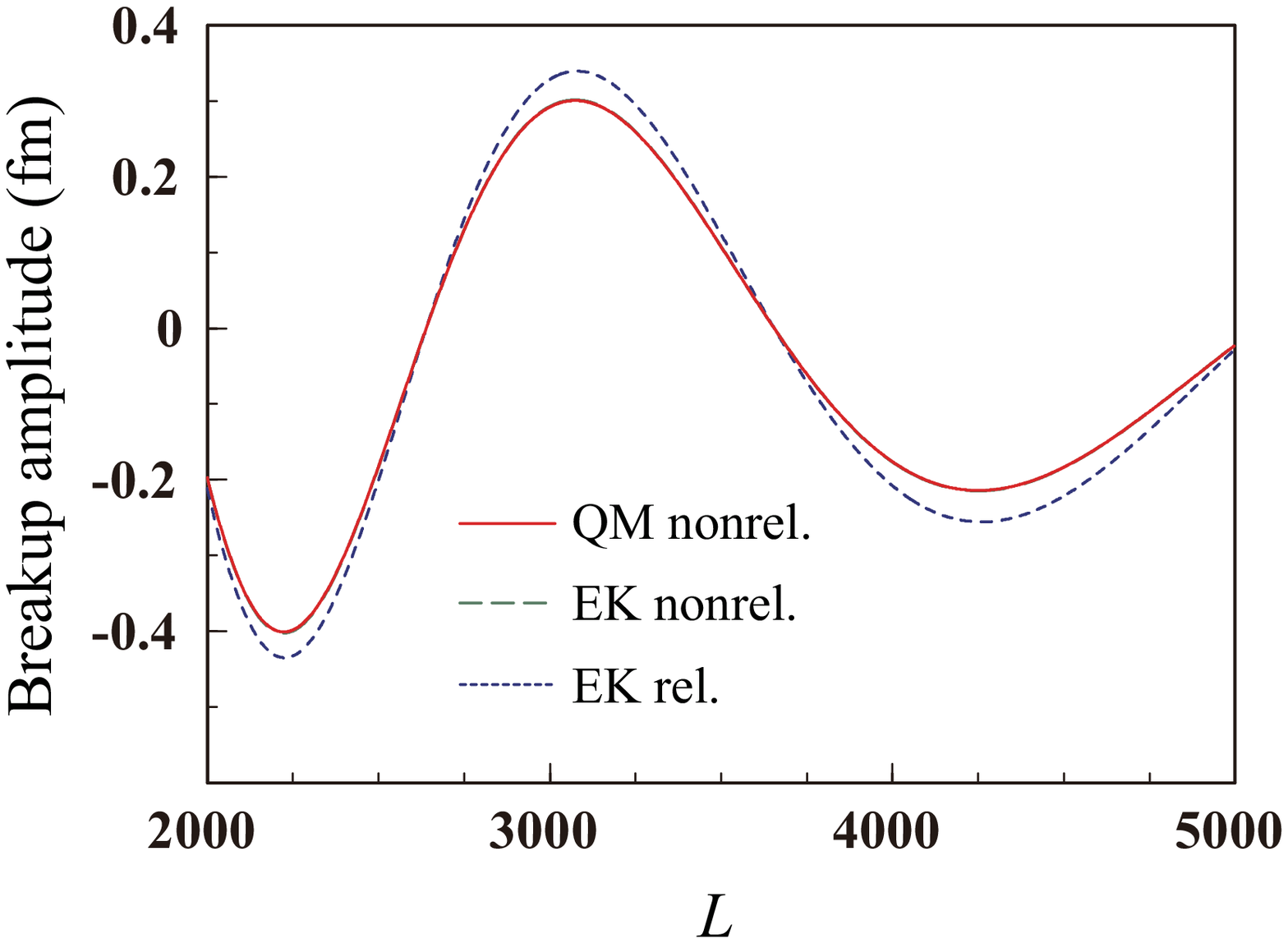}
}
\caption{\label{fig2-s4}
Real part of breakup amplitude for $^8$B$+^{208}$Pb at
250 MeV/nucleon as a function of the orbital angular momentum $L$
between $^8$B and $^{208}$Pb. The final state of $^8$B is chosen to
be the s-wave 6th bin state, and the $z$-component $m_0$ of the spin
of $^8$B in the incident channel chosen as 1. The solid, dashed, and
dotted lines show the results of nonrelativistic QM
CDCC, nonrelativistic E-CDCC, and relativistic E-CDCC,
respectively.
}
\end{figure}

Figure~\ref{fig2-s4}
shows the real part of the breakup amplitude for $^8$B$+^{208}$Pb at
250 MeV/nucleon as a function of $L$. As a breakup channel,
we choose the s-wave 6th bin state, whose breakup amplitude has
the largest value. For more details of numerical input, see
Ref.~\citen{OB10}. The solid
line represents the result of nonrelativistic QM CDCC,
which adopts Eq.~\eqref{fCq-s4} for all $L$ and
has no dynamical relativistic corrections.
The dotted and dashed lines are the
results of E-CDCC, based on Eq.~\eqref{fC7-s4},
with and without dynamical relativistic corrections, respectively.
One sees from the figure that at small
$L$, i.e., $L \la 300$, the dashed and dotted lines agree very well
with each other, and deviate from the solid line. On the other hand,
at large $L$, the solid and dashed lines show a very good agreement,
as desired,
and differ from the dotted line.
Thus, using the amplitude obtained by
nonrelativistic QM CDCC for small $L$ and that by relativistic
E-CDCC for large $L$ allows one to construct an accurate coupled-channel
framework that includes dynamical relativistic corrections and
QM effects, i.e., {\it relativistic CDCC}.

%
%%%%%%%%%%%%%%%%%%%%%%%
%%%  Figure 3
%%%%%%%%%%%%%%%%%%%%%%%
\begin{figure}[htpb]
%\begin{figure}[b]
\centerline{
\includegraphics[width=70mm,keepaspectratio]{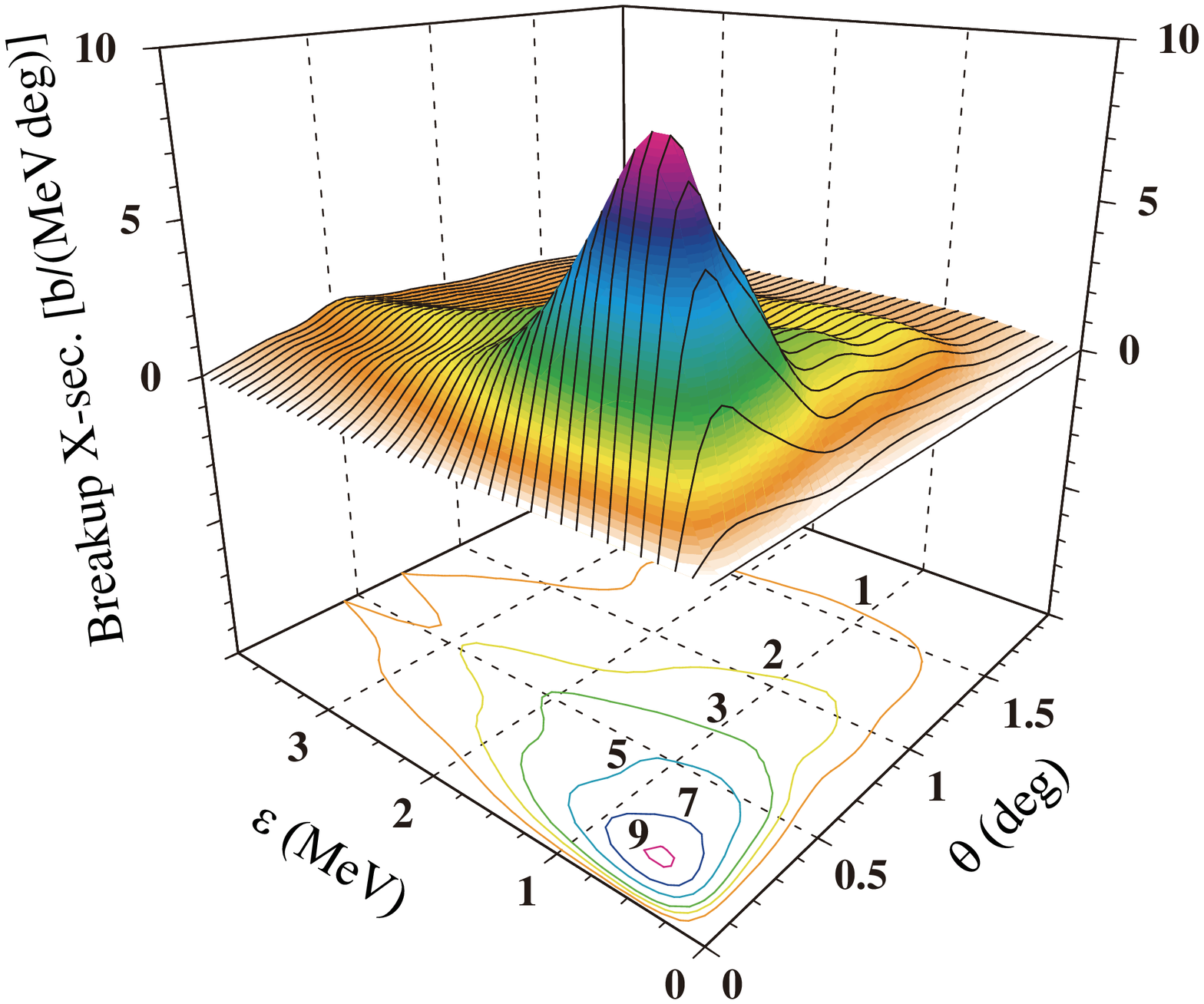}
\includegraphics[width=70mm,keepaspectratio]{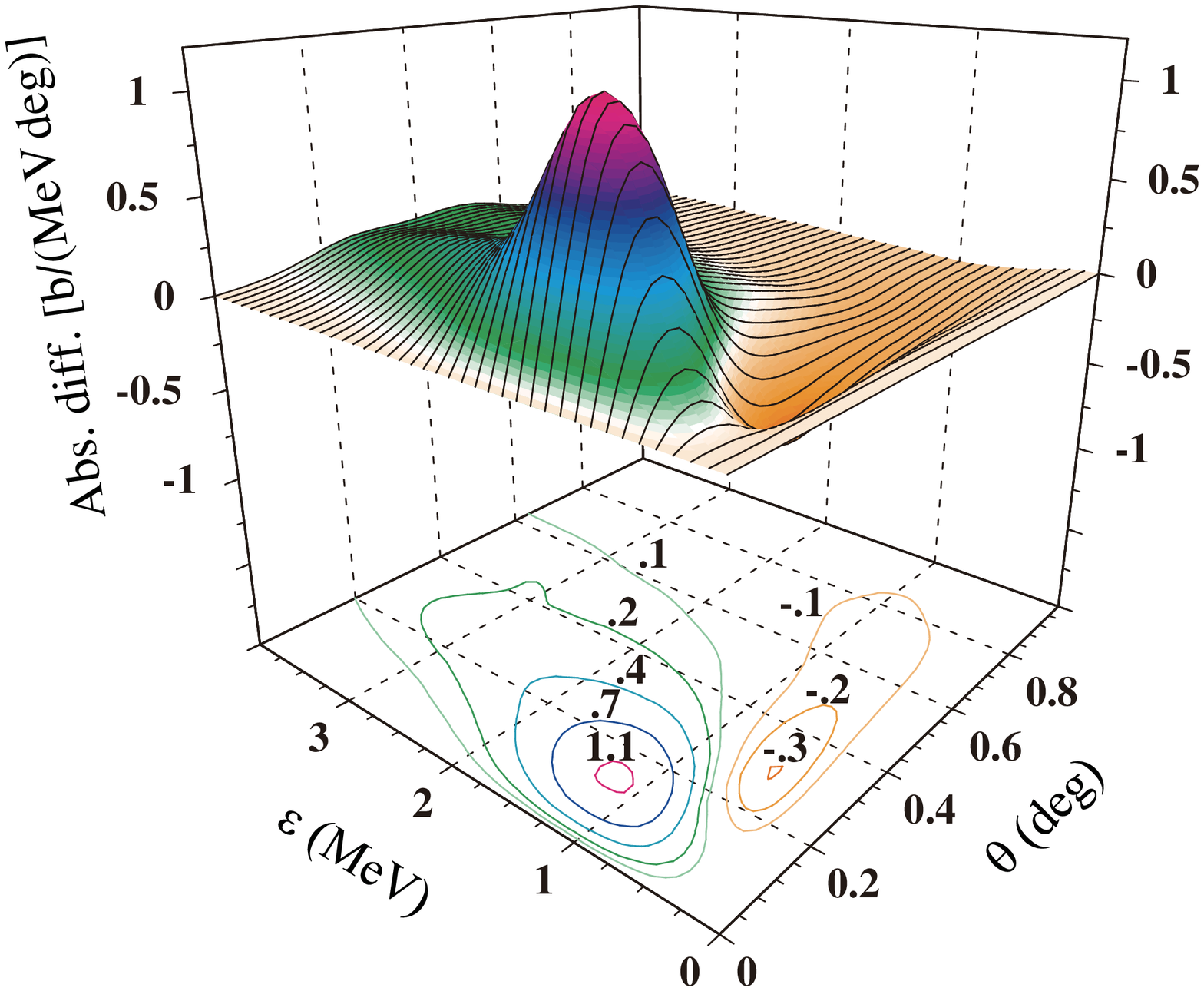}
} \caption{\label{fig3-s4} Double differential breakup cross section
(DDBUX) for $^8$B$+^{208}$Pb at 250 MeV/nucleon including dynamical
relativistic corrections (left panel), and its difference from the
calculation without relativistic corrections (right panel). }
\end{figure}
In the left panel of Fig.~\ref{fig3-s4}
we show the double differential breakup
cross section (DDBUX), $d^2 \sigma_{\rm BU}/(d\epsilon d \theta)$,
for the $^8$B$+^{208}$Pb reaction at 250 MeV/nucleon calculated with
relativistic CDCC based on Eq.~\eqref{fh-s4}.
$\epsilon$ is the relative energy of the two
fragments, $p$ and $^7$Be, of $^8$B after the breakup, and $\theta$ is the
scattering angle of the c.m. of $^8$B (the $p$ and $^7$Be system).
The right
panel shows the difference of the DDBUX
in the left panel from that
calculated with nonrelativistic CDCC, which also is
based on Eq.~\eqref{fh-s4} but including
no dynamical relativistic corrections.
Note that a relativistic treatment of the
kinematics is adopted in both calculations.
One observes a sizable
increase in the DDBUX due to relativity at forward angles ($\theta
\la 0.2^\circ$) and around $\epsilon=1$ MeV, where the DDBUX has
quite large values as shown in the left panel.
The increase is indeed large, i.e., several tens of \%, which
shows the importance of including dynamical relativistic corrections
in analysis of Coulomb breakup at intermediate energies.

\vspace{3mm}

Thus, we have developed E-CDCC for treating both nuclear
and Coulomb breakup very accurately and efficiently.
The scattering amplitude $f_{c0}$ to each channel $c$ is expressed
as a sum of the partial amplitudes $f'_{L;c0}$ with $L$ the
{\lq\lq}classical'' orbital angular momentum.
The QM corrections in
$f_{c0}$ can be made by replacing the $f'_{L;c0}$
with the corresponding CDCC results, and the replacement is necessary
only for small $L$.
E-CDCC calculations with the QM corrections are
much faster than CDCC calculations, with keeping
the same accuracy as of the latter. In addition to that,
E-CDCC makes it clear
the relation between the eikonal picture and the QM picture.
Furthermore, E-CDCC can incorporate a relativistic Li\'{e}nard-Wiechert potential
as a Coulomb interaction and the dynamical relativistic effects
generated.
E-CDCC has thus become a standard method
for describing breakup reactions in which the Coulomb breakup contribution
is essential.

\section{Proposal of four-body CDCC as a method of treating breakup
of three-body projectile}
\label{sec5}

In this section, we present a new version of CDCC to treat
four-body breakup reactions, in which the projectile breaks up
into three constituents, i.e., {\it four-body CDCC}.
In principle, CDCC is a method of treating many-body reaction
systems. However, except for three-body reactions
that involve two-body projectiles, it is difficult to solve
the CDCC equation. The main problem of the difficulty is how to
calculate a set of discretized continuum states for a many-body
projectile.

For the discretization procedure, as mentioned in \S\ref{sec2},
three methods have been proposed so far: the average (Av) method,
the midpoint (Mid) method, and the pseudostate (PS) method.
The Av method has been widely used and successful in describing
three-body breakup reactions.
The Mid method is shown to give the same results as of the Av
method.~\cite{Piy89,Piy99}
These two methods are, however, difficult to apply to
many-body reaction systems, where the projectile is to be
described as at least a three-body system.
The main difficulty lies on obtaining
the discretized continuum states of such projectiles;
both the Av method and the Mid method
require the exact many-body continuum wave functions of projectile.
An alternative to the Av and Mid methods is the PS
method~\cite{Mat03,Ega04,Mat04,Mat06,Mor01,Rod08}, in
which the continuum states are replaced by pseudostates obtained by
diagonalizing the internal Hamiltonian of the projectile in a space
spanned by $L^2$-type basis functions. One can adopt the transformed
harmonic oscillator (THO)~\cite{Mor01} or the
Gaussian~\cite{Mat03,Ega04} form as the basis functions. The
validity of the PS method was confirmed for scattering of two-body
projectiles by the agreement between CDCC solutions calculated with the
Av and PS methods~\cite{Mat03, Ega04}.

A great advantage of the PS method is that we do not need exact continuum
wave functions to obtain discretized continuum states, which is an
essential point to extend CDCC for many-body reaction systems.
CDCC with the PS method based on
Gaussian basis
functions was shown to be highly successful in describing the elastic
scattering of $^6$He, a typical three-body ($\alpha+n+n$) projectile,
at not only high energies but also low energies near the
Coulomb barrier~\cite{Mat04,Mat06}; see \S\ref{sec7-3} for the details.
Thus, we have developed a new framework, {\it four-body CDCC}, for
describing a reaction process of a three-body projectile and a target
nucleus, explicitly taking account of breakup channels of the
four-body reaction system.
Later, a four-body CDCC calculation with the THO basis
was also successfully carried out for low energy $^6$He elastic
scattering.~\cite{Rod08}

There is another difficulty in describing many-body breakup reactions,
i.e., how to obtain a smooth breakup cross section with respect to
the breakup energy $\varepsilon$ of the many-body system.
Because of the discretization of the continuum states of the projectile,
solving CDCC equations provides breakup $S$-matrix elements to the
discretized-continuum states, which give discrete
breakup cross sections. On the other hand, the breakup cross section
in reality is of course continuous as a function of $\varepsilon$.
Thus, one needs a way of smoothing the discretized
breakup $S$-matrix elements.

If we could adopt the Av method, the smoothing procedure
was straightforward, because of the clear relation between the
continuum state and the discretized continuum one.
In fact, there has been an attempt~\cite{Rod09} to
apply the Av method to $^6$He breakup processes by directly
calculating the three-body continuum state with
utilizing the hyperradial continuum wave functions.~\cite{Tho04}
Though this alternative four-body CDCC worked very well
for the elastic scattering of $^6$He, convergence of the breakup cross
section $d\sigma/d\ve$ was not obtained as indicated in
Fig.~5 of Ref.~\citen{Rod09}.

With the PS discretization, we can smooth the breakup cross
section by calculating the smoothing function
$\langle\psi^{(-)}(\varepsilon)|\Phi_{\gamma}\rangle$, where
$\psi^{(-)}(\varepsilon)$ is the exact continuum wave function
with the incoming boundary condition
and $\Phi_\gamma$ is the $\gamma$th discretized continuum state.
Details of this smoothing procedure are
shown in Refs.~\citen{Mat03} and \citen{Mat09} for three-body reactions
and in Ref.~\citen{Ega09} for four-body reactions.
This method is much more practical than the aforementioned
Av method for the three-body projectile, because we need only
the overlap of $\psi^{(-)}(\varepsilon)$ with the
spatially damping (compact)
wave function $\Phi_\gamma$. To obtain a convergence of
$d\sigma/d\ve$ with very high accuracy is, however, still
difficult at present.

In this situation, very recently,
we proposed a new smoothing procedure~\cite{Mat10}
with the complex-scaling method.~\cite{Agu71}
The advantage of this
method is that we do not need the exact continuum wave function
 at all, i.e., only the overlap between two compact wave
functions is required.
Below we describe this new method for the scattering of $^6$He 
by a target A, as a typical example of four-body reactions.

The Schr\"odinger equation for the four-body system of our
interest is given by
\begin{align}
[H-E_{\rm tot}]|\Psi^{(+)}\rangle =0,
\label{eq:4b-Schr-ex-s5}
\end{align}
where $(+)$ indicates the outgoing boundary condition and
$E_{\rm tot}$ is the total energy of the system.
The total Hamiltonian $H$ is defined by
\begin{align}
 H &=K_R+h_{\rm P}+U_{n\rm A}+U_{n\rm A}+U_{\alpha \rm A},
\end{align}
where the internal Hamiltonian $h_{\rm P}$ of $^6$He is
given by
\begin{align}
 h_{\rm P}&=K_{y}+K_{r}+V_{nn}+V_{n\alpha}+V_{n\alpha}
\end{align}
within the $\alpha+n+n$ three-body model adopted.
The relative coordinate between $^6$He and A is denoted by
$\bm{R}$ and the internal coordinates of $^6$He by
a set of Jacobi coordinates, $\bm{\xi}=(\bm{y},\bm{r})$.
The kinetic energy
operator associated with $\bm{R}$ ($\bm{\xi}$) is represented
by $K_{R}$ ($K_\xi$), $V_{xx'}$
is a nuclear interaction between $x$ and $x'$,
and $U_{x\rm A}$ is the sum of the nuclear and Coulomb potentials
between $x$ and A.

In CDCC with the PS method,
the reaction process is assumed to take place
in a model space
\begin{eqnarray}
 {\cal P}'=\sum_{\gamma}|\Phi_{\gamma}\rangle \langle \Phi_{\gamma}|,
 \label{eq:com-set-s5}
\end{eqnarray}
where the ${\Phi}_{\gamma}$ are the eigenstates obtained by
diagonalizing $h_{\rm P}$ by $L^2$-type basis functions.
Details of the calculation for $\Phi_\gamma$ are shown later
in this section.
The four-body Schr\"odinger equation is then solved
in the model space
\begin{align}
{\cal P}'[H-E_{\rm tot}]{\cal P}'|\Psi^{(+)}_{\rm CDCC} \rangle =0.
\label{eq:4b-Schr-s5}
\end{align}
This so-called model space assumption has already been justified
as mentioned in \S\ref{sec2}; there is no essential difference
between three-body CDCC and four-body CDCC in this point.
In fact, we have confirmed that
elastic and breakup cross sections calculated with
three-body CDCC or four-body CDCC converged with respect to 
expanding the model space~\cite{Mat03,Ega04,Mat04,Mat06}.

The exact $T$-matrix element to a breakup state
of the $\alpha+n+n$ system is given by
\begin{eqnarray}
T_{\ve}(\bm{p},\bm{k},{\bm{P}})=
 \langle \psi^{(-)}_\ve(\bm{p},\bm{k}) \chi^{(-)}_\ve(\bm{P})|
U-U^{\rm (Coul)}_{\rm ^6He}
|\Psi^{(+)}\rangle
, \label{exact-T-s5}
\end{eqnarray}
where $U=U_{n\rm A}+U_{n\rm A}+U_{\alpha\rm A}$ and
$U_{\rm ^6He}^{\rm (Coul)}$ is the Coulomb interaction between $^6$He
and A.
$\bm{P}$, $\bm{p}$, and $\bm{k}$ are the
asymptotic relative momenta conjugate to the coordinate
$\bm{R}$, $\bm{y}$, and $\bm{r}$, respectively. These three
momenta specify the kinematics of the four particles to be
detected in measurements.
The final-state wave functions,
$|\psi_\ve^{(-)}(\bm{p},\bm{k})\rangle$ and
$|\chi_\ve^{(-)}(\bm{P})\rangle$, with the incoming boundary condition,
are defined by
\begin{equation}
 \left[T_R+U^{\rm (Coul)}_{\rm ^6He}-(E_{\rm tot}-\ve)\right]
  |\chi^{(-)}_\ve(\bm{P})\rangle =0,
  \label{eq:Sch-A-s5}
\end{equation}
\begin{equation}
 \left[h_{\rm P}-\ve\right]|\psi_\ve^{(-)}(\bm{p},\bm{k}) \rangle=0,
  \label{eq:Sch-B-s5}
\end{equation}
where
$E_{\rm tot}-\ve=(\hbar P)^2/(2\mu_R)$
and
$\ve=(\hbar p)^2/(2\mu_{y})+(\hbar k)^2/(2\mu_{r})$;
the $\mu$ are the reduced masses regarding the relative coordinate
given in the subscript.
Inserting the approximate complete set Eq.~\eqref{eq:com-set-s5}
into Eq.~\eqref{exact-T-s5}, one can get with high accuracy
the $T$-matrix element~\cite{Mat03}
\begin{eqnarray}
T_{\ve}(\bm{p},\bm{k},{\bm{P}})
&\approx& \sum_{\gamma \ne 0}
\langle \psi_\ve^{(-)}(\bm{p},\bm{k}) |\Phi_{\gamma}\rangle
T_{\gamma}
\label{approx-T-s5}
\end{eqnarray}
with the CDCC $T$-matrix element
\begin{eqnarray}
T_{\gamma}&=&\langle \Phi_{\gamma}\chi^{(-)}_{\ve_\gamma}(\bm{P}_\gamma)|
 U-U^{\rm (Coul)}_{\rm ^6He}
 |\Psi^{(+)}_{\rm CDCC}\rangle
\end{eqnarray}
to the $\gamma$th discrete breakup state $\Phi_{\gamma}$
that has the eigenenergy $\ve_\gamma$.
Equation~\eqref{approx-T-s5} has been derived by replacing $\bm{P}$ with
$\bm{P}_\gamma$ in $\chi^{(-)}_\ve(\bm{P})$.
The $T_{\gamma}$ are obtainable by CDCC, but
it is quite hard to calculate
the smoothing function
$\langle\psi_\ve^{(-)}(\bm{p},\bm{k})|\Phi_{\gamma}\rangle $ directly
as mentioned above.
Fortunately, however, one can find a way of circumventing this difficulty
for the calculation of the double differential breakup
cross section (DDBUX) $d^2\sigma /(d\ve d\Omega_{\bm{P}})$.

Using Eq.~\eqref{approx-T-s5}, one can rewrite
the DDBUX as
\begin{eqnarray}
\frac{d^2\sigma}{d\ve d\Omega_{\bm{P}}} =
\frac{(2\pi)^4\mu_R^2}{\hbar^4}\frac{P}{P_0}
\int d \vp' d \vk'
\delta(\ve-\ve')
|T_{\ve'}(\vp',\vk',{\bm{P}'})|^2
\approx \frac{1}{\pi}{\cal R}(\ve,\Omega_{\bm{P}})
\label{xsec-1-s5}
\end{eqnarray}
with the generalized response function
\begin{eqnarray}
 {\cal R}(\ve,\Omega_{\bm{P}})
&=& \frac{(2\pi)^4\mu_R^2}{\hbar^4}\frac{P}{P_0}
{\rm Im} \Big[
\sum_{\gamma,\gamma'\ne 0} T_{\gamma}^{*}
\langle \Phi_{\gamma}|G^{(-)}|\Phi_{\gamma'}\rangle
T_{\gamma'}
\Big]
,\label{response-fun-s5}
\end{eqnarray}
where $G^{(-)}$ is the propagator given by
\begin{eqnarray}
  G^{(-)}&=&\lim_{\eta \to +0} \frac{1}{\ve-h_{\rm P} - i\eta}.
\end{eqnarray}
The propagator $G^{(-)}$ operates only
on spatially damping functions $\Phi_{\gamma}$.
This makes the
calculation of $\langle \Phi_{\gamma}|G^{(-)}|\Phi_{\gamma'}\rangle$ feasible.

We now use the complex-scaling method in which
the scaling transformation operator $C(\theta)$ and its inverse are defined by
\begin{eqnarray}
\langle \vrr,\vy | C(\theta)|f \rangle
&=& e^{3i\theta}f(\vrr e^{i\theta},\vy e^{i\theta}), \\
\langle f | C^{-1}(\theta)| \vrr,\vy \rangle
 &=& \{e^{-3i\theta}f(\vrr e^{-i\theta},\vy e^{-i\theta}) \}^{*}.
\end{eqnarray}
Using these operators, one can get
\begin{eqnarray}
\langle \Phi_{\gamma}|G^{(-)}|\Phi_{\gamma'}\rangle
=\langle \Phi_{\gamma}|C^{-1}(\theta) G_{\theta}^{(-)}C(\theta)|
\Phi_{\gamma'}\rangle ,
\label{C-G-1-s5}
\end{eqnarray}
where
\begin{eqnarray}
G_{\theta}^{(-)}=\lim_{\eta \to +0} \frac{1}{\ve-h^{\theta}_{\rm P} - i\eta}
\end{eqnarray}
with
$h_{\rm P}^\theta=C(\theta)h_{\rm P}C^{-1}(\theta)$.
When $-\pi<\theta<0$,
the scaled propagator $\langle\bm{\xi}|G^{(-)}_\theta|\bm{\xi}'\rangle$
is a damping function of $\bm{\xi}$ and $\bm{\xi}'$.
Note that positive $\theta$ has been taken so far in the
complex-scaling method, but in the present situation
it must be negative since $G^{(-)}$ has the incoming boundary condition. 
The scaled propagator $G_{\theta}^{(-)}$ is a $L^2$-type 
function~\cite{Mat09}, and is then expanded with $L^2$-type basis functions: 
\begin{eqnarray}
G^{(-)}_\theta &\approx& \sum_i
  \frac{|\phi^\theta_i\rangle\langle\tilde{\phi}_i^\theta|}
  {\ve-\ve_i^\theta},
  \label{Gtheta2-s5}
\end{eqnarray}
where $\phi_i^\theta$ is the $i$th eigenstate of
$h_{\rm P}^\theta$ in a model space spanned by $L^2$-type basis
functions, i.e.,
$\langle\tilde{\phi}_i^\theta|h_{\rm P}^\theta
|\phi_{i'}^\theta\rangle=
\ve_i^\theta \delta_{ii'}$.
Inserting Eq.~\eqref{C-G-1-s5} with Eq.~\eqref{Gtheta2-s5} into
Eq.~\eqref{response-fun-s5}
leads to a desired form
\begin{eqnarray}
 \frac{d^2\sigma}{d\ve d\Omega_{\bm{P}}}&\approx&
  \frac{1}{\pi}
  \frac{(2\pi)^4\mu_R^2}{\hbar^4}\frac{P}{P_0}
  {\rm Im}\sum_i\frac{T_{i}^\theta \tilde{T}_i^\theta}
  {\ve-\ve_i^\theta}
  \label{sigma-CSM-s5}
\end{eqnarray}
with
\begin{equation}
 \tilde{T}_i^\theta \equiv
 \sum_{\gamma'} \langle \tilde{\phi}_i^\theta|C(\theta)
 |\Phi_{\gamma'} \rangle
 T_{\gamma'}, \quad
  T_i^\theta\equiv
  \sum_\gamma
  T_\gamma^*   \langle \Phi_\gamma|C^{-1}(\theta)|\phi_i^\theta \rangle.
  \label{R-s5}
\end{equation}
Equation~\eqref{sigma-CSM-s5} contains no exact three-body continuum states
$\psi_{\ve}^{(-)}(\vk,\vp)$.
This is the most important advantage of this new smoothing procedure.

It should be noted that the use of the response
function in the complex-scaling method to obtain smooth breakup
spectra has already been done in some previous studies.~\cite{Myo99,Suz10}
The essential feature of the present study is
the extension of the method for genuine reaction calculations.
As mentioned above, the $T$-matrix element $T_\gamma$
in Eq.~\eqref{R-s5}
is the solution of the CDCC calculation for the
scattering of a many-body projectile by a target nucleus,
which thus contains all degrees of
freedom of reaction dynamics within the model space adopted.

\vspace{3mm}

%
%%%%%%%%%%%%%%%%%%%%%%%
%%%  Figure 1 (Sec. 5)
%%%%%%%%%%%%%%%%%%%%%%%
\begin{figure}[htbp]
\begin{center}
 \includegraphics[width=0.4\textwidth,clip]{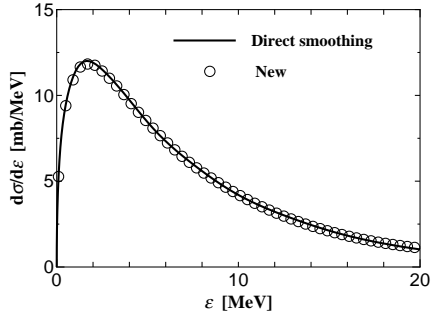}
 \caption{Breakup cross section for the $d+^{58}$Ni scattering
 at 80 MeV. The solid line is the result of the direct smoothing
 method and the open circles show the result of the new method. }
 \label{fig1-s5}
\end{center}
\end{figure}
We tested first the validity of Eq.~\eqref{sigma-CSM-s5} for a
three-body breakup
reaction, for which the exact
breakup $T$-matrix element
$T(\vk,\bm{P})=\sum_{n}\langle\psi^{(-)}(\bm{k})|{\Phi}_{n}\rangle{T_{n}}$
is obtainable
by taking the overlap
$\langle \psi^{(-)}(\bm{k})|{\Phi}_{n}\rangle$ directly.~\cite{Mat03}
This approach is referred to as
{\it the direct smoothing method} below.
As an example, we consider the $^{58}$Ni($d$, $pn$) reaction at 80~MeV;
see Ref.~\citen{Mat03} for the details of the CDCC calculation.
Figure~\ref{fig1-s5} shows the differential breakup cross
section $d\sigma/d\ve$, which is obtained by integrating the
DDBUX of
Eq.~\eqref{sigma-CSM-s5} over the solid angle
$\Omega_{\bm P}$. The new method (open circles)
yields the same result as the direct smoothing method (solid line).
The validity of the new method is thus confirmed.

Next, the new method is applied to the $^{12}$C$(^6$He, $\alpha nn)$
reaction at 229.8~MeV.
As for the interactions $V_{nn}$ and $V_{n\alpha}$ in $h_{\rm P}$, we take the
GPT~\cite{Gog70} and KKNN~\cite{Kan79} potentials, respectively.
These potentials with a Gaussian form
reproduce well experimental data of low-energy nucleon-nucleon and
nucleon-$\alpha$ scattering.
The particle exchange between valence neutrons and neutrons
in $\alpha$ is treated approximately with the orthogonality condition
model~\cite{Sai69}.
\begin{figure}[htbp]
\begin{center}
 \includegraphics[width=0.6\textwidth,clip]{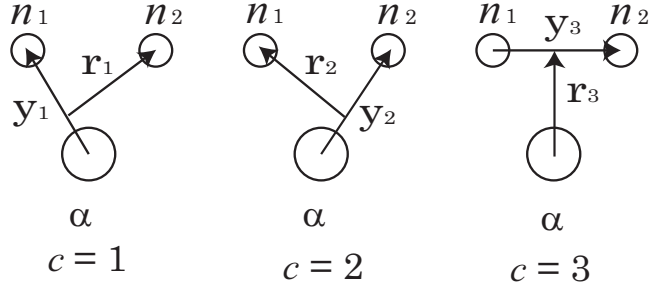}
 \caption{Jacobi coordinates of three rearrangement
 channels ($c=1\mbox{--}3$) of $^6$He.}
 \label{fig2-s5}
\end{center}
\end{figure}

For the diagonalization of $h_{\rm P}$ and $h_{\rm P}^\theta$, we adopt
the Gaussian expansion method (GEM)~\cite{Hiy03}. In GEM,
the $\alpha+n+n$ system is described by
a superposition of three channels, each of which corresponds to
a different set of Jacobi coordinates $(\vy_c, \vrr_c)$, as shown in
Fig.~\ref{fig2-s5}. For each channel ($c$),
the radial parts of the internal wave functions are
expanded by a finite number of Gaussian basis functions
\begin{equation}
 \varphi_{j\lambda}(\bm{y}_c)=y_c^\lambda e^{-(y_c/\bar{y}_j)^2}
  Y_\lambda(\Omega_{y_c}), \quad
 \varphi_{i\ell}(\bm{r}_c)=
  r_c^\ell e^{-(r_c/\bar{r}_i)^2}Y_\ell(\Omega_{r_c}),
\end{equation}
where $\lambda$ ($\ell$)
is the orbital angular momentum regarding $\bm{y}_c$ ($\bm{r}_c$).
The range parameters $\bar{y}_j$ and $\bar{r}_i$ are
taken to lie in geometric progression
\begin{equation}
 \bar{y}_j=(\bar{y}_{\rm max}/\bar{y}_{\rm 1})^
  {(j-1)/j_{\rm max}}, \quad
 \bar{r}_i=(\bar{r}_{\rm max}/\bar{r}_{\rm 1})^
  {(i-1)/i_{\rm max}}.\label{para-s5}
\end{equation}
The parameters actually depend on $c$, but we omitted the dependence
in Eq.~\eqref{para-s5} for simplicity.
\begin{table}[htbp]
 \caption{Gaussian range parameters.}
 \begin{center}
 \begin{tabular}{c|ccccccc}
  \hline
  Set&$c$&$j_{\rm max}$&
  $\bar{y}_1$ (fm)&$\bar{y}_{\rm max}$ (fm)&$i_{\rm max}$
  &$\bar{r}_1$ (fm)&$\bar{r}_{\rm max}$ (fm)
  \\ \hline\hline
  \multirow{2}{*}{I} &3&10&   0.1&       10.0& 10&   0.5&  10.0   \\
  &1, 2&10&   0.5&       10.0&10&    0.5&  10.0 \\ \hline
 \multirow{2}{*}{II} &3 &15&   0.1&       20.0& 15&   0.5&  20.0   \\
  &1, 2&15 &  0.5&       20.0& 15&   0.5&  20.0 \\ \hline
  \multirow{2}{*}{III} &3 &20&  0.1&       50.0& 20&   0.5&  50.0   \\
  &1, 2&20&  0.5&       50.0&20&    0.5&  50.0 \\ \hline
 \end{tabular}
 \end{center}
\label{basis-para}
\end{table}
In order to confirm the convergence of the breakup cross section
with respect to the size of the model space, we prepare the
three sets
of basis functions shown in Table~\ref{basis-para}.
For the $0^+$ and ${1}^-$ states,
maximum internal angular momenta
$\ell_{\rm max}$ and $\lambda_{\rm max}$ are both set to unity.
For the $2^+$ states, we take
$\ell_{\rm max}=\lambda_{\rm max}=1$ for $c=1$ and 2,
and $\ell_{\rm max}=\lambda_{\rm max}=2$ for $c=3$.
%\bf
For other numerical inputs, see Ref.~\citen{Mat10}.

%
%%%%%%%%%%%%%%%%%%%%%%%
%%%  Figure 3
%%%%%%%%%%%%%%%%%%%%%%%
\begin{figure}[htbp]
\begin{center}
 \includegraphics[width=0.32\textwidth,clip]{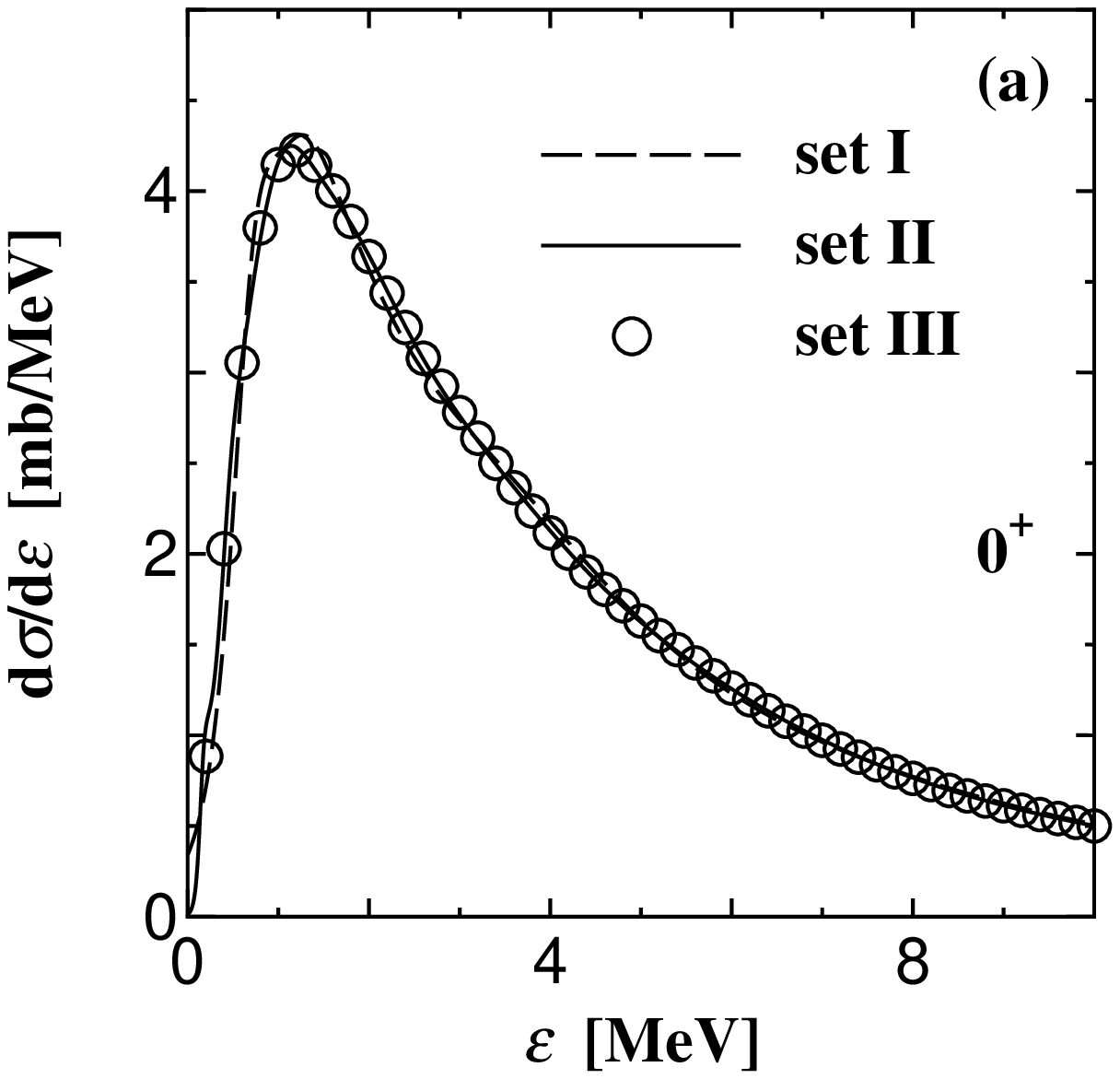}
 \includegraphics[width=0.32\textwidth,clip]{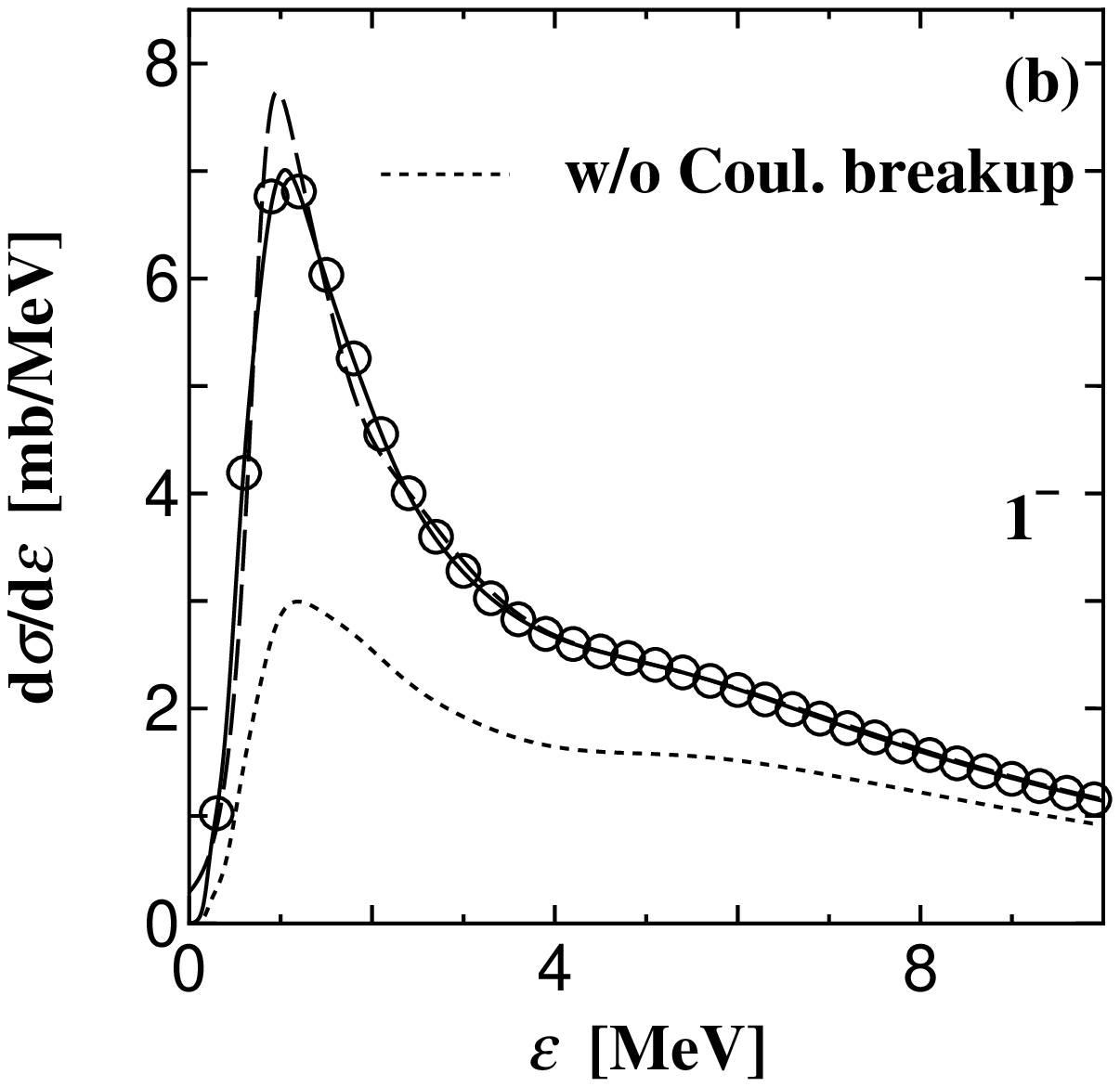}
 \includegraphics[width=0.32\textwidth,clip]{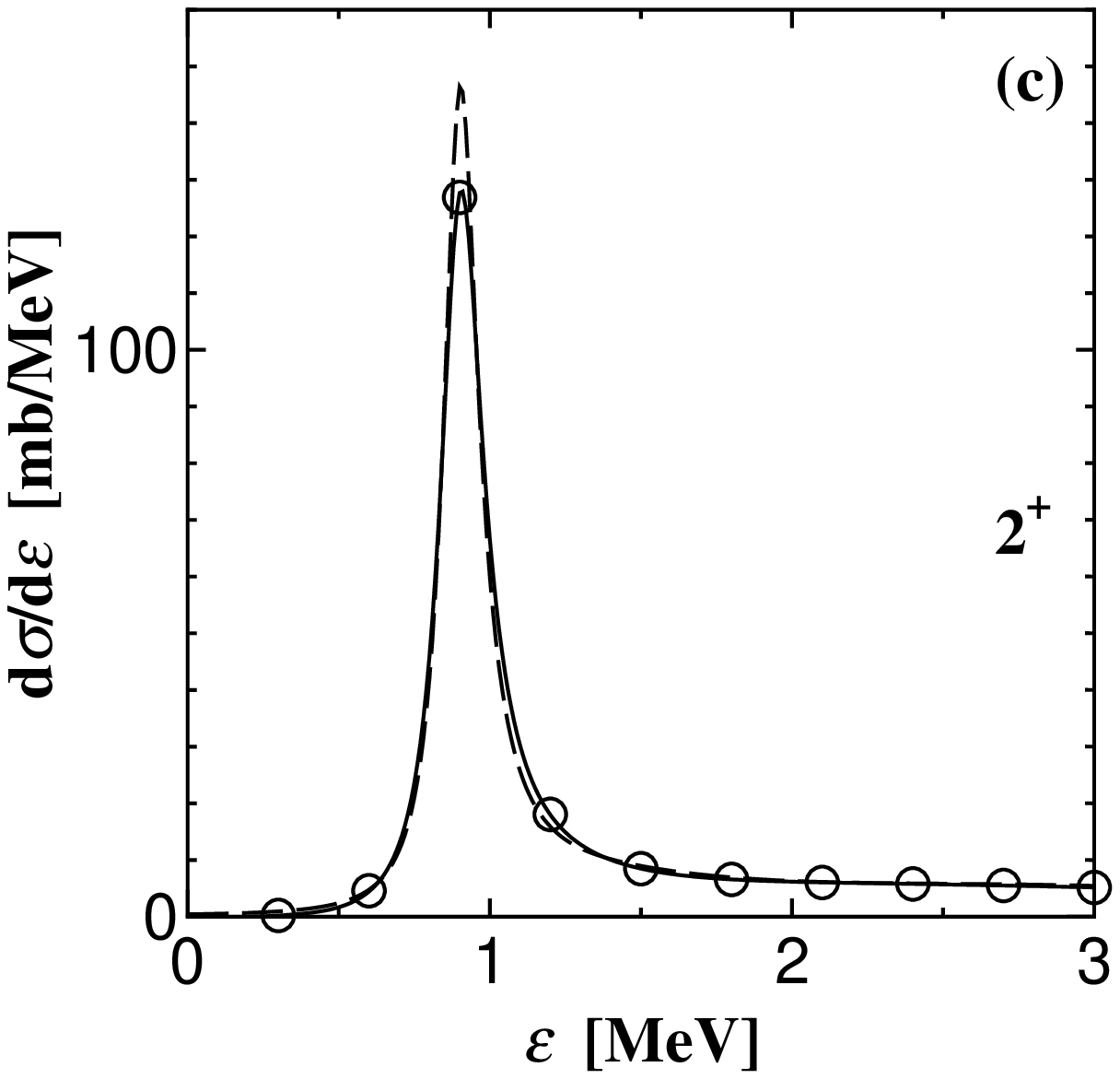}
 \caption{Convergence of the breakup cross sections to the $0^+$ (a),
 $1^-$ (b), and $2^+$ (c) continua. In each panel, the dashed
 line, the solid  line, and the open circles correspond to results of
 sets I, II, and III, respectively. The dotted line in (b) shows the
 result when Coulomb breakup processes are switched off.}
 \label{fig3-s5}
\end{center}
\end{figure}
\begin{figure}[htbp]
\begin{center}
 \includegraphics[width=0.32\textwidth,clip]{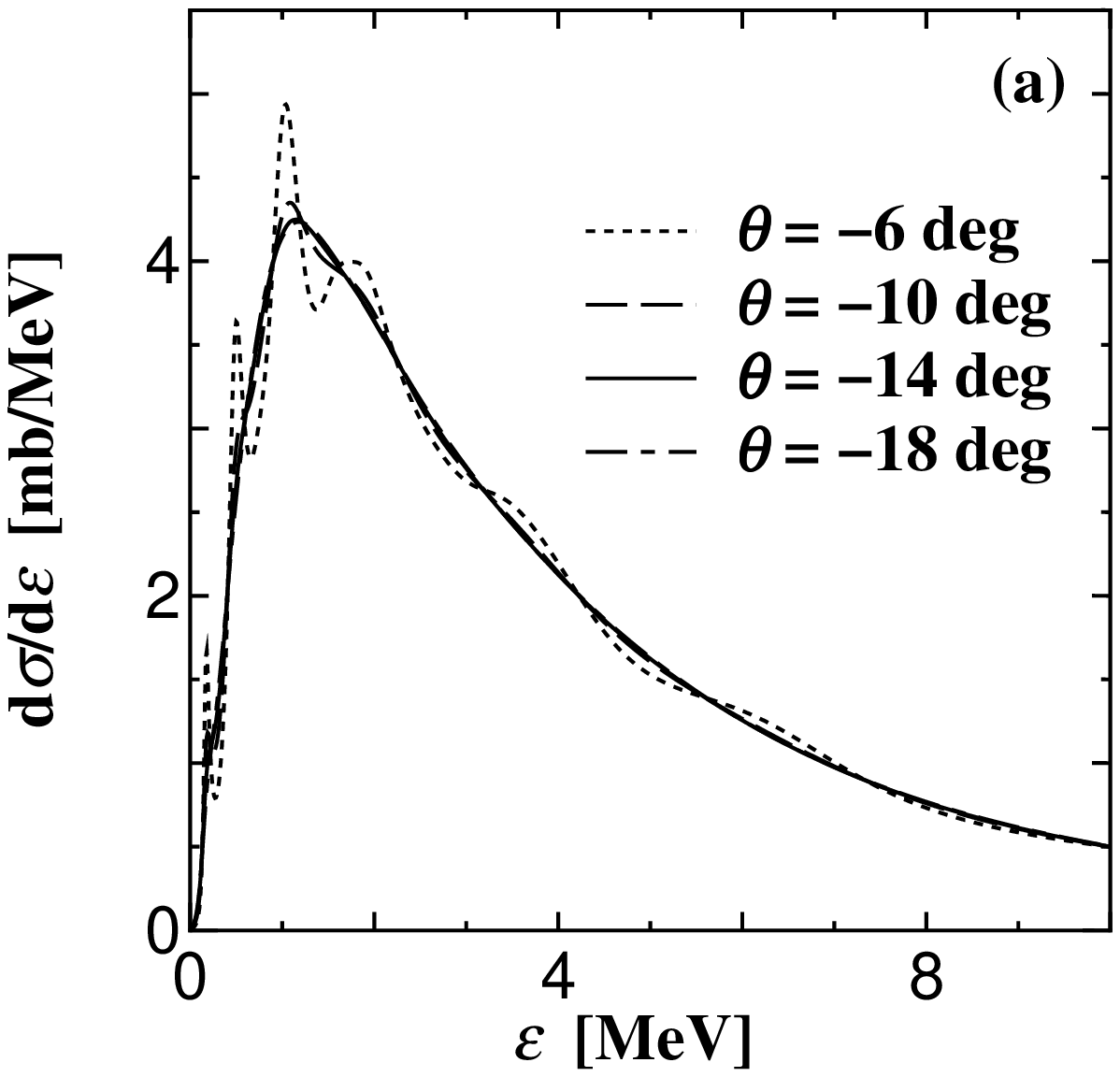}
 \includegraphics[width=0.32\textwidth,clip]{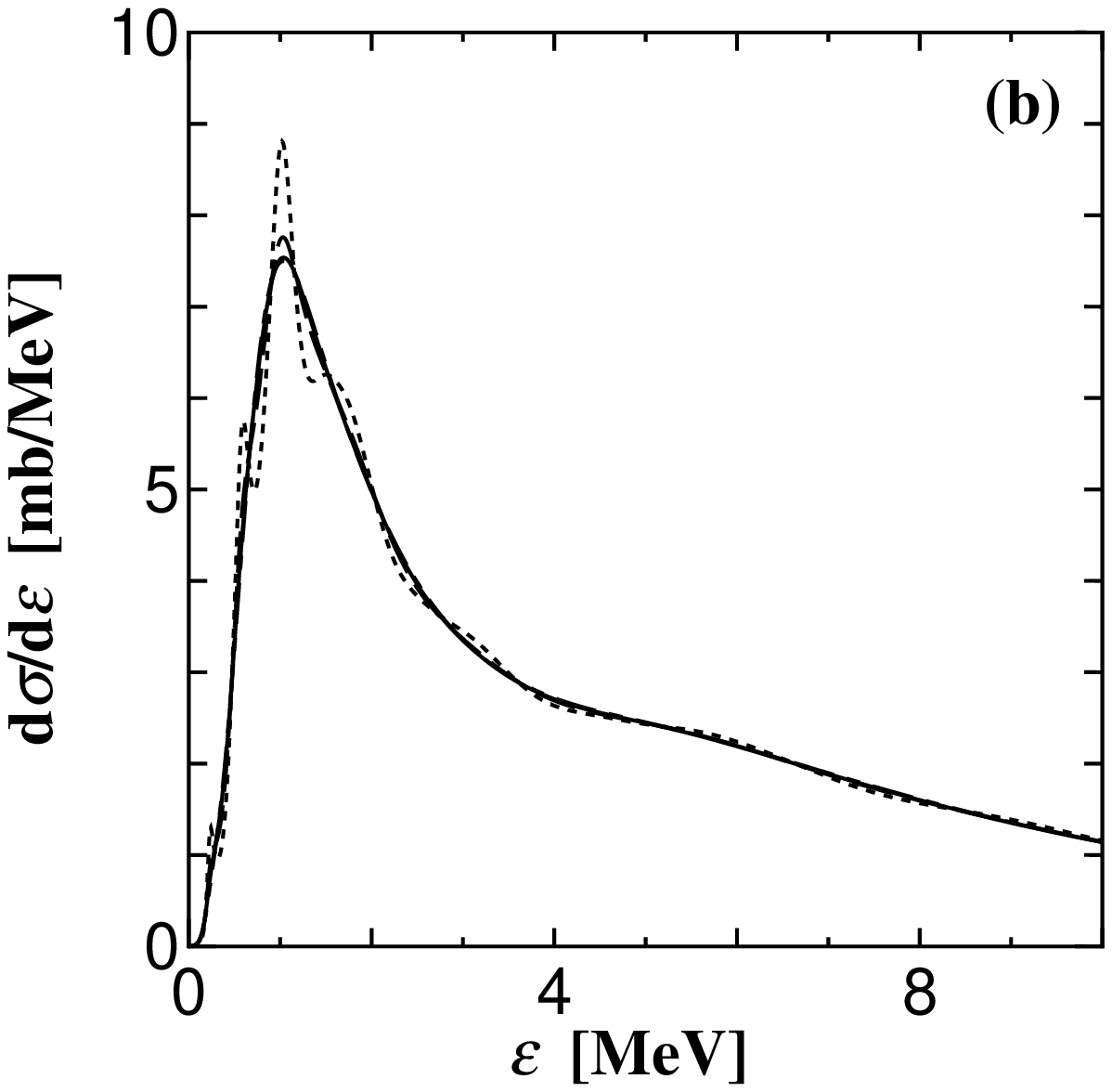}
 \includegraphics[width=0.32\textwidth,clip]{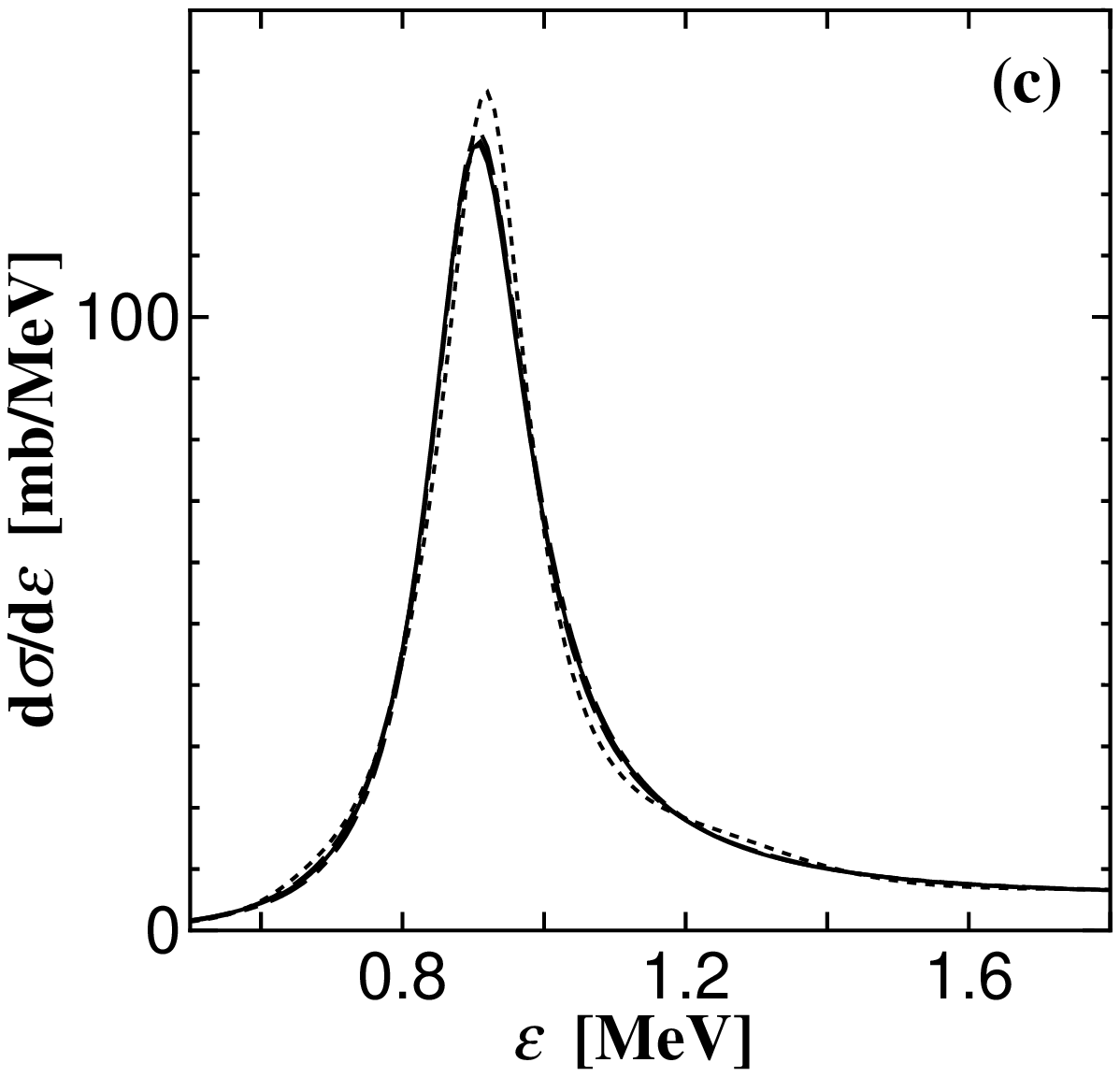}
 \caption{Dependence of the breakup cross section
 to the $0^+$ (a), $1^-$ (b), and $2^+$ (c) continua on
 complex-scaling angle $\theta$.}
 \label{fig4-s5}
\end{center}
\end{figure}

Figure~\ref{fig3-s5} shows the breakup cross sections $d \sigma/d\ve$
to the $0^+$, $1^-$, and $2^+$ continua. For all the cross sections, sets~II
and III yield the same result, whereas the result of set~I is 
somewhat different from it. 
The convergence of CDCC solution with respect to expanding 
the model space is thus obtained with set~II. 
Figure~\ref{fig4-s5} shows the breakup cross sections for four values of 
the complex-scaling angle $\theta$.
All the breakup cross sections converge at
$\theta=-14^\circ$, when $\theta$ decreases from $-6^\circ$ to $-18^\circ$.
%
%%%%%%%%%%%%%%%%%%%%%%%
%%%  Figure 4
%%%%%%%%%%%%%%%%%%%%%%%

The present calculation includes nuclear and Coulomb breakup processes.
The contribution of the latter is indeed large
for the breakup to the $1^-$ continuum, as indicated by the dotted line in
Fig.~\ref{fig3-s5}(b). It increases the breakup cross section by a
factor of 2 from the result without Coulomb breakup.
On the other hand,
the Coulomb breakup effect turns out to be
negligible for the $0^+$ and $2^+$
breakup spectra, and the latter dominates the breakup
cross section. Consequently, the Coulomb breakup effect in
the present reaction system is not significant for
both the breakup reaction and the elastic scattering.
For heavier targets, however, Coulomb breakup processes
become dominant.
Four-body CDCC can treat both nuclear and Coulomb breakup
processes and hence their interference. 
In \S\ref{sec7-3}, we apply four-body CDCC to 
$^6$He scattering from both light and heavy targets 
in a wide range of incident energies.

\section{Proposal of the eikonal reaction theory (ERT)
that makes CDCC applicable to inclusive reactions}
\label{sec6}

CDCC is highly reliable for describing exclusive reactions
but not applicable to inclusive cross sections
such as a neutron removal cross section.
In this section, we present an accurate method of treating
inclusive reactions within the framework of CDCC.
According to the method, the nuclear and Coulomb breakup processes
are consistently treated by CDCC
without making the adiabatic approximation to the Coulomb interaction,
so that the removal cross section calculated never diverges.
This method is referred to as {\it the eikonal reaction theory (ERT)}.
This section consists of a brief review of
Refs.~\citen{Yah11} and \citen{Has11},
and a new application of ERT to two-neutron removal reactions.

\subsection{Formulation of ERT}
\label{sec6-1}
Let us consider as a projectile (P) the two-body system composed of 
a core nucleus (c) and a valence neutron ($n$), 
and hence the three-body (c+$n$+A) system 
for the scattering of P from a target A. 
We start with the $S$-matrix operator as a formal solution to
the coupled-channel (E-CDCC) equations~\eqref{cceq-s4}
in {\S\ref{sec4}},
\bea
S=\exp\Big[
-i{\cal C}\int_{-\infty}^{\infty} dz {\hat O^{\dagger}}U{\hat O}
\Big]
\label{S-matrix-operator-s6}
\eea
with the interaction
\beq
U=U_n^{\rm (Nucl)}+U_{\rm c}^{\rm (Nucl)}+U_{\rm c}^{\rm (Coul)},
\label{pot-s6}
\eeq
where $U_n^{\rm (Nucl)}$ is the nuclear part of the optical potential
between $n$ and A, and $U_{\rm c}^{\rm (Nucl)}$ and $U_{\rm c}^{\rm (Coul)}$
are, respectively,
the nuclear and Coulomb parts of the optical potential between c and A.
${\cal C}$ is the path ordering operator
which describes the multistep scattering processes accurately.
The operator ${\hat O}$ is defined by
\bea
{\hat O} =\frac{1}{\sqrt{\hbar {\hat v}}} e^{i {\hat K} \cdot z},
\label{O-s6}
\eea
where the wave-number operator ${\hat K}$ is given
by ${\hat K}=\sqrt{2\mu_R(E-h)}/{\hbar}$
with the internal Hamiltonian $h$ of P,
and ${\hat v}={\hbar {\hat K}}/{\mu_R}$ is the velocity operator.
In the definition of
${\hat O}$ we assume that there is no Coulomb monopole interaction between
P and A for simplicity.
In the actual calculation, the exponent in Eq.~\eqref{Coulwf-s4},
with $K_c$ and $\eta_c$ replaced with the operator form,
is used as an exponent in ${\hat O}$; the form of ${\hat v}$ is changed
accordingly.

In the Glauber model, the adiabatic approximation is made,
in which $h$ is replaced with the ground-state
energy $\epsilon_0$, and hence ${\hat O^{\dagger}}U{\hat O}$ and
${\cal C}$ in Eq.~\eqref{S-matrix-operator-s6} are reduced to
$U/(\hbar v_0)$ and 1, respectively, where $v_0$ is the velocity of $P$
in the ground state relative to A.
At intermediate energies, this treatment is known to be valid for
short-range nuclear interactions, but not for the long-range Coulomb
interactions. Therefore, we make the adiabatic approximation in the
evaluation of only ${\hat O^{\dagger}}U_n^{\rm (Nucl)} {\hat O}$,
i.e., we use
\beq
    {\hat O^{\dagger}}U_n^{\rm (Nucl)} {\hat O} \rightarrow
     U_n^{\rm (Nucl)}/(\hbar v_0).
\label{replacement-N-s6}
\eeq
For the scattering of $^{31}$Ne from a $^{208}$Pb target at 240~MeV/nucleon,
it is shown by CDCC that the effect due to
the replacement~\eqref{replacement-N-s6}
is small; it is 0.2\% for the reaction cross section $\sigma_{\rm R}$,
1.9\% for the elastic-breakup cross section $\sigma_{\rm EB}$,
4.1\% for the one-neutron stripping cross section $\sigma_{n\rm :STR}$.
Using this replacement,
$S$ can be separated into the neutron part $S_n$ and
the core part $S_{\rm c}$,
\beq
  S = S_n S_{\rm c}
  \label{S-separation-s6}
\eeq
with
\bea
  S_n&=&
    \exp\Big[ - \frac {i}{\hbar v_0} \int_{-\infty}^{\infty} dz
    U_n^{\rm (Nucl)} \Big] ,
  \label{Sn-s6} \\
  S_{\rm c}&=&\exp\Big[
-i{\cal C}\int_{-\infty}^{\infty} dz
{\hat O^{\dagger}}(U_{\rm c}^{\rm (Nucl)}+U_{\rm c}^{\rm (Coul)})
{\hat O} \Big].
\label{Sc-s6}
\eea
This is the most important result of ERT.
We can derive several kinds of cross sections with the product form
Eq.~\eqref{S-separation-s6}, following the formulation of
the cross sections in
the Glauber model~\cite{HM85,Hen96};
see, e.g., Ref.~\citen{Has11} for the explicit form of them.
It should be noted that one cannot evaluate $S_{\rm c}$
directly with Eq.~\eqref{Sc-s6},
since it includes the operators ${\hat O}$ and ${\cal C}$.
However, one may find that
$S_{\rm c}$ is the formal solution to the Schr\"odinger equation
\beq
   \left[ -\frac{\hbar^2}{2\mu_R}\nabla_R^2 + h + U_{\rm c}^{\rm (Nucl)}
   + U_{\rm c}^{\rm (Coul)}-E \right]\Psi_{\rm c}=0,
\label{Schrodinger-eq-core-s6}
\eeq
when the eikonal approximation is made. We can thus obtain
$S_{\rm c}$ by solving Eq.~\eqref{Schrodinger-eq-core-s6}
with E-CDCC~\cite{Oga03,Oga06}.
As mentioned above, $S_n$ is obtained by Eq.~\eqref{Sn-s6}.

\subsection{One-neutron removal reaction of $^{31}$Ne}
\label{sec6-2}

We apply ERT to the one-neutron removal reactions
for the $^{31}$Ne$+^{12}$C scattering at 230~MeV/nucleon and
the $^{31}$Ne$+^{208}$Pb scattering at 234~MeV/nucleon
with a $^{30}$Ne$+n+$A three-body model.
The optical potentials for the $n$-target and $^{30}$Ne-target subsystems
are obtained by folding the effective nucleon-nucleon
interaction~\cite{Abu08} with one-body nuclear densities.
The densities of P and A are constructed by
the same method as in Ref.~\citen{Hor10}.
We assume the ground state of $^{31}$Ne to be
either the $^{30}{\rm Ne}(0^+) \otimes 1{\rm p}3/2$
or the $^{30}{\rm Ne}(0^+) \otimes 0{\rm f}7/2$, with
the one-neutron separation energy of 0.33~MeV.
As for the breakup states, we include s-, p-, d-, f-, and g-waves
up to the relative momentum between $^{30}$Ne and $n$ of $0.8$~fm$^{-1}$.
For more detailed numerical inputs, see Ref.~\citen{Yah11}.

Table \ref{tab1-s6} presents the integrated elastic-breakup
cross section $\sigma_{\rm EB}^{}$,
the one-neutron stripping cross section $\sigma_{n\rm :STR}^{}$,
the one-neutron removal cross section $\sigma_{-n}^{}$,
and the spectroscopic factor
${\cal S}=\sigma^{\rm exp}_{-n}/\sigma^{\rm th}_{-n}$.
${\cal S}$ calculated with the 1p3/2 ground-state neutron configuration
little depends on the target and less than unity,
but that with the 0f7/2 configuration does not satisfy these conditions.
Therefore, we can infer that
the major component of the ground state of $^{31}{\rm Ne}$
is $^{30}{\rm Ne}(0^+) \otimes 1{\rm p}3/2$ with ${\cal S} \sim 0.69$.
This value is consistent with the result shown
in Sec.~\ref{sec3-5}.
We adopt this configuration in the following.

\begin{table}
\begin{center}
\caption
{Integrated cross sections and the spectroscopic factor for
the $^{31}$Ne-$^{12}$C scattering at 230~MeV/nucleon and
the $^{31}$Ne-$^{208}$Pb scattering at 234~MeV/nucleon.
The cross sections are presented in unit of mb and the data are taken from
Ref.~\citen{Nak09}.
}
\label{tab1-s6}
\begin{tabular}{cccccccc}
\hline \hline
 & \multicolumn{3}{c}{$^{12}$C target} & \hspace{2mm} &
 \multicolumn{3}{c}{$^{208}$Pb target} \\ \cline{2-4}  \cline{6-8}
 {} & ${\rm p}_{3/2}$ & ${\rm f}_{7/2}$ & Exp. &
 {} & ${\rm p}_{3/2}$ & ${\rm f}_{7/2}$ & Exp. \\ \hline
% $\sigma_{\rm R}^{}$ & 1572.5 & 1489.9 & & & 5518.0 & 4589.5 & \\
 $\sigma_{\rm EB}^{}$ & 23.3 & 3.3 & & & 799.5 & 73.0 & (540) \\
% $\sigma_{\rm R}^{}$(-$n$) & 1463.5 & 1458.6 & & & 5151.5 & 4524.2 & \\
% $\sigma_{\rm bu}^{}$(-$n$) & 4.5 & 1.0 & & & 677.2 & 60.5 & \\ \hline
 $\sigma_{n\rm :STR}^{}$ & 90 & 29 & & & 244 & 53 & \\ \hline
 $\sigma_{-n}^{}$ & 114 & 32 & 79 & & 1044 & 126 & 712 \\
 ${\cal S}$ & 0.693 & 2.47 & & & 0.682 & 5.65 & \\ \hline \hline
\end{tabular}
\end{center}
\end{table}

Since the potential between $^{30}$Ne and $n$ is not well known,
we change each of the potential parameters
by 30\% and see how this ambiguity of the potential affects the
resulting value of ${\cal S}$.
We obtain
${\cal S}= 0.693 \pm 0.133 \pm 0.061$
for a $^{12}$C target and
${\cal S}=0.682 \pm 0.133 \pm 0.062$
for a $^{208}$Pb target; the second and third numbers
following the mean value
stand for the theoretical and experimental uncertainties, respectively.
Thus, ${\cal S}$ includes a sizable theoretical uncertainty.
This situation completely changes if we look at the
asymptotic normalization coefficient $C$~\cite{MT90}, i.e.,
$C= 0.320 \pm 0.010 \pm 0.028$~fm$^{-1/2}$
for a $^{12}$C target and
$C= 0.318 \pm 0.008 \pm 0.029$~fm$^{-1/2}$
for a $^{208}$Pb target.
Thus, $C$ has a much smaller theoretical uncertainty than ${\cal S}$.
This means that the one-nucleon removal reaction is quite peripheral.

\subsection{Two-neutron removal reaction of $^{6}$He}
\label{sec6-3}

ERT is applied to two-neutron removal reactions
of $^6$He from $^{12}$C and $^{208}$Pb targets at 240 MeV/nucleon.
In this case, the projectile is treated as a three-body ($\alpha+n+n$)
system and hence four-body CDCC is used.
The optical potentials for the $n$-target and $\alpha$-target subsystems
are calculated by folding
the Melbourne nucleon-nucleon $g$-matrix interaction~\cite{Amo00}
with the densities obtained by
the spherical Hartree-Fock (HF) calculation with
the Gogny D1S interaction.~\cite{DG80,Ber91}
The model space of the following calculation is the same as
in Ref.~\citen{Mat10}, with which good convergence is achieved.

\begin{table}[htb]
\begin{center}
\caption{
Integrated cross sections for two-neutron removal reaction of $^6$He
on $^{12}$C and $^{208}$Pb targets at 240~MeV/nucleon.
The cross sections are presented in unit of mb and
the experimental data are taken from Ref.~\citen{Aum99}. }
\label{tab2-s6}
\begin{tabular}{cccccc}
\hline \hline
 & \multicolumn{2}{c}{$^{12}$C target} & \hspace{2mm} &
 \multicolumn{2}{c}{$^{208}$Pb target}
\\ \cline{2-3} \cline{5-6}
 {} & Calc. & Exp. &
 {} & Calc. & Exp.
\\ \hline
% $\sigma_{\rm EB}^{}$ & 16.1 & 30 $\pm$ 5 & & 514.1 & 650 $\pm$ 110 \\
 $\sigma_{n:\rm STR}^{}$ & 153.4 & 127 $\pm$ 14 & & 353.6 & 320 $\pm$ 90 \\
 $\sigma_{2n:\rm STR}^{}$ & 29.0 & 33 $\pm$ 23 & & 148.9 & 180 $\pm$ 100 \\ \hline
 $\sigma_{-2n}^{}$ & 198.5 & 190 $\pm$ 18 & & 1016.6 & 1150 $\pm$ 90 \\
\hline \hline
\end{tabular}
\end{center}
\end{table}
Table \ref{tab2-s6} shows the one- and two-neutron stripping
cross sections, $\sigma_{n:\rm STR}^{}$ and $\sigma_{2n:\rm STR}^{}$,
respectively,
and the two-neutron removal cross section $\sigma_{-2n}^{}$.
The present framework reproduces well the experimental data~\cite{Aum99}
with no adjustable parameters.
Thus, we can clearly see the reliability of ERT for two-neutron
removal reactions on both light and heavy targets.

\subsection{Comparison of ERT and the Glauber model}
\label{sec6-4}

In this subsection, the accuracy of the Glauber model is tested
for deuteron induced reactions at 200~MeV/nucleon.
We estimate the relative difference
\bea
\delta_{X}=[X({\rm CDCC})-X({\rm GL})]/X({\rm CDCC}) ,
\eea
where $X({\rm CDCC})$ and $X({\rm GL})$ correspond to integrated
cross sections calculated by CDCC with ERT and the Glauber model, respectively.
For the Glauber-model calculation in which the eikonal and
adiabatic approximations are made, the Coulomb interaction is neglected.
The numerical inputs are given in Ref.~\citen{Has11}.

%%%%%%%%%%%%%%%%%%%%%%%
%%%  Figure 1
%%%%%%%%%%%%%%%%%%%%%%%
\begin{figure}[htbp]
\begin{center}
 \includegraphics[width=0.45\textwidth,clip]{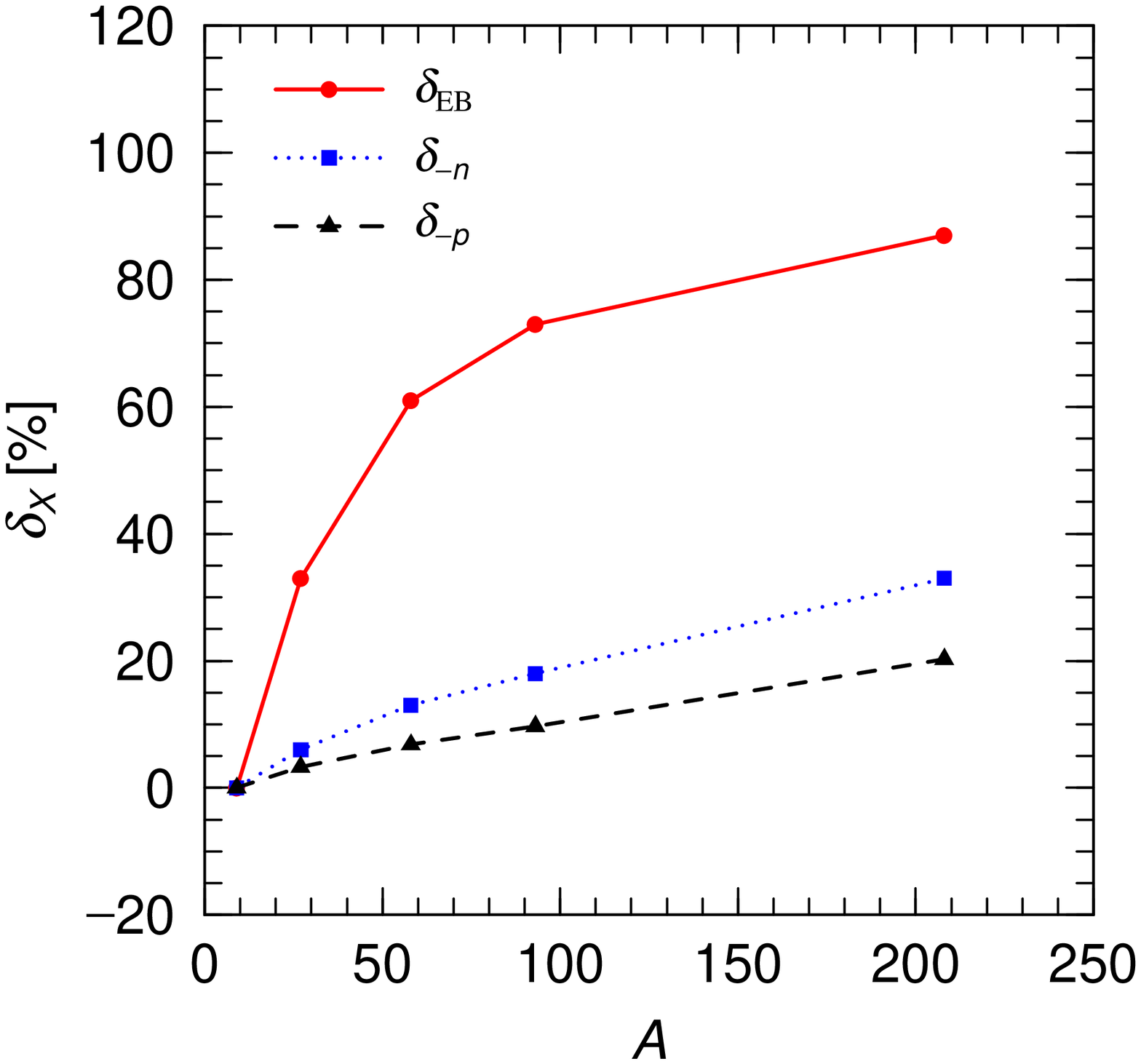}
 \includegraphics[width=0.45\textwidth,clip]{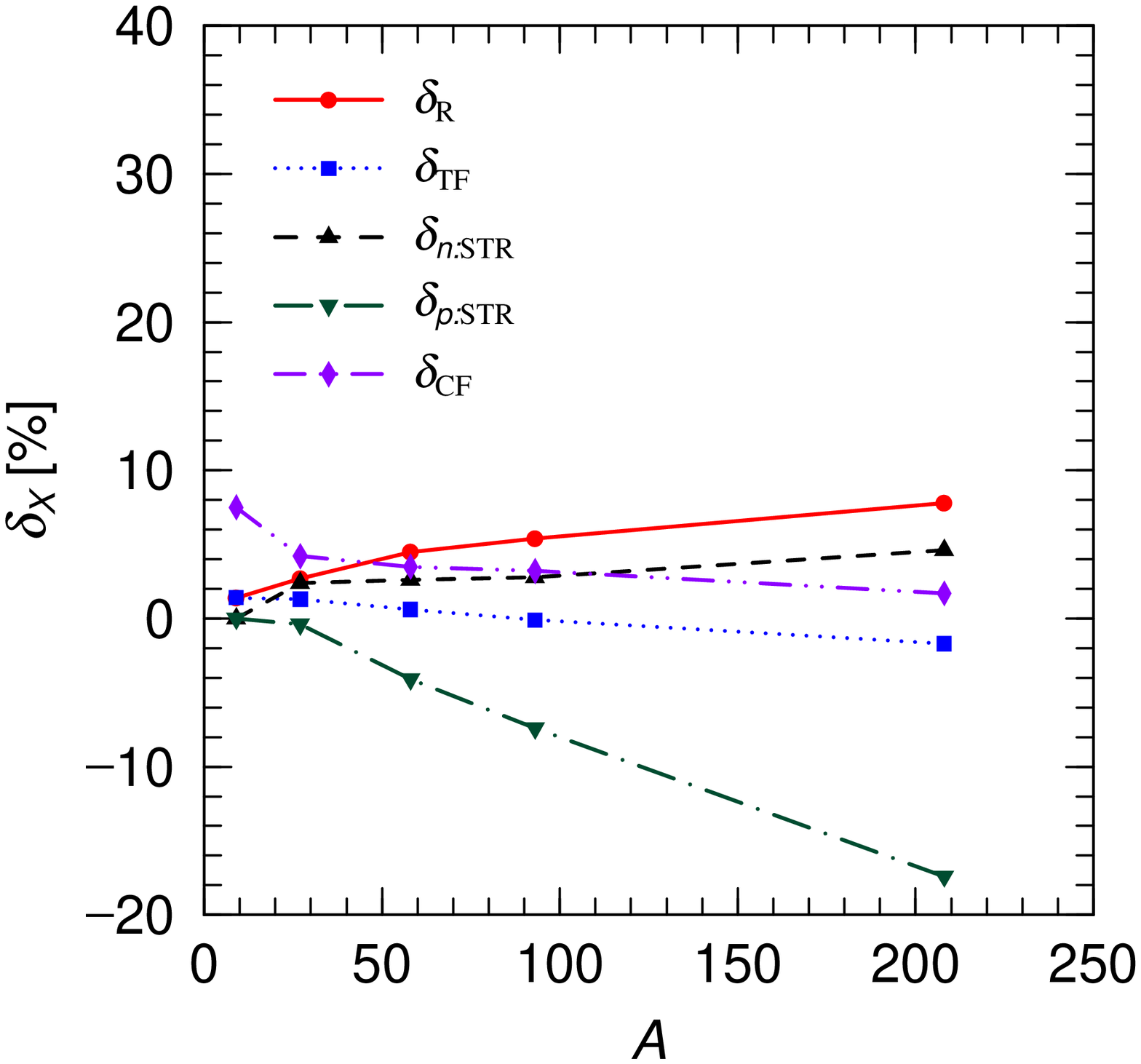}
 \caption{
 Accuracy of the Glauber model for various integrated cross sections.
 }
 \label{fig1-s6}
\end{center}
\end{figure}
%%%%%%%%%%%%%%%%%%%%%%%
%
Figure~\ref{fig1-s6} shows $\delta_{X}$ as a function of target mass
number $A$ for $\sigma_{\rm EB}$,
$\sigma_{\rm R}$, $\sigma_{n:{\rm STR}}$, $\sigma_{-n}$,
the proton stripping ($\sigma_{p:{\rm STR}}$) and removal
($\sigma_{-p}$) cross sections,
the total fusion cross section $\sigma_{\rm TF}$, and
the complete fusion cross section $\sigma_{\rm CF}$.
The Glauber model is good for light targets, as expected.
For heavier targets where the Coulomb breakup is essential,
however, the Glauber model is not good for the elastic-breakup
and proton stripping cross sections.

\section{Applications of CDCC to scattering of unstable nuclei}
\label{sec7}

In this section, we review some recent
applications of CDCC to scattering of unstable nuclei. See the
references cited in the following subsections for the details of the
formalism, numerical inputs, other results, and further discussion.

\subsection{One-neutron removal reactions of $^{18}$C and $^{19}$C on
  proton target}
\label{sec7-1}

One-neutron removal reactions have played a key role in investigating
exotic properties of neutron-rich nuclei. For light targets such as Be
and C, the so-called stripping model~\cite{Han96,Han03} based on the eikonal
approximation has been successful in describing the
removal process. On the other hand,
for a proton target, the removal process can
be interpreted as an elastic breakup, since there is no excited state
in the target, and also a neutron transfer process
producing deuteron hardly occurs. Therefore, we should
treat accurately the elastic breakup process for the one-neutron removal
reactions on a proton target, to which the stripping model is not
applicable.

%%%%%%%%%%%%%%%%%%%%%%%
%%%  Figure 1
%%%%%%%%%%%%%%%%%%%%%%%
\begin{figure}[htbp]
 \begin{center}
  \includegraphics[width=0.8\textwidth,clip]{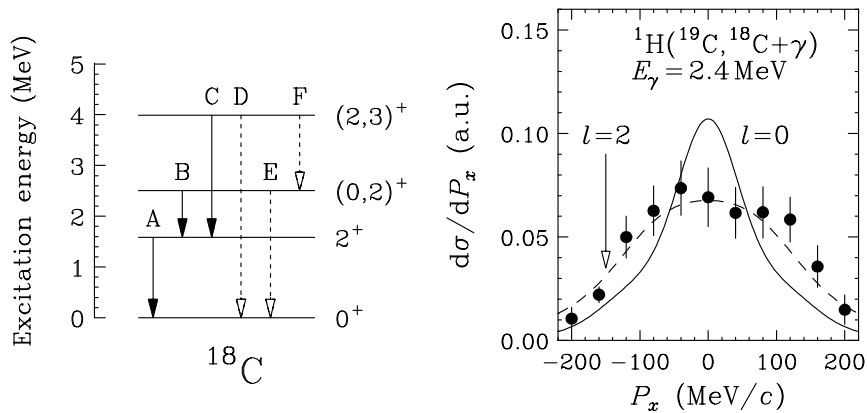}
 \end{center}
\caption{Level scheme and transitions of $^{18}$C (left panel) and
 transverse-momentum distribution of $^{18}$C in coincidence with the
 2.4~MeV $\gamma$-ray (right panel). The solid and dashed lines
 represent the CDCC calculations by assuming neutron removal from the
 $\ell=0$ and $\ell=2$ orbits in $^{19}$C, respectively.}
\label{fig1-s7}
\end{figure}
In Ref.~\citen{Kon09}, one neutron removal cross sections
of $^{18}$C and $^{19}$C on a proton target were measured,
and CDCC was applied to analyze the data to determine the
neutron configuration of these nuclei.
Figure~\ref{fig1-s7} (right panel)
shows the transverse-momentum distribution of
$^{18}$C after the breakup of $^{19}$C in coincidence with the
2.4~MeV $\gamma$-ray, which corresponds to the
population of the 4.0~MeV state of $^{18}$C; see the level
scheme in the left panel of Fig.~\ref{fig1-s7}.
The experimental data are consistent with the CDCC
calculation for neutron breakup from the 0d orbit in
$^{19}$C(1/2$^+$). From this analysis, we have successfully confirmed the
spin-parity
$I^\pi=(2,3)^+$ assignment for the 4.0~MeV state in $^{18}$C;
see Ref.~\citen{Kon09} for more detailed discussion.

For both reactions induced by $^{18}$C and $^{19}$C, the theoretical
cross sections calculated by CDCC with the shell-model spectroscopic
factors obtained by the WBP interaction~\cite{WB92}
explain the most of the
relative values, as shown in the Tables of Ref.~\citen{Kon09}.
As for the absolute values, the measured and calculated results
agree very well with each other for the breakup of $^{19}$C
that has a small neutron separation energy $S_n$.
On the other hand, the measured cross section
is significantly smaller than the calculated one
for the breakup of $^{18}$C that has a quite large $S_n$.
This is consistent with the finding of a systematic analysis
of nucleon removal cross sections,~\cite{Gad08} though the mechanism
of this {\lq\lq}quenching'' of the shell-model spectroscopic
factors is still under discussion.
Thus, CDCC is shown to be a powerful method for determining
configurations of valence neutrons in exotic nuclei.

\subsection{Study on the low-lying states in the unbound nucleus $^{13}$Be}
\label{sec7-2}

Shell evolution of neutron-rich nuclei is one of the hottest
topics of nuclear physics.
In Ref.~\citen{Kon10}, we investigated the shell structure of
$^{13}$Be, which is unbound and has a neutron number $N$ = 9.
We did the same analysis as in \S\ref{sec7-1} of the
one-neutron removal cross section of $^{14}$Be by a proton target.

%%%%%%%%%%%%%%%%%%%%%%%
%%%  Figure 2
%%%%%%%%%%%%%%%%%%%%%%%
\begin{figure}[htbp]
 \begin{center}
  \includegraphics[width=0.8\textwidth,clip]{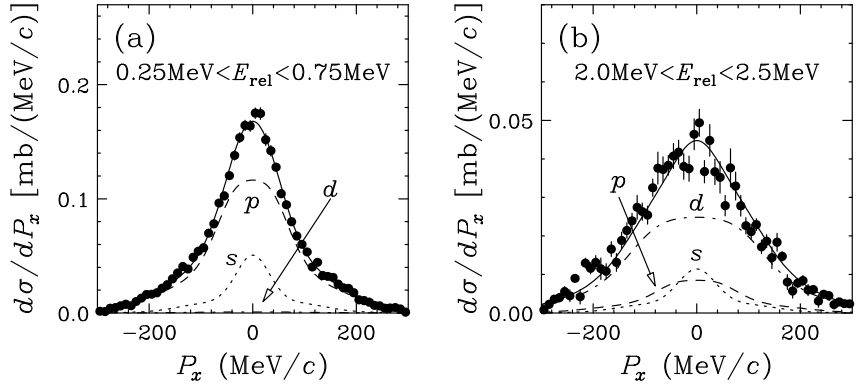}
 \end{center}
 \caption{Transverse momentum distributions of the $^{12}$Be+$n$ system
 for (a)  0.25~MeV $<$ $E_{\rm rel}$ $<$ 0.75~MeV and (b) 2.0~MeV $<$
 $E_{\rm rel}$ $<$ 2.5~MeV. The s-, p-, and d-wave components are
 shown by  the dotted, dashed, and dash-dotted lines, respectively. The
 solid  lines are the sums of the three components.}
 \label{fig2-s7}
\end{figure}
We show in Fig.~\ref{fig2-s7} the transverse momentum distributions of
the $^{12}$Be+$n$ system for 0.25~MeV$< E_{\rm rel} <$0.75~MeV (left panel)
and 2.0~MeV$< E_{\rm rel} <$2.5~MeV (right panel),
where $E_{\rm rel}$ is the relative energy of $^{12}$Be and $n$.
The dotted, dashed, and dash-dotted
lines correspond to the s-, p-, and d-wave contributions
calculated by CDCC, respectively.
We adopted a $^{13}$Be+$n$ two-body model with
each of these relative angular momenta as an initial state
structure of $^{14}$Be, and then added all the results incoherently.
The strengths
of the individual components as well as the resonance parameters
assumed were treated as free adjustable parameters to reproduce
the experimental data; see Ref.~\citen{Kon10} for more details.
From this analysis, we confirmed
the low-lying p-wave resonance at $E_{\rm rel}$ = 0.51(1)~MeV
as well as the d-wave resonance at $E_{\rm rel}$ =
2.39(5)~MeV on the broad s-wave distribution with the s-wave
scattering length $a_{\rm s}=-3.4(6)$~fm in the invariant mass spectrum
of $^{13}$Be. The spin-parity of the
p-wave resonance state was assigned to $I^\pi$ =
1/2$^-$. This intruder 1/2$^-$ state indicates the disappearance of the
$N$ = 8 shell closure and the shell evolution in the vicinity of the
neutron drip line.

\subsection{Elastic and breakup cross sections of $^6$He}
\label{sec7-3}

Two-neutron halo nuclei  such as $^6$He and $^{11}$Li have exotic
properties, i.e., soft dipole excitation and a di-neutron
correlation. These properties can be investigated via
breakup reactions, where the projectile breaks up into three
(core+$n$+$n$) fragments. For this purpose,
accurate description of the four-body systems is
essential, and four-body CDCC proposed in \S\ref{sec5} is one of
the most reliable methods for treating such four-body breakup
processes.

%%%%%%%%%%%%%%%%%%%%%%%
%%%  Figure 3
%%%%%%%%%%%%%%%%%%%%%%%
\begin{figure}[htbp]
 \begin{center}
  \includegraphics[width=0.4\textwidth,clip]{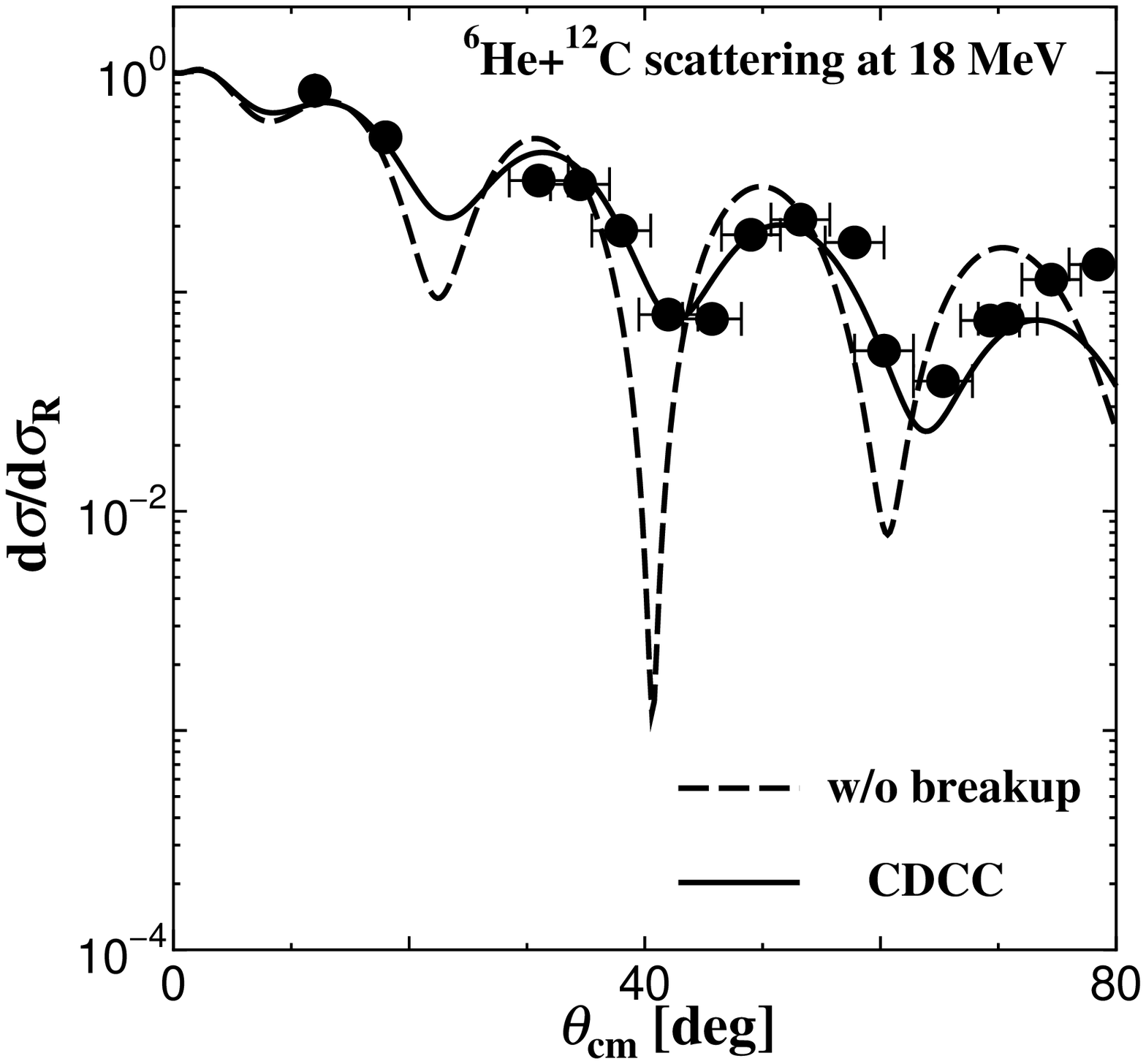}
  \includegraphics[width=0.4\textwidth,clip]{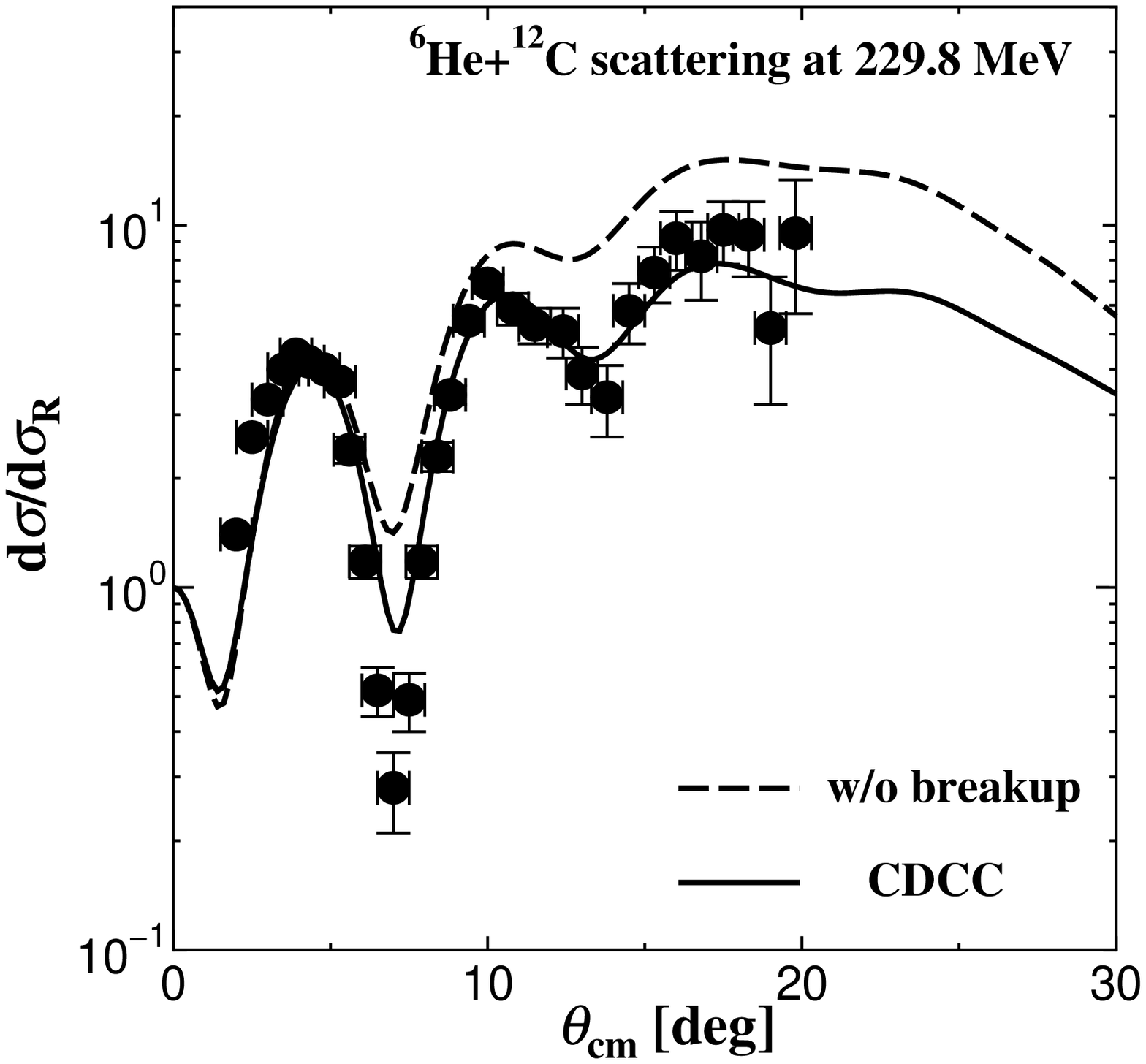}
 \end{center}
\caption{Angular distributions of the elastic differential cross section
 for the $^6$He$+^{12}$C scattering at
 18~MeV (left panel) and 229.8~MeV (right panel). The solid (dashed)
 line shows the result with (without) breakup effects.
 The experimental data in the left and right panels are taken
 from Refs.~\citen{Mil04} and \citen{Lap02}, respectively.}
\label{fig3-s7}
\end{figure}
First, we applied four-body CDCC to the elastic scattering
of $^6$He.~\cite{Mat04,Mat06} Figure~\ref{fig3-s7}
shows the results of the four-body CDCC calculation compared with the
experimental data for the $^6$He$+^{12}$C scattering at 18~MeV~\cite{Mil04}
and 229.8~MeV.~\cite{Lap02} In this analysis, coupling potentials in
four-body CDCC are
calculated by a double-folding model with the density-dependent M3Y
interaction~\cite{Kob82} that has only the real part. The imaginary part of each
coupling potential is assumed to have the same form as of
the real part with a normalization constant $N_{\rm I}$
to be optimized to reproduce the experimental data.
{We took $N_{\rm I}=0.3$ and 0.5 for the scattering
at 18~MeV and 229.8~MeV, respectively.
One sees clearly from Fig.~\ref{fig3-s7} the importance of the
breakup effects; note that $N_{\rm I}$
does not change the oscillation pattern.}
In this work, we omitted couplings to
$1^-$ continuum states, because the Coulomb breakup effect is negligible
in this reaction system.

\begin{figure}[htbp]
 \begin{center}
  \includegraphics[width=0.4\textwidth,clip]{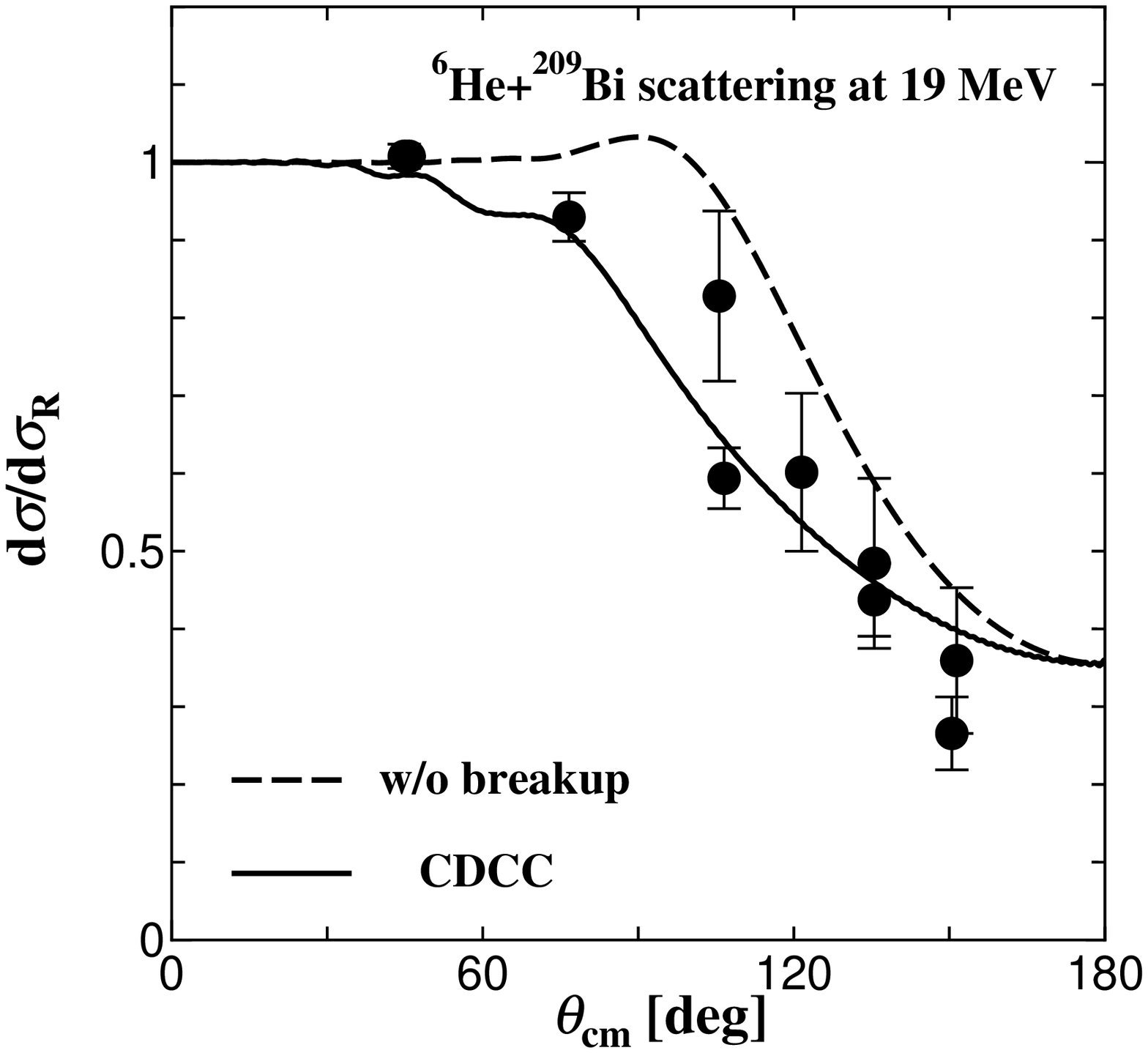}
  \includegraphics[width=0.4\textwidth,clip]{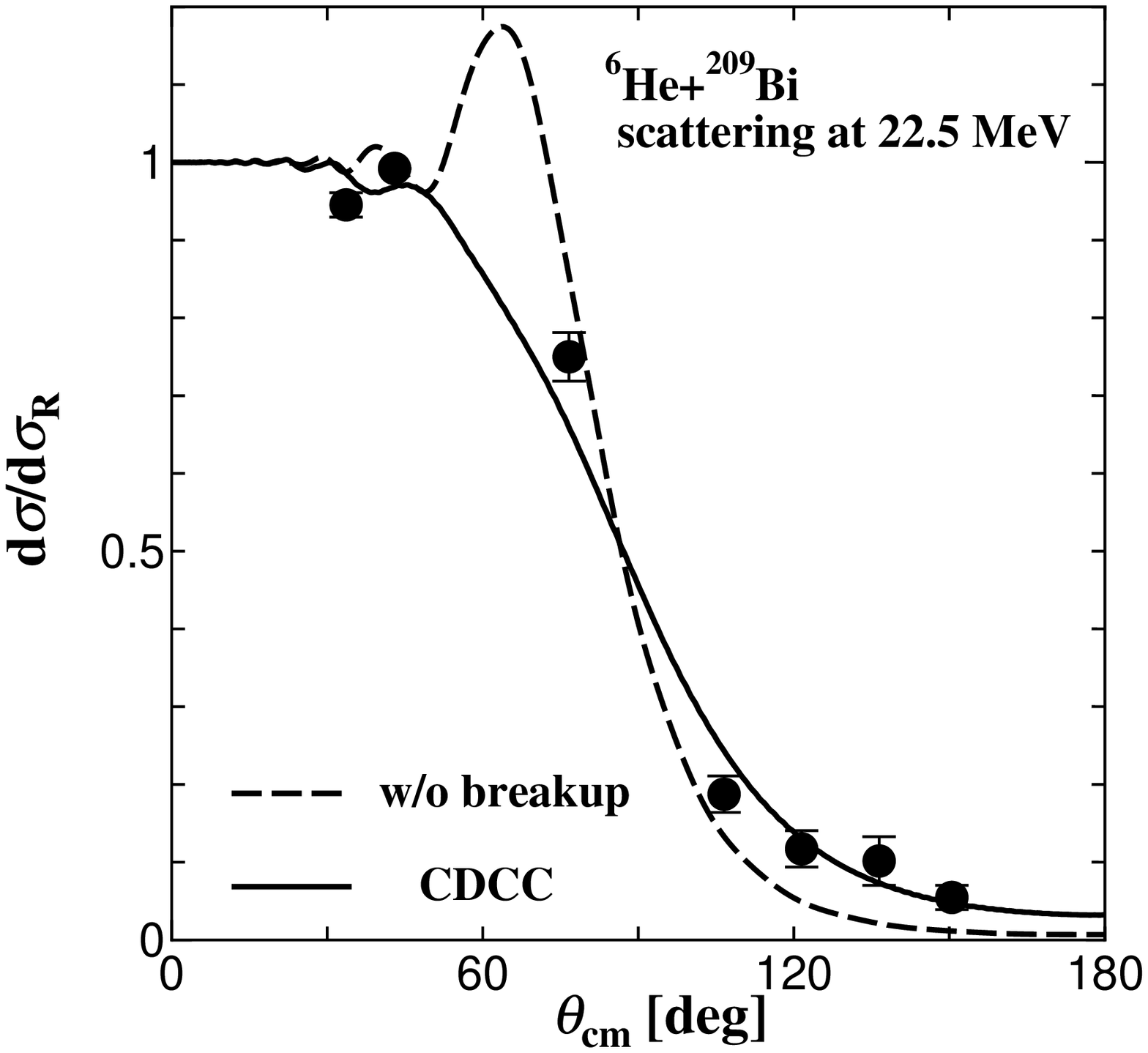}
 \end{center}
\caption{
 Same as in Fig.~\ref{fig3-s7} but
 for the $^6$He$+^{209}$Bi scattering at
 19~MeV (left panel) and 22.5~MeV (right panel).
 The experimental data are taken from Refs.~\citen{Agu00}
 and \citen{Agu01}.}
\label{fig4-s7}
\end{figure}
In Fig.~\ref{fig4-s7}, angular distributions of the elastic
differential cross section for the $^6$He+$^{209}$Bi scattering at
19~MeV and 22.5 MeV~\cite{Agu00,Agu01} are shown. Since Coulomb breakup
effects are dominant in this system, we take into account couplings to
$1^-$ continuum states.
In this calculation, we adopt phenomenological optical
potentials for the interactions between the $^{209}$Bi target
and the constituents of $^6$He.
One sees that the four-body CDCC calculation
reproduces well the experimental data. Again, the breakup
effects are significant.
It was found in
Ref.~\citen{Mat06} also that three-body CDCC, in which a
di-neutron plus $\alpha$ structure was assumed for $^6$He,
could not reproduce the data. This shows importance of 
the accurate description of $^6$He by means of the three-body model.

\begin{figure}[htbp]
\begin{center}
 \includegraphics[width=0.4\textwidth,clip]{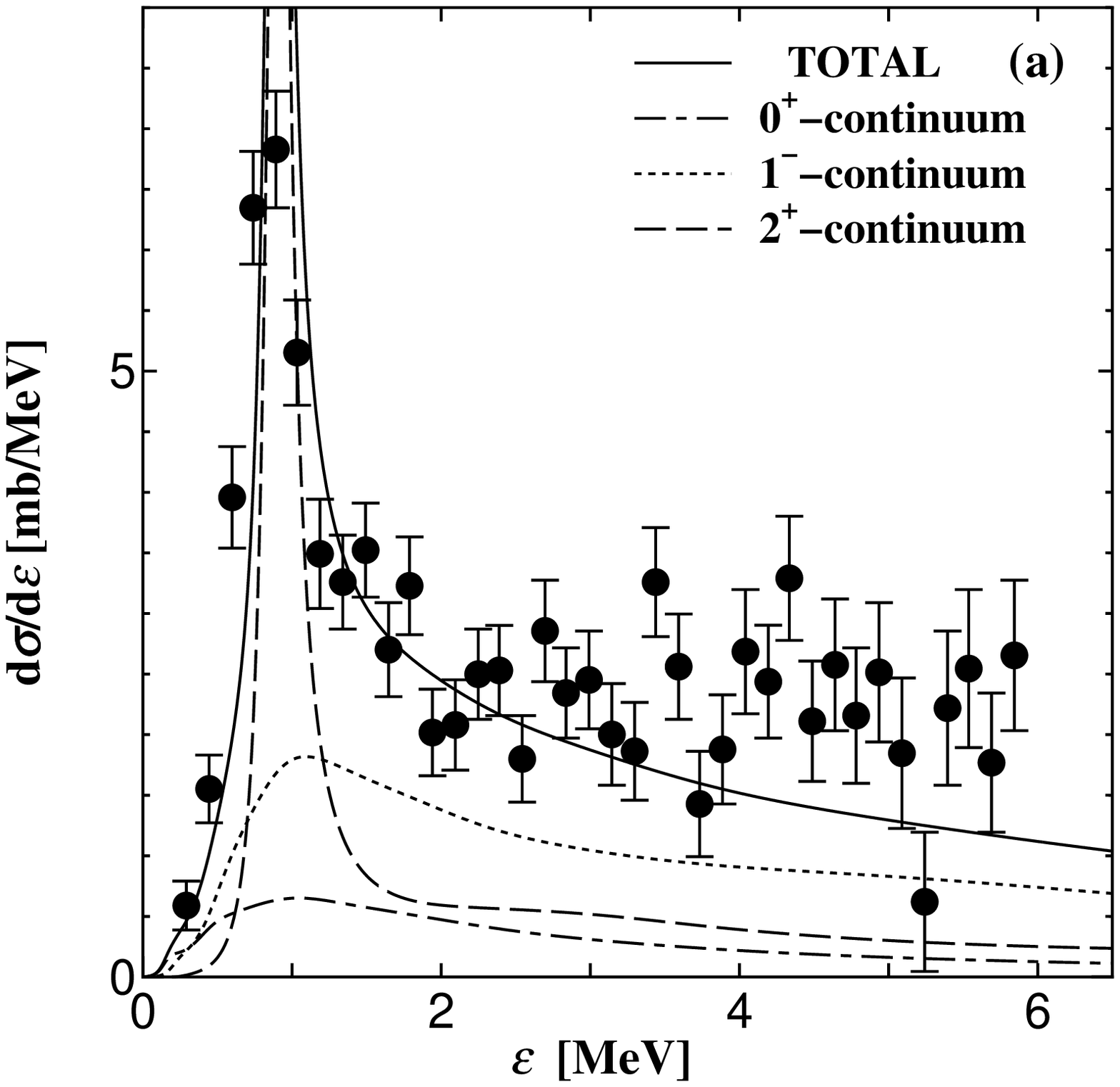}
 \includegraphics[width=0.4\textwidth,clip]{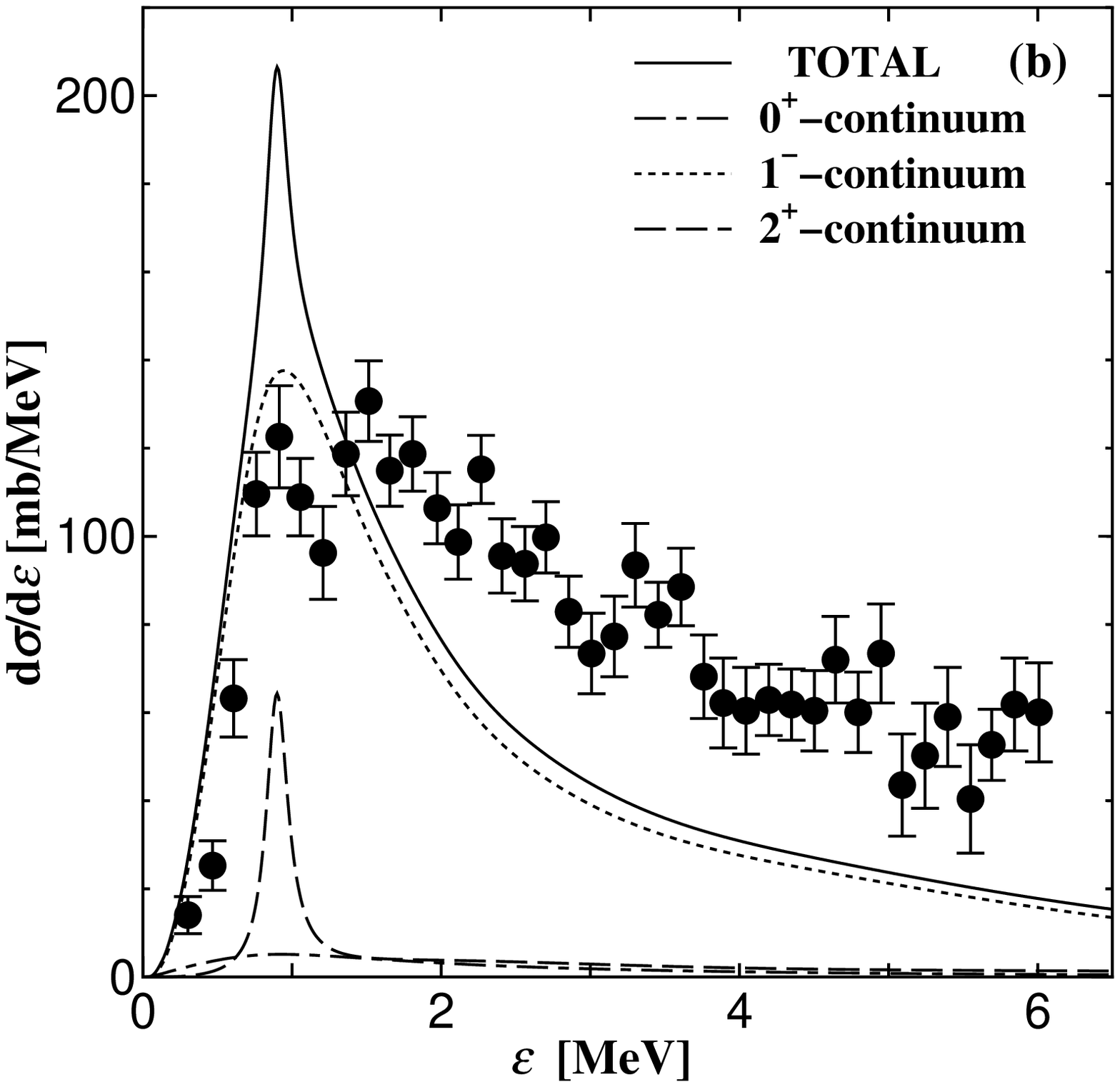}
 \caption{Comparison of the breakup cross section calculated by CDCC
 (solid line) with experimental data for (a) the $^6$He+$^{12}$C
 scattering at 240~MeV/nucleon and (b) the $^6$He+$^{208}$Pb scattering at
 240~MeV/nucleon. The dash-dotted, dotted, and dashed lines correspond to
 the contributions of the $0^+$, $1^-$, and $2^+$ breakup, respectively,
 and the solid line is the sum of them. The
 experimental data are taken from Ref.~\citen{Aum99}.}
 \label{fig5-s7}
\end{center}
\end{figure}
Next, four-body CDCC with the new smoothing method described in
\S\ref{sec5} is applied to
analyses of breakup reactions of $^6$He. In Fig.~\ref{fig5-s7}, the
calculated breakup cross section $d\sigma/d\ve$ is compared with the
experimental data for the $^6$He$+^{12}$C and $^6$He$+^{208}$Pb
reactions at 240~MeV/nucleon~\cite{Aum99}.
We found that nuclear breakup was dominant in the former,
whereas Coulomb breakup to the $1^-$ continuum was dominant
in the latter.
For the $^{12}$C target, the present theoretical result is consistent
with the experimental data
except for the peak of the 2$^+$ resonance around
$\ve=1$ MeV.
For the $^{208}$Pb target,
the present method undershoots the experimental data
for $\ve \ga 2$~MeV. A possible origin of this undershooting is
the contribution of
the inelastic breakup reactions not included in the present calculation.
It was reported in Ref.~\citen{Ers00} that
the inelastic breakup
effect was not negligible, and the elastic breakup cross
section calculated with four-body DWBA
also underestimated the data.

One of the most important features of these analyses with CDCC
is the treatment of the nuclear and Coulomb breakup amplitudes as
well as their interference on the same footing.
Through a systematic analysis of $^8$B breakup reactions,~\cite{Oga09b}
we showed the importance of
the nuclear-Coulomb interference
which has been neglected in many studies on breakup of unstable nuclei.

\section{Applications of CDCC to reactions essential in cosmology and
astrophysics}
\label{sec8}

In this section, we review our recent applications of CDCC to
some reactions important for cosmology and astrophysics.
Readers are invited to the references cited in the following
subsections for more detailed description of the calculation
and complete discussion.

\subsection{Determination of the astrophysical factor $S_{17}$ for
the $^7$Be($p,\gamma$)$^8$B reaction}
\label{sec8-1}

Intensive measurements of $^8$B breakup at intermediate energies
\cite{Kik9798,Dav01,Iwa99} have been done to indirectly determine the
astrophysical factor $S_{17}(0)$,
the value in the zero-energy limit,
for the $^7$Be($p,\gamma$)$^8$B
reaction, which is important for neutrinophysics \cite{Bah01}.
The results of the indirect measurements
of $S_{17}(0)$ are, however, significantly smaller than the
result of the precise direct measurement of
$^7$Be($p,\gamma$)$^8$B~\cite{Jun03}.
This may cast doubt on the reliability of the indirect measurements
of astrophysical factors through breakup reactions of unstable nuclei,
which are planned to be done
at forthcoming RI beam facilities such as FAIR at GSI and FRIB at MSU,
and at the brand-new facility RIBF at RIKEN.
Thus, to clarify the reason for this discrepancy is
a very important subject of nuclearastrophysics.
For this purpose we
accurately analyzed the $^8$B breakup by $^{208}$Pb at
52~MeV/nucleon with full coupled-channel calculation, i.e., CDCC,
taking account of both nuclear and Coulomb
breakup with all higher-order processes.~\cite{Oga06}

%
%%%%%%%%%%%%%%%%%%%%%%%
%%%  Figure 1
%%%%%%%%%%%%%%%%%%%%%%%
\begin{figure}[htpb]
%\begin{figure}[b]
\centerline{
\includegraphics[width=13.4cm]{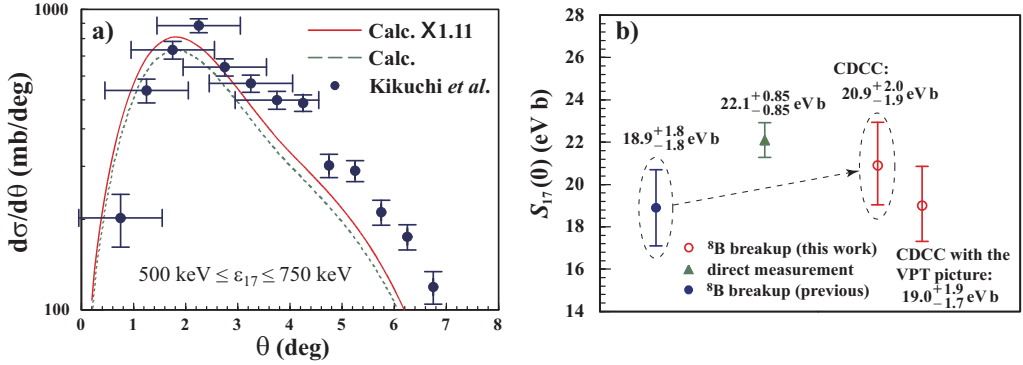}
} \caption{\label{fig1-s8}
a) Result of the CDCC analysis of the $^8$B breakup cross section
at 52~MeV/nucleon.~\cite{Kik9798}
b) Summary of the present study on $S_{17}(0)$ compared with the previous
results. See the text for details.
}
\end{figure}
In the left panel of Fig.~\ref{fig1-s8}
we show the result of the CDCC analysis
of the $^8$B breakup cross section
by $^{208}$Pb at 52~MeV/nucleon.~\cite{Kik9798} The cross section, which is
integrated over the relative energy $\epsilon_{17}$ of $^7$Be and $p$
between 500 and 750 keV, is shown as a function of the scattering
angle $\theta$ of the center-of-mass of $^8$B (the $^7$Be$+p$ system).
Numerical inputs for the CDCC
calculation are described in Ref.~\citen{Oga06} in detail. The dashed
line is the result of the calculation using a normalized $^8$B internal
wave function, whereas the solid line is the result of the $\chi^2$ fitting
to the experimental data.
The theoretical results are smeared out by using a filtering
table \cite{Mot04} to take account of the experimental resolution
and efficiency. Since the filtering table was made with assuming the
s-state breakup of $^8$B, which turns out to be valid for
$\theta \la 4^\circ$, we use the data in that region in the
$\chi^2$-fitting procedure. Note that each horizontal bar put on the
data points below $4^\circ$ does not represent a statistical error
but it shows the range of $\theta$ in which the breakup cross sections
contribute to each data point \protect\cite{Kik9798}.
The purpose of the $\chi^2$ fitting is to determine the tail of the
overlap function $I({\bm r})$
between the ground states of $^7$Be and $^8$B, i.e.,
the asymptotic normalization coefficient (ANC) $C$;
the resulting value of $C$ is 0.74 fm$^{-1/2}$.

It is shown in Ref.~\citen{MT90} that $S_{17}(0)$ for the
$^7$Be($p,\gamma$)$^8$B reaction is accurately determined
if $C$ is obtained from an alternative reaction in which
only the tail region of the overlap function $I({\bm r})$ has
a significant contribution.
We use this so-called ANC method and
$S_{17}(0)=20.9_{-0.6}^{+1.0}$ (theor) $\pm 1.8$ (expt) eV b
is obtained; the theoretical uncertainties are carefully evaluated
by changing the numerical inputs as described in Ref.~\citen{Oga06}.
This value of $S_{17}(0)$ is significantly
larger than the previous one, $19.0 \pm 1.8$~eV b,~\cite{Kik9798}
which was obtained from the same experimental data as used
in the present analysis
with the VPT. It is also found that
if we assume simple one-step E1 transition in our analysis of
the $^8$B breakup process as in the VPT analysis,
$19.0_{-1.7}^{+1.9}$~eV b
is obtained \cite{Oga06}, which agrees very well with the previous result.
This clearly shows the importance of the accurate description of
the breakup process, taking account of both nuclear and Coulomb breakup
with all higher-order processes.

In the right panel of Fig.~\ref{fig1-s8} we summarize
the results of the present analysis of the $^8$B breakup reaction,
compared with the results of the precise direct $^7$Be($p,\gamma$)$^8$B
measurement~\cite{Jun03} and the previous indirect
$^8$B breakup measurement~\cite{Kik9798}. One sees
that the agreement between the values of $S_{17}(0)$ obtained from the
direct and indirect measurements is significantly improved.
It should be noted that very recently,
$S_{17}(0)=20.9\pm0.7$ (theor) $\pm 0.6$ (expt)~eV~b
was obtained~\cite{Jun10}, as a compilation result
of direct measurements,
with a small correction to the result of Ref.~\citen{Jun03}.

Thus, we conclude that the indirect $^8$B breakup
measurement indeed enables accurate determination of the astrophysical factor
$S_{17}(0)$ if the breakup reaction is analyzed accurately.
This conclusion will be important for a future project to
extract nuclearastrophysics information from breakup reactions
of unstable nuclei.

\subsection{Three-body model analysis of subbarrier $\alpha$ transfer reaction}
\label{sec8-2}

Recently, indirect measurements of $^{13}$C($\alpha,n$)$^{16}$O
using subbarrier $\alpha$ transfer reaction
$^{13}$C($^{6}$Li$,d$)$^{17}$O(6.356~MeV, $1/2^+$) has been
performed by Florida State University group~\cite{Joh06};
DWBA was used for the analysis of the transfer cross section.
$^6$Li is known to have a $\alpha+d$ two-body
structure with a small binding energy of 1.47~MeV. Furthermore,
the final state of $^{17}$O(6.356~MeV, $1/2^+$), which is
denoted by $^{17}$O$^*$ below for simplicity, locates
just 3~keV below the $\alpha$-$^{13}$C threshold. Therefore,
roles of the breakup states of $^6$Li and $^{17}$O$^*$
in the $\alpha$ transfer process $^{13}$C($^{6}$Li$,d$)$^{17}$O$^*$
should be clarified, to obtain a more proper value of the
cross section of $^{13}$C($\alpha,n$)$^{16}$O at very low
energies.

In this study,~\cite{Fuk11} we first calculate with CDCC the elastic
cross sections of $^6$Li-$^{13}$C at 3.6~MeV and $d$-$^{17}$O$^*$
at 1.1~MeV corresponding to the initial and
final channels, respectively, of the
$^{13}$C($^{6}$Li$,d$)$^{17}$O$^*$ reaction.
It is found numerically that
the breakup effects of $^6$Li are important
but those of $^{17}$O$^*$ are small. Therefore, we evaluate
the $^{13}$C($^{6}$Li$,d$)$^{17}$O$^*$ cross section at
3.6~MeV including the breakup states of $^6$Li with CDCC;
the transition of the transfer process is described by
Born approximation, i.e., we adopt the so-called CDCC-BA framework.
%
%%%%%%%%%%%%%%%%%%%%%%%
%%%  Figure 2
%%%%%%%%%%%%%%%%%%%%%%%
\begin{figure}[htpb]
%\begin{figure}[t]
\centerline{
\includegraphics[width=70mm,keepaspectratio]{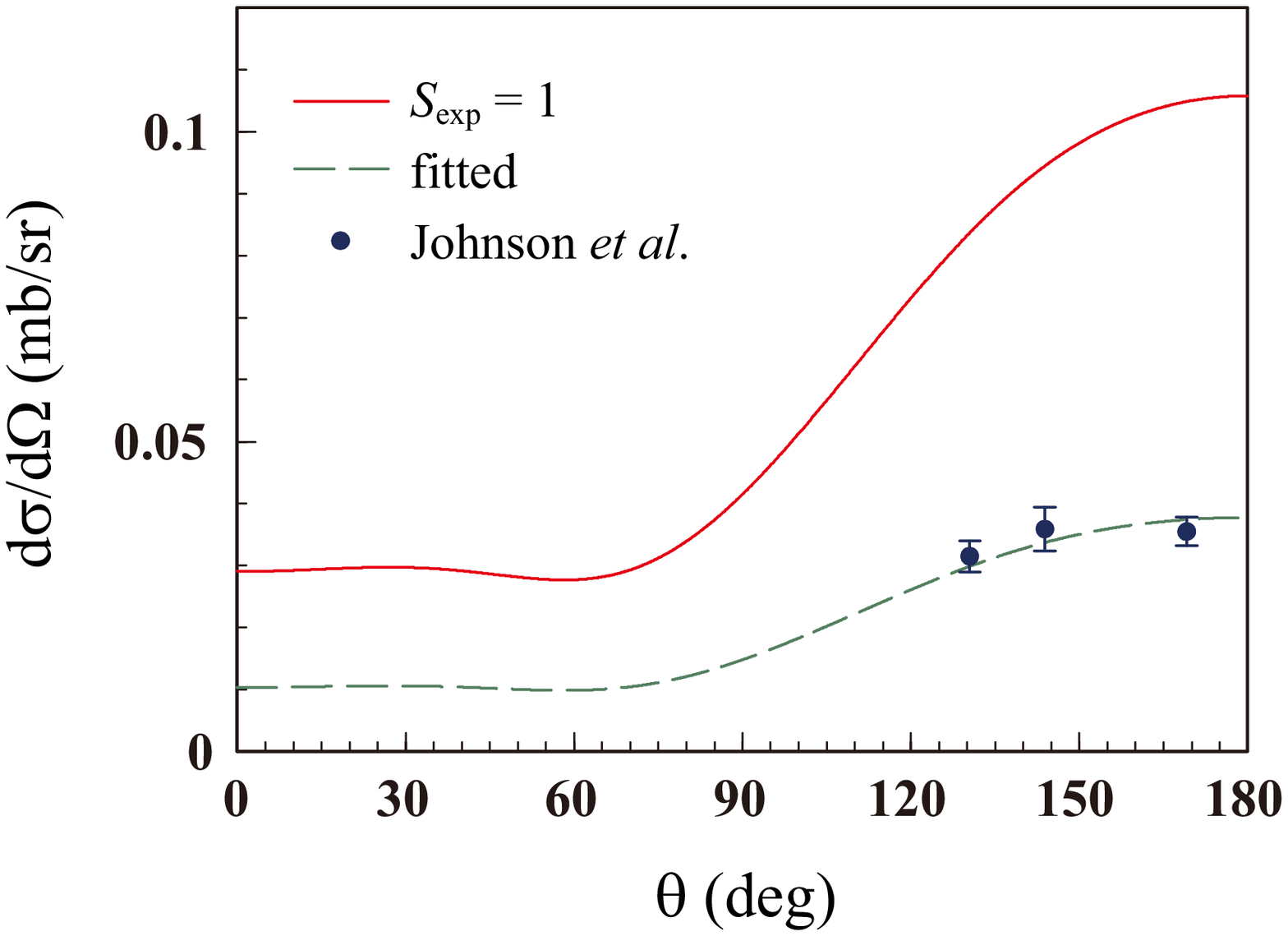}
\includegraphics[width=70mm,keepaspectratio]{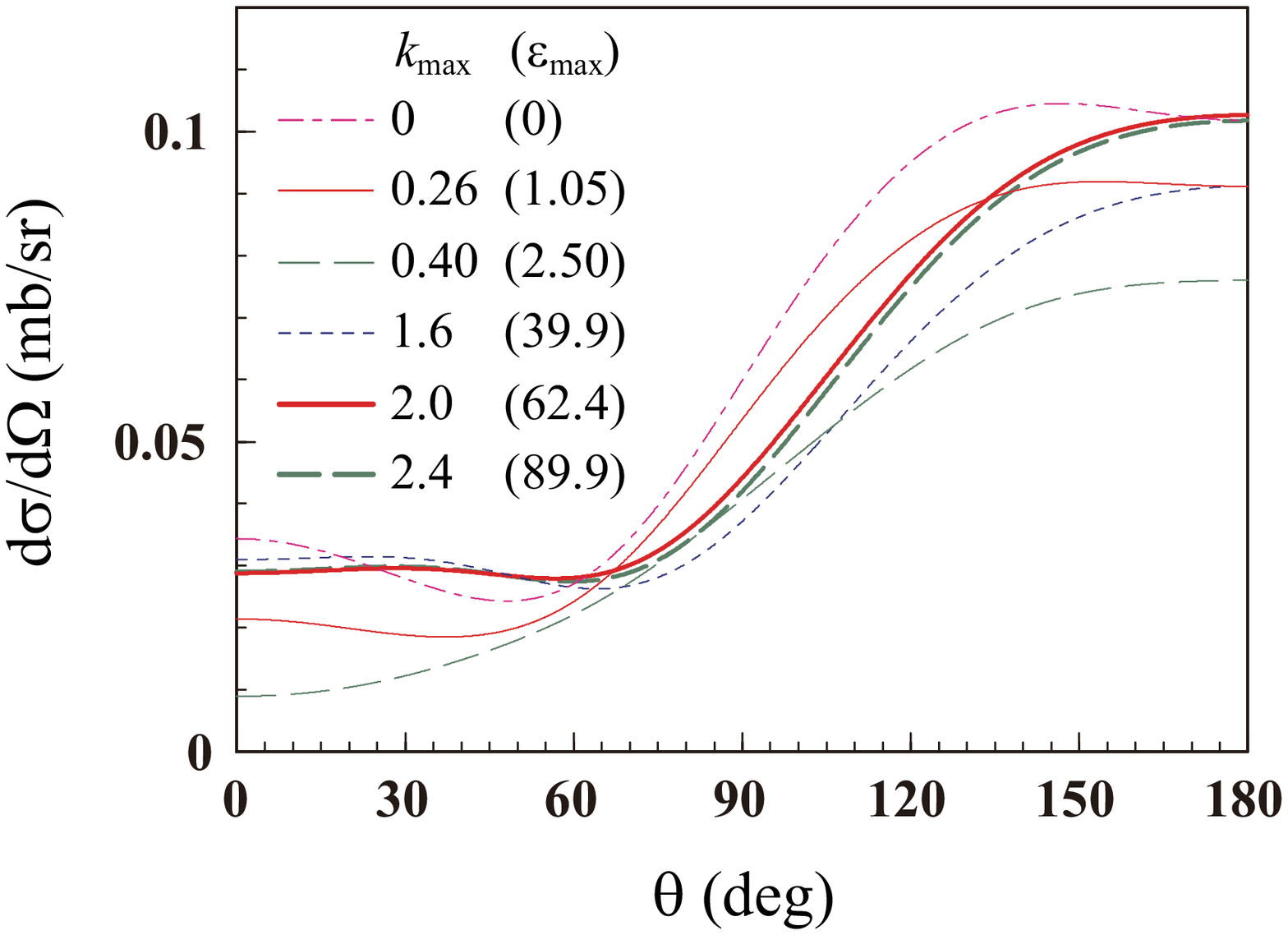}
}
 \caption{
(left panel)
Cross section of the transfer reaction
$^{13}$C($^{6}$Li$,d$)$^{17}$O$^*$
at 3.6~MeV. The solid line is the result of
the calculation with a normalized $\alpha$-$^{13}$C wave function
and the dashed line is the result of the $\chi^2$ fitting to the experimental
data.~\cite{Joh06}.
(right panel)
Dependence of the cross section on the model space of CDCC; see the
text for details.
}
\label{fig2-s8}
\end{figure}
In the left panel of Fig.~\ref{fig2-s8}, we show
the transfer cross section calculated by CDCC-BA with a
normalized $\alpha$-$^{13}$C
wave function (solid line) and that after the $\chi^2$ fitting
(dashed line) to the experimental data.~\cite{Joh06}
Through this analysis, we obtained
$1.03\pm 0.29$~fm$^{-1}$
for the square of the Coulomb-modified ANC~\cite{Joh06}
of $^{17}$O$^*$ with the $\alpha$-$^{13}$C configuration,
where the second number shows the theoretical and experimental
uncertainties together.
This result is consistent with the previous result
$0.89\pm 0.23$~fm$^{-1}$ obtained by the DWBA analysis.~\cite{Joh06}
It is found that the transfer processes through the breakup
states of $^6$Li are negligible and only the so-called
back-couplings from the breakup states to the ground state of
$^6$Li are important. This means only a proper description
of the scattering wave between $^6$Li and $^{13}$C in the
elastic channel is necessary. Therefore, use of a reliable
optical potential between $^6$Li and $^{13}$C, as in the
previous DWBA analysis, may give a proper value of the ANC
including the back-coupling effects implicitly.
However, this will not always be the case, since usually the optical
potential is determined only phenomenologically. Furthermore, breakup
transfer processes might be important in other subbarrier $\alpha$
transfer reactions.
The present three-body approach, therefore, should
be applied systematically to these reactions.

Before ending this subsection, we discuss the convergence of the
subbarrier transfer cross section with CDCC.
In the right panel of Fig.~\ref{fig2-s8} we show the dependence of
the cross section on the maximum value $k_{\rm max}$ of the relative
wave number of the $d$-$\alpha$ continuum adopted in CDCC.
The values of $k_{\rm max}$ are shown in unit of fm$^{-1}$ and
the corresponding values of the $d$-$\alpha$ relative energy
$\epsilon_{\rm max}$ are given in the parentheses in unit of MeV.
One sees clearly that the convergence is very slow and obtained
eventually at $k_{\rm max}=2.0$~fm$^{-1}$ ($\epsilon_{\rm max}=62.4$~MeV).
In the usual CDCC calculation,
one takes only the open channels, in which the relative energy
$E_{\rm rel}$
between the projectile and the target is positive.
 The result thus obtained (thin solid line)
is, however, sizably different from the converged one (thick dotted
line).
Therefore, inclusion of the closed channels,
in which $E_{\rm rel}$ is negative,
is necessary for the accurate description
of the $^{13}$C($^{6}$Li$,d$)$^{17}$O$^*$ reaction at 3.6~MeV with CDCC.

\subsection{Nonresonant triple $\alpha$ process at low temperatures}
\label{sec8-3}

Understanding of the formation of $^{12}$C, an essential element of
life, is one of the most important subjects in physics.
It is well known that the so-called triple-$\alpha$ process, i.e.,
the sequence of the following two reactions
\beq
\alpha+\alpha \longrightarrow {}^8\mbox{Be} \; \Longrightarrow \;
\alpha+{}^8{\rm Be}\longrightarrow
{}^{12}{\rm C}(0^+_2)
\longrightarrow
{}^{12}{\rm C}(2^+_1)+\gamma,
\label{reac1-s8}
\eeq
is the path to $^{12}$C. Particularly, the Hoyle resonance
(the second $0^+$ state) at 7.65~MeV of $^{12}$C plays an
essential role in this triple-$\alpha$ process.
At low temperatures, say, $T\la~10^8$~K, however, the energy of
three $\alpha$ particles cannot reach the Hoyle resonance.
In this case, a direct fusion process of three $\alpha$ particles
not through the Hoyle resonance becomes dominant. Thus,
evaluation of the reaction rate of
\beq
\alpha+\alpha+\alpha \longrightarrow {}^{12}{\rm C}(2^+_1)+\gamma
\label{reac2-s8}
\eeq
is essential for understanding the formation of $^{12}$C.

%
%%%%%%%%%%%%%%%%%%%%%%%
%%%  Figure 3
%%%%%%%%%%%%%%%%%%%%%%%
\begin{figure}[htpb]
%\begin{figure}[t]
\centerline{
\includegraphics[width=70mm,keepaspectratio]{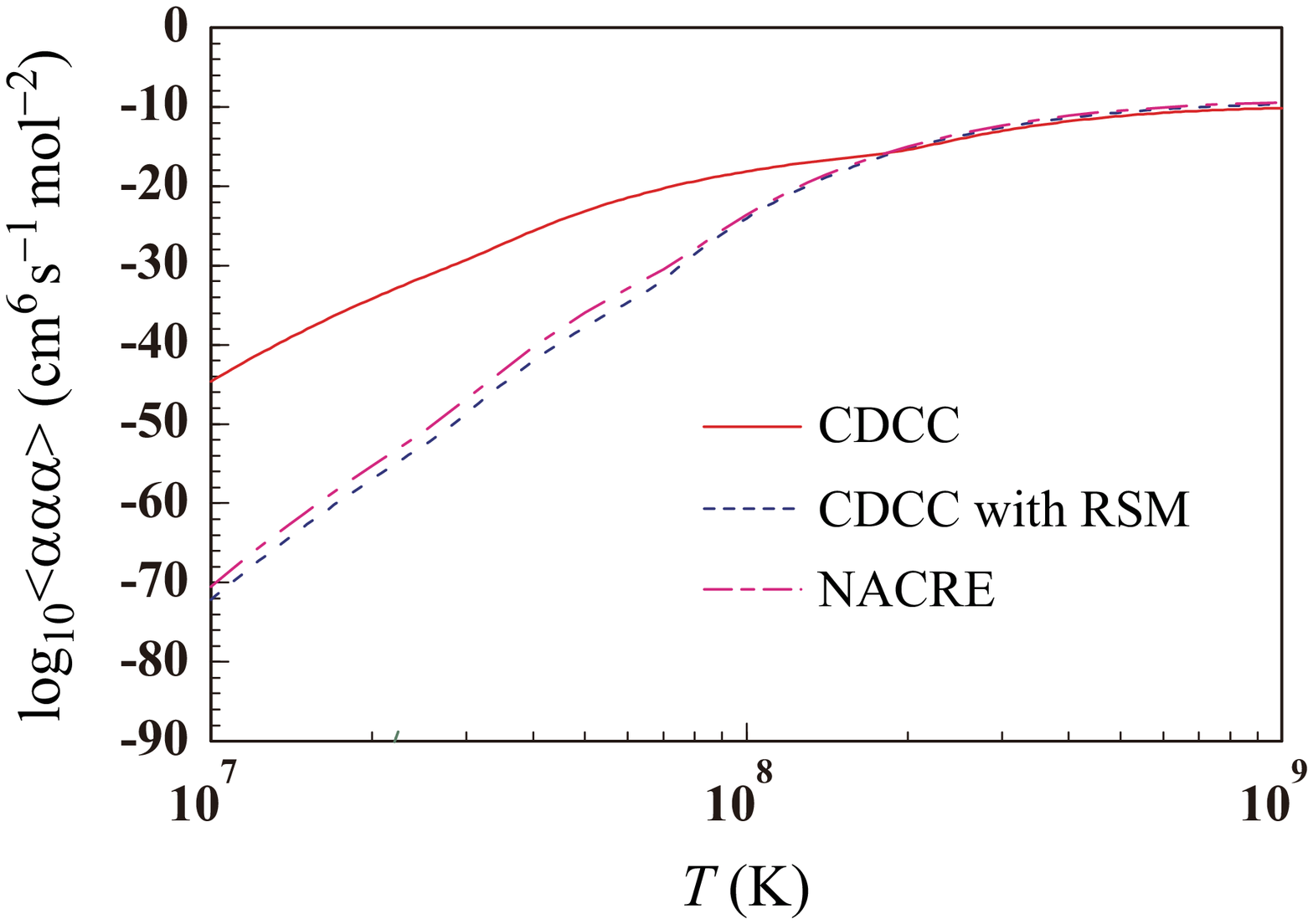}
\includegraphics[width=70mm,keepaspectratio]{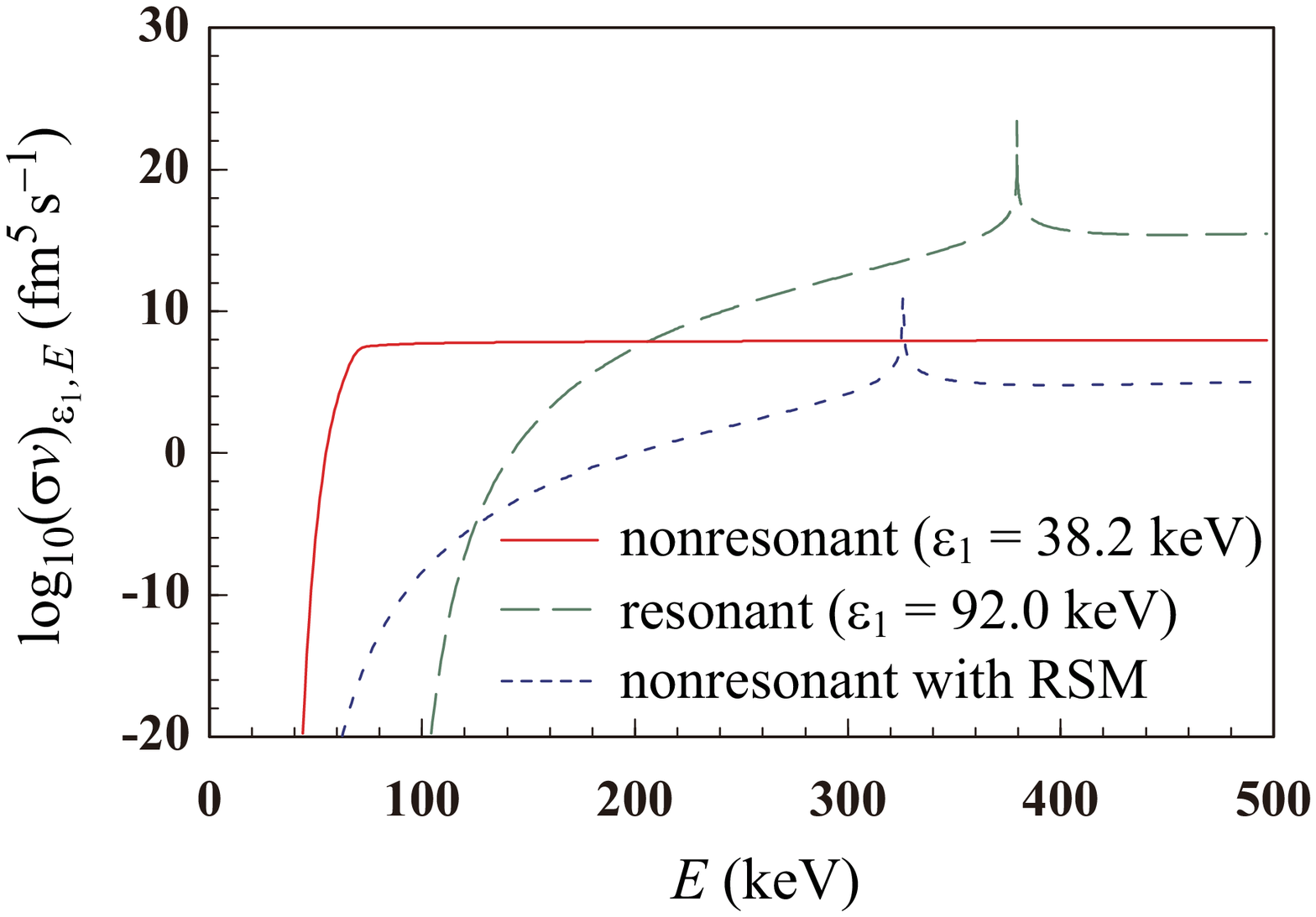}
}
\caption{\label{fig3-s8}
(left panel)
Triple-$\alpha$ reaction rate as a function of
temperature. The solid line represents the result of CDCC
and the dotted line is the result of CDCC with the RSM.
The dash-dotted line shows the reaction rate of NACRE~\cite{Ang99}.
(right panel)
The solid and dashed lines show
the reaction probability $\left( \sigma v \right)_{\epsilon_1,E}$
of the nonresonant ($\epsilon_1=38.2$~keV) and resonant
($\epsilon_1=92.0$~keV) processes, respectively.
The dotted line shows the result for the nonresonant process with
the RSM in the CDCC calculation.
}
\end{figure}
In Ref.~\citen{Oga09},
formulation of this ternary fusion process (TFP) based on CDCC
was developed, and applied to the study of the nonresonant
triple-$\alpha$ reaction.
We describe resonant and nonresonant processes on
the same footing. The left panel of Fig.~\ref{fig3-s8} shows
the resulting reaction rate.
The horizontal axis is temperature
and the vertical axis is the order of the rate. The solid line is the new reaction rate calculated
with CDCC, which is much larger than the rate of NACRE \cite{Ang99}
shown by the dash-dotted line.
The difference is up to about 20 orders of magnitude around $10^7$ K.
We stress that it is
shown in Ref.~\citen{Oga09}
that
the approximate method to mimic the nonresonant
triple-$\alpha$ process, the resonance shift method (RSM), proposed
by Nomoto~\cite{Nom82} and used in NACRE, has no theoretical foundation.
This can be understood clearly if
one sees the reaction probability $(\sigma v)_{\epsilon_1,E}$,
where $\epsilon_1$ is the $\alpha$-$\alpha$ relative energy and $E$
is the total energy of the three-$\alpha$ system.
For clear discussion, we show the results of $(\sigma v)_{\epsilon_1,E}$
obtained with the coupled-channel effects switched off;
the same model space as in the CDCC calculation converged is adopted.
The dashed line in the right panel of Fig.~\ref{fig3-s8} shows
$(\sigma v)_{\epsilon_1,E}$ for $\epsilon_1=92.0$~keV, i.e., the resonant
capture probability. As for nonresonant capture process,
we show by the solid line the result for $\epsilon_1=38.2$~keV calculated
with CDCC, which has completely different energy dependence from that of
the dashed line.
On the other hand, in the RSM, the probability shown by the dotted line is used.
One sees that it has a resonance peak at different energy from that of the
Hoyle state ($E=387$ keV).
This is also the case with other values of $\epsilon_1$.
Therefore, one finds that
the RSM implicitly assumes that
there are infinite number of resonances around the Hoyle resonance,
which is obviously inconsistent with the experimental information
on $^{12}$C.
If we adopt in our CDCC calculation the assumption used in
the RSM (and also in NACRE), we obtain the reaction rate shown by the
dotted line in the left panel of Fig.~\ref{fig3-s8} that agrees well
with the rate of NACRE.

It should be noted
that the calculation of the triple-$\alpha$ reaction at low energies requires
extremely large model space as described in Ref.~\citen{Oga09}.
We have confirmed a clear convergence
of the reaction rate with respect to the model space of CDCC;
the maximum value $r_{\rm max}$ of the $\alpha$-$\alpha$ relative coordinate $r$
is set to 5,000 fm.
If $r$ is truncated at a smaller value, say, 200~fm, we have a reaction
rate much smaller than the solid line in the left panel of
Fig.~\ref{fig3-s8}.
This is also the case when a screened Coulomb interaction with
a screening radius of a few tens of fm is used~\cite{Ish11}.
It is also found that single-channel calculation never converges,
and an adiabatic description
of the three-particle system at low energies does not work at all.

Our new reaction rate has been applied to several astrophysics
studies.~\cite{DP09,Sud11,SH10,PO10,Mat11b,Mor10,Kik12} Some of them,
Refs.~\citen{DP09,Sud11,SH10,PO10},
show that our reaction rate is incompatible with observation,
whereas others, Refs.~\citen{Mat11b,Mor10,Kik12},
claim that our rate is consistent with observation;
a slight modification on our reaction rate is necessary in Ref.~\citen{Mor10}.
On the nuclear physics side, very recently, a new
evaluation of the nonresonant triple-$\alpha$ reaction rate
with the hyperspherical harmonic R-matrix method was performed
by Nguyen and collaborators~\cite{Ngu11}. They
showed that the nonresonant contribution was indeed large,
i.e., about 20 orders of magnitude larger than the rate of NACRE
at $10^7$~K, which is qualitatively consistent with our finding.
At temperature higher than $7\times 10^7$~K, however, their
result agrees very well with the rate of NACRE. Further investigation
on the nonresonant triple-$\alpha$ reaction rate, as well as experimental
challenges to reveal the features of the low-energy three-$\alpha$
continuum, will be very important for our understanding of the origin
of $^{12}$C.

\section{Applications of CDCC to nuclear engineering}
\label{sec9}

One of the advantages of CDCC is its fully quantum-mechanical
aspect. Therefore, it is applicable to reactions at low incident
energies, which are important for nuclear engineering.
In this section, we show some results of applications of CDCC to
studies in this field.

\subsection{Deuteron induced reactions on $^{6,7}$Li}
\label{sec9-1}

{%\bf
In the international fusion material irradiation facility
(IFMIF) project,~\cite{Mat04b}
deuteron induced reactions on $^{6,7}$Li are considered as one
of the most promising reactions to generate
high-intensity neutrons. Understanding of the
$^{6,7}$Li($d,xn$) processes, where $x$ indicates all possible
reaction products, at incident energies up to 50~MeV is highly important.
Neutron spectra observed at forward angles show a broad peak with
approximately half the incident energy~\cite{Hag05}. This suggests
an importance of deuteron breakup processes, namely, deuteron
dissociation and proton stripping due to nuclear fields. In the past
works~\cite{Kon01,Fis07}, these processes
were treated using semiclassical models such as the modified
intra-nuclear cascade (INC) model~\cite{Kon01} and the
Serber model~\cite{Ser47}. Since the incident energy of interest
here is relatively low, more sophisticated quantum mechanical
approaches will be necessary for quantitative understanding
of the neutron production rate.

The $^{6,7}$Li($d,xn$) cross sections are expected to consist of
the contributions from the deuteron elastic breakup and proton
stripping processes.
In Ref.~\citen{Ye09} we applied CDCC and the Glauber
model~\cite{Gla59} to estimate the former and the latter,
respectively, of the $^{7}$Li($d,xn$) reaction at 40~MeV.
As for the nucleon optical model potential for $^{7}$Li,
we took the parameter set determined in
Ref.~\citen{Ye08} through a systematic analysis
of the nucleon elastic scattering on $^{6,7}$Li.}

\begin{figure}
 \includegraphics[width=1.0\textwidth]{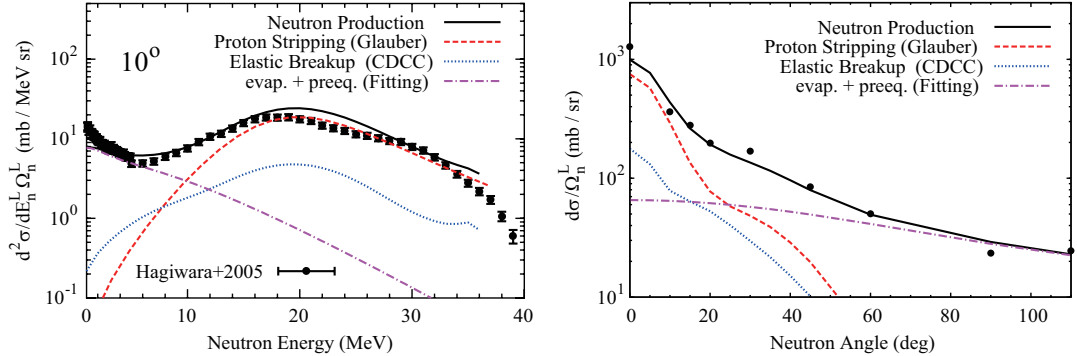}
 \caption{Comparison between the experimental data and the calculated
 result of $d^2 \sigma / (dE_n^{\rm L} d\Omega_n^{\rm L})$ at the neutron
 emission angle 10$^\circ$ (left panel) and
 the neutron angular distribution (right panel)
 of $^7$Li$(d, xn)$ at 40 MeV. For the latter,
 the cross section is integrated over the neutron outgoing energies.
 The proton stripping and elastic breakup components are plotted by
 the dashed and dotted lines, respectively, and
 the sum of the evaporation and pre-equilibrium components is
 shown by the dash-dotted line.
 The solid line represents the total neutron production.}
 \label{fig1-s9}
\end{figure}
In Fig.~\ref{fig1-s9} we show the comparison between the calculated
result and the experimental data.
The left panel shows the double differential breakup cross
section $d^2 \sigma / (dE_n^{\rm L} d\Omega_n^{\rm L})$ with
respect to the energy $E_n^{\rm L}$ and
the angle $\Omega_n^{\rm L}$ of the outgoing neutron in the
laboratory frame; the result corresponds to the neutron scattering angle
at 10$^\circ$.
The angular distribution $d \sigma / d\Omega_n^{\rm L}$
is shown in the right panel. In each panel,
the solid and dashed lines show, respectively, the contributions of
the elastic breakup and the stripping. The dotted line is
the sum of the evaporation and pre-equilibrium components, which are
evaluated by an empirical method, i.e., the moving source model.~\cite{Ye09}

The calculation reproduces the experimental data quite well.
As expected, the deuteron breakup processes are strongly
involved with the formation
of the bump around 20~MeV of the outgoing neutron energy.
The proton stripping process is more predominant than the elastic
breakup process. It is worth noting
that the strong contribution from the stripping process is also
indicated by the DWBA calculation for ($d,px$) reactions on
medium-heavy nuclei at 25.5 MeV~\cite{Pam78}.  The relative
contribution from the stripping process is reduced as the angle
increases. It is found also
that the deuteron breakup processes dominate neutron production
at forward angles and are negligible at backward  angles, while the
evaporation and pre-equilibrium processes have a major contribution
at backward angles.  Consequently, the present analysis suggests that
accurate description of the deuteron breakup reactions including both
the elastic breakup and proton stripping processes is essential for
reliable design of the neutron sources using the $^7$Li$(d, xn)$
reaction.

Although this model calculation reproduces successfully well the
experimental data at small angles, we found some problems of the
Glauber model calculation for the neutron stripping. The Glauber
model calculation shows that the peak position in the emission
spectra shifts to high energy as the emission angle increases, and
fails to reproduce the experimental spectra at large angles; see
Ref.~\citen{Ye09}. Furthermore, inclusive ($d,xp$) spectra for heavy
target nuclei are underestimated by the model calculation~\cite{Ye11}
because the Glauber model cannot treat Coulomb breakup effects accurately.
These may suggest a limitation of
applying the Glauber model.
Therefore, it is desirable to apply the eikonal reaction theory (ERT)
described in \S\ref{sec6} to the evaluation of
the stripping process. This may eventually make it feasible
to {\it predict} nucleon production from deuteron induced reactions.

\subsection{Neutron induced reactions on $^{6}$Li}
\label{sec9-2}

Lithium isotopes will be used as a tritium-breeding
material in $d$-$t$ fusion reactors, and accurate nuclear data are
required for $n$- and $p$-induced reactions. Indeed,
the international atomic energy agency (IAEA) is organizing
a research coordination meeting to prepare nuclear data libraries for
advanced fusion devices, FENDL-3~\cite{FENDL}, and the maximum incident energy is set
to 150~MeV to comply fully with the requirements for the IFMIF
project. Lithium isotopes are some of the most important materials in
these libraries. In fact, understanding of the
breakup properties of lithium isotopes by neutron is crucial to determine
the neutron energy spectra in blankets of fusion
reactors. The tritium breeding ratio, nuclear heating distributions,
and radiation damage of structural materials are affected by the
$n+$Li reactions significantly.

In spite of the importance of the $n+^6$Li reaction mentioned
above, experimental data of the breakup processes
leading to the $^6$Li continuum
are extremely rare for the neutron energy region above
20~MeV. Furthermore, the statistical model,
often used in evaluation of nuclear data for medium to heavy nuclei, cannot
be applied to the $^6$Li($n,n'$) reactions. As mentioned
in Refs.~\citen{Chi97} and \citen{Fuk97}, the mechanisms leading to
the three-body ($n+d+\alpha$) and four-body ($n+n+p+\alpha$)
final states are particularly important.
Therefore, more reliable theoretical calculations for the cross sections are
highly desirable.

In Ref.~\citen{Mat11}, we performed a microscopic analysis of $n+^6$Li
scattering by means of CDCC. In this analysis, $^6$Li is
described as a $d+\alpha$ system, and the interaction between $n$ and
$^6$Li is calculated by a folding model with the JLM effective
interaction~\cite{Jeu76}, where the normalization of the imaginary part,
$\lambda_w$, is optimized to reproduce the elastic cross
sections. Details of the calculation are shown in Ref.~\citen{Mat11}.

\begin{figure}[htbp]
 \begin{center}
  \includegraphics[width=0.45\textwidth,clip]{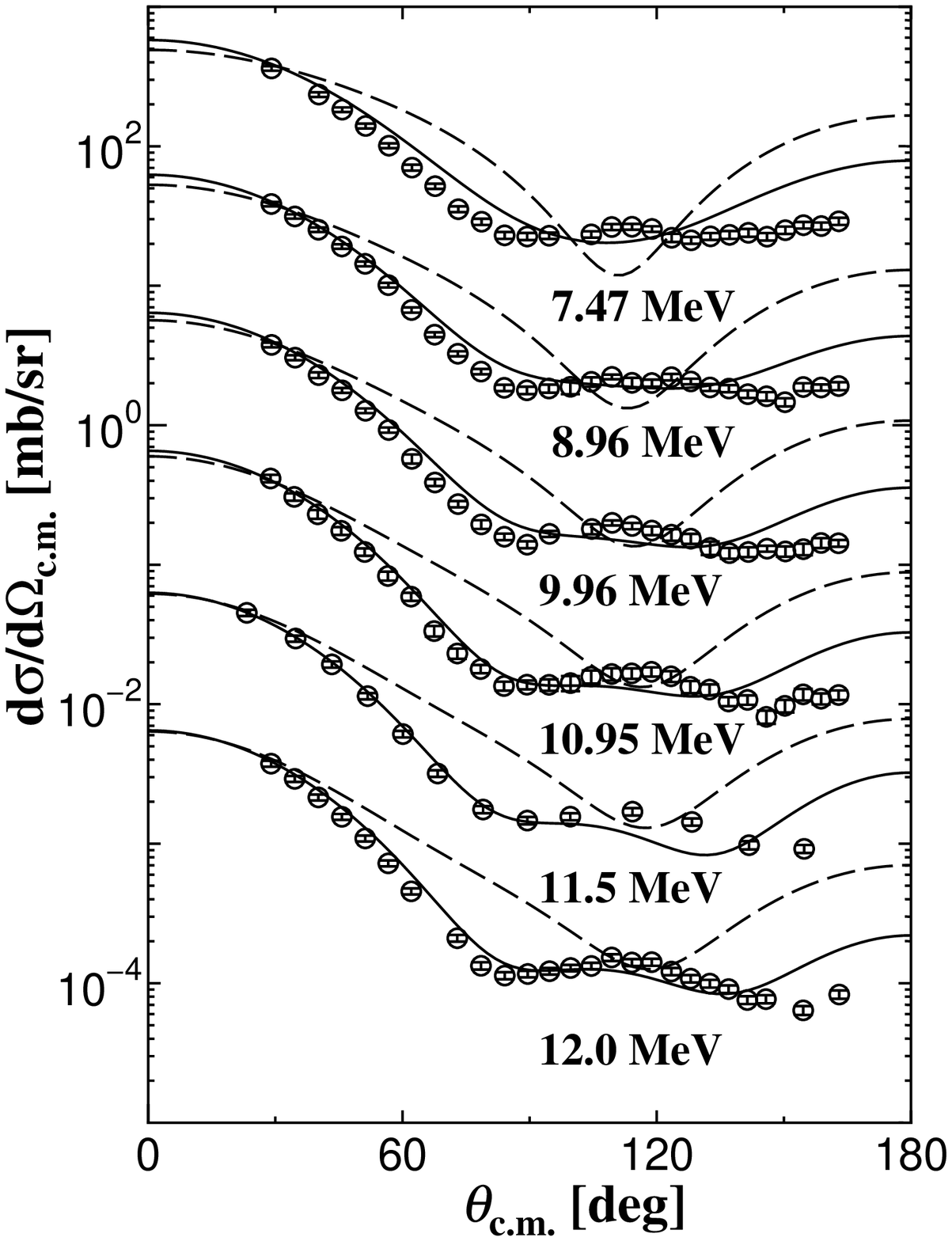}
  \includegraphics[width=0.45\textwidth,clip]{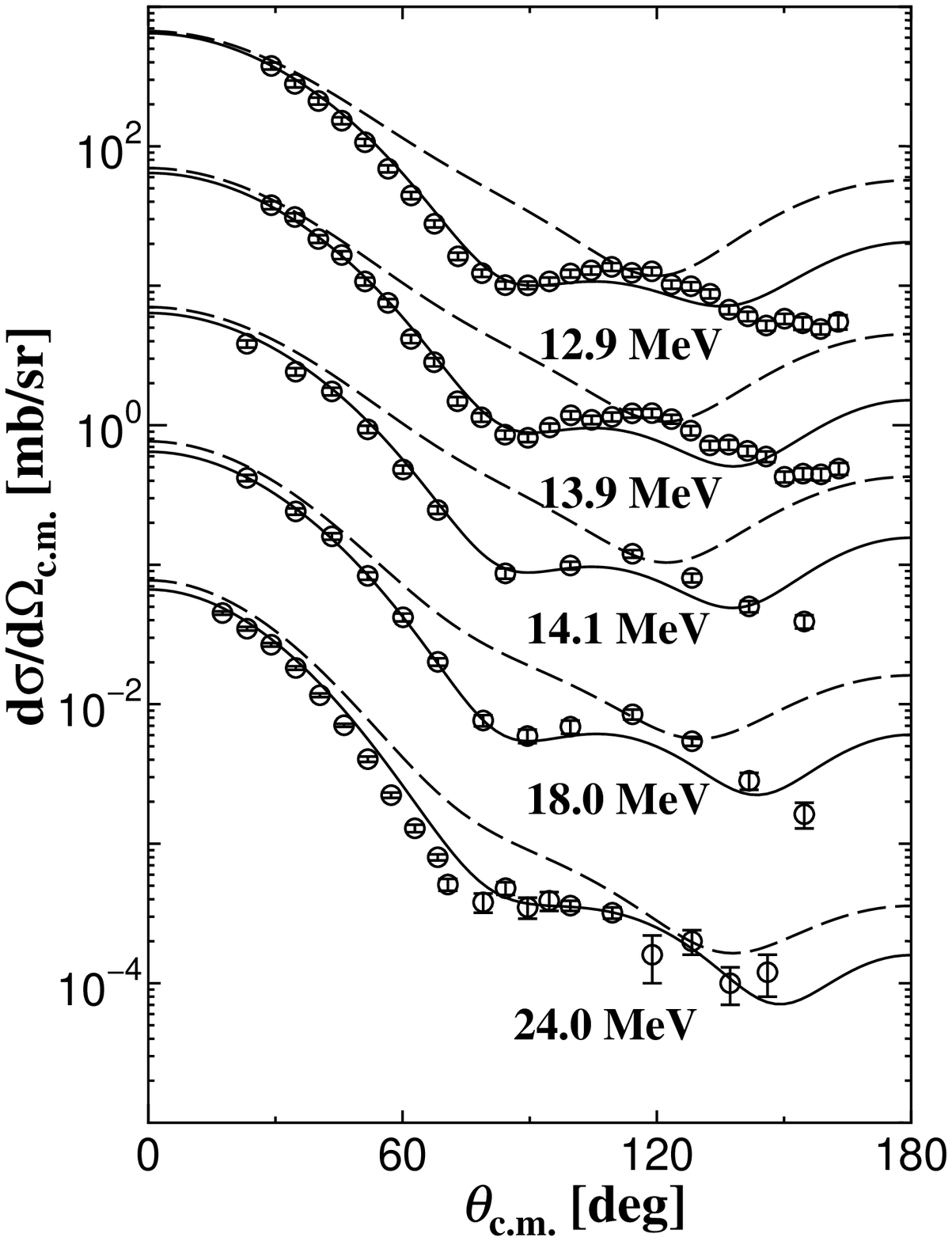}
 \end{center}
 \caption{Angular distribution of the elastic differential cross section
 of $n+^6$Li scattering for incident energies between 7.47 and 24.0~MeV.
 The solid and dashed lines correspond to the result with and without
 couplings to breakup states of $^6$Li, respectively.
 Experimental data are taken from Refs.~\citen{Chi98,Hog79,Han88}.
 The results are subsequently shifted downward by a factor of
 $10^{-1}$--$10^{-5}$ from 8.96 MeV to 12.0 MeV on the left panel, and
 $10^{-1}$--$10^{-4}$ from 13.9 MeV to 24.0 MeV on the right panel,
 respectively.}
 \label{fig2-s9}
\end{figure}
Figure~\ref{fig2-s9} shows the differential elastic cross sections of
$n+^6$Li for incident energies between 7.47 and 24.0 MeV. One sees that
the results of the CDCC calculation (the solid lines) are
in good agreement with the experimental data. The dashed lines represent
the results of a single-channel calculation, in which couplings to the
breakup states are omitted. It is found that breakup effects shown by the
difference between the dashed and solid lines are significant to
reproduce the angular distributions of the elastic scattering.
For all incident energies, we take $\lambda_w=0.1$ to reproduce
the data.
It should be noted that the single-channel calculation
cannot reproduce the experimental data with any values of $\lambda_w$.
The very small value of $\lambda_w$ obtained may indicate that
most of the reaction channels that cause loss of the incident
flux are treated explicitly in the CDCC calculation.

\begin{figure}[htbp]
 \begin{center}
  \includegraphics[width=\textwidth,clip]{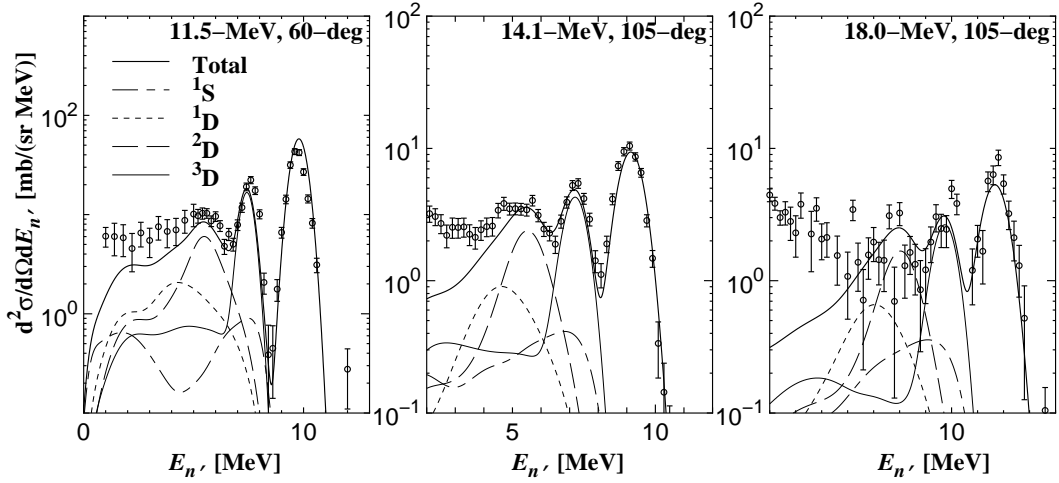}
 \end{center}
 \caption{The neutron spectra calculated by CDCC
 with the JLM interaction comparing with measured data at
 selected angular points in the laboratory system. The experimental data
 are taken from Ref.~\citen{Chi98}.}
 \label{fig3-s9}
\end{figure}
In Fig.~\ref{fig3-s9}, the calculated neutron spectra are
compared with the experimental data at selected scattering
angles in the laboratory
frame and incident energies; $\lambda_w$ is fixed at 0.1.
Components of the breakup states of $^1$S, $^1$D, $^2$D, and $^3$D
are represented by the
dash-dotted, dashed, dotted, and thin solid lines, respectively.
These results are broadened by considering the finite resolution of the
experimental apparatus~\cite{Chi98}.
In each panel, three peaks of the experimental data correspond
to the elastic scattering (right), the inelastic scattering to
the $3^+$ resonance (center), and the inelastic scattering
to the $2^+$ resonance (left).
The CDCC calculation gives a good agreement with experimental data in
the high neutron energy region. On the other hand, in the low neutron
energy region, which corresponds to highly exited states of $^6$Li,
the calculated cross section considerably undershoot
the experimental data.
This indicates that experimental
data contain contribution from the ($n,2n$) process due to
a four-body breakup reaction $^6$Li($n,nnp$)$\alpha$, as indicated in
Ref.~\citen{Chi98}.
In the present calculation, the four-body breakup
effects are treated only implicitly  by using a complex optical potential.
Estimate of these effects with four-body CDCC will be very important.

Thus, CDCC with the JLM interaction is expected to be
a powerful framework for the data evaluation of the $^6$Li($n,n'$)
reactions. Once the JLM parameter is determined by an analysis of
elastic scattering, evaluation of the inelastic cross sections and
neutron breakup spectra can be done with no free adjustable parameters.
This is a very important feature of the present framework that
enables a quantitative calculation of the cross sections
for nuclear engineering studies.

\section{Summary}
\label{sec10}

This review has shown recent developments of the continuum discretized
coupled-channels method (CDCC) and its
applications to nuclear physics, cosmology and astrophysics, and
nuclear engineering, after the previous review
articles of Refs.~\citen{Kam86} and \citen{Aus87}.

First, we have shown the theoretical foundation of CDCC.
The primary approximation in CDCC is the truncation  of
the angular momentum $\ell$ between two constituents of the projectile.
This $\ell$-truncation changes the two-body potentials
between the target and the constituents
to three-body potentials.
The CDCC solution is the zeroth-order
solution to the distorted Faddeev equations with the three-body potentials,
the first-order correction to which is largely suppressed.
The CDCC solution is thus a good solution to the three-body
Schr\"odinger equation, when the maximum value of $\ell$ is large enough.
This theoretical statement has recently been
confirmed numerically by directly
comparing the CDCC and Faddeev solutions.~\cite{Del07}
In CDCC calculations, it is necessary to confirm
the convergence of the solution with
respect to expanding the model space, so that
the CDCC solution has no model space dependence and hence
is a good solution to the three-body Schr\"odinger equation.

As an underlying theory of CDCC, we have constructed a microscopic
reaction theory for nucleus-nucleus scattering based on
the multiple scattering theory.
The input of the theory is either the $t$- or the $g$-matrix 
NN interaction instead of the realistic one. The former has much milder
$r$ dependence, so that the Glauber model becomes reliable 
for the systems in which the Coulomb interactions are weak, when 
either the $t$- or the $g$-matrix is used as an input of the model. 
In the scattering of lighter projectiles from lighter targets at intermediate
energies, effects induced by finite nucleus such as
projectile breakup and target collective excitations are found to be
small. Then the double-folding model becomes reliable for the scattering.
Using the model, one can construct the microscopic optical potential with
projectile and target densities calculated by fully microscopic
structure theories such as antisymmetrized molecular dynamics (AMD),
the Hartree-Fock (HF) method, and the Hartree-Fock-Bogoliubov (HFB) method.
This fully microscopic framework has been applied to
the scattering of stable nuclei and unstable neutron-rich Ne isotopes
at intermediate energies with
success in reproducing experimental data.
The reliable microscopic optical potential can be used as an input of CDCC
calculations, since the nonlocal potential can be localized with
the Brieva-Rook (BR) method accurately.
This microscopic version of CDCC is quite useful to
analyze scattering of unstable nuclei systematically.

Then we have extended CDCC in three directions, and developed
the following new frameworks for breakup reactions:
i) {\it eikonal-CDCC (E-CDCC)} for Coulomb-dominated breakup
processes,
ii) {\it four-body CDCC} for breakup of a three-body projectile,
and iii) {\it the eikonal reaction theory (ERT)} for inclusive breakup
processes.

E-CDCC treats both nuclear and Coulomb
breakup very accurately and efficiently.
The quantum-mechanical (QM) corrections, if necessary, can be
made by replacing the partial scattering amplitudes corresponding
to small impact parameters $b$ with the solutions of a QM calculation,
i.e., CDCC.
E-CDCC calculations with the QM corrections are
much faster than CDCC calculations, with keeping the same accuracy
as the latter.
E-CDCC has thus become a standard method
for describing breakup reactions in which the Coulomb breakup contribution
is essential.
Recently, inclusion of a relativistic Li\'{e}nard-Wiechert potential
as a Coulomb interaction was accomplished in E-CDCC, and
dynamical relativistic effects generated by the potential were
found to give a sizable increase in the $^8$B and $^{11}$Be
breakup cross sections at 250~MeV/nucleon.

Four-body CDCC adopts pseudostates, which are obtained by
diagonalizing the internal Hamiltonian of a three-body projectile,
as discretized-continuum states of the projectile. The
four-body wave function is expanded by the pseudostates together
with the bound state(s); they are assumed to form a complete set
in a space in which the reaction takes place.
Additionally, a new method
for obtaining a smooth breakup energy spectrum from discrete
breakup cross sections given by CDCC was proposed.
Four-body CDCC together with
this new smoothing method based on the complex-scaling method
has completed the description four-body breakup processes.
Clear convergence with respect to expanding the model space is seen
in both the elastic and the breakup cross section of the $^6$He scattering,
and the results converged are consistent with experimental data.
Four-body CDCC with the new smoothing method is indispensable for
systematic studies of unstable nuclei that have a three-body structure.

ERT is a framework that makes CDCC applicable to inclusive breakup
processes such as neutron removal reactions, in which only
the core nucleus of the projectile is detected.
ERT gives separation of the scattering matrix into the contribution
of each constituent of the projectile, without making the
adiabatic approximation to the Coulomb potential. This is an essential
point of ERT that eliminates the well-known shortcoming of the
Glauber model, i.e., the divergence problem in Coulomb breakup.
ERT has been successfully applied to the one-neutron removal from $^{31}$Ne, 
and one- and two-neutron removal from $^{6}$He.
With ERT, the accuracy of the Glauber model was systematically
investigated for deuteron induced reactions at 200~MeV/nucleon.
The Glauber model is good for light targets, but not for heavier targets,
because of the Coulomb breakup contributions for the latter.

CDCC, E-CDCC, four-body CDCC, ERT, and the microscopic version of CDCC
have extensively and successfully been applied to studies of various reactions
essential in nuclear physics, cosmology and astrophysics, and nuclear
engineering. These methods treat nuclear and Coulomb
breakup processes
nonperturbatively and nonadiabatically, and have no free adjustable
parameters. These features are essential for {\it quantitative}
studies such as the investigation of four-body breakup properties
of $^6$He at the Coulomb barrier energy, determination of
astrophysical $S$ factors, evaluation of the contribution from
the nonresonant triple-$\alpha$ process, the accurate estimation of
neutron yields for the nuclear engineering design, and so forth.
Validation of findings or conjectures obtained by nuclear structural
studies through direct comparison with experimental cross sections
is also highly important. The microscopic reaction theory plays
definitely important roles in this subject, as we have confirmed
$^{31}$Ne to be a halo nucleus with large deformation.

There are several interesting future works based on CDCC and the extended
theories.
One is systematic applications of these theories to
nuclear physics and the related fields.
Another interesting work is to clarify the relation between
CDCC and the dynamical eikonal approximation (DEA).~\cite{Bay05,Gol06}
Many nuclei can be described well by
a core nucleus and extra nucleon(s).
In the scattering of these nuclei, excitations of the core nucleus
are important in general.~\cite{Lou11,Cre11}
It is interesting to extend CDCC to treat the core excitations
during the breakup.
Use of other three-body structural models in four-body CDCC is also an
important future work. The cluster-orbital shell model
(COSM),~\cite{SI88,Aoy06}
which has been applied to up to five-body systems,
is one of the most promising models.
Extension of four-body CDCC to five- and
six-body CDCC with COSM will be a very attractive subject to explorer
drip line nuclei. Finally, it will be very important to
develop more accurate description of transfer reactions at relatively
low energies, by treating the rearrangement channels nonperturbatively.
This will be essential to investigate excited states of stable and
unstable nuclei, which are to be accessed via transfer reactions.
These future works are important
in further development of nuclear physics and the related fields.

\section*{Acknowledgement}

This review is based on the collaboration with
C.~A.~Bertulani,
S.~Chiba,
T.~Egami,
T.~Fukui,
S.~Hashimoto,
Y.~Hirabayashi,
E.~Hiyama,
D.~Ichinkhorloo,
Y.~Iseri,
M.~Kamimura,
T.~Kamizato,
M.~Kan,
K.~Kat\=o,
M.~Kawai,
M.~Kimura,
M.~Kohno,
Y.~Kondo,
T.~Nakamura,
R.~A.~D.~Piyadasa,
Y.~R.~Shimizu,
T.~Sumi,
S.~Tagami,
S.~Watanabe,
Y.~Watanabe,
and
T.~Ye.
The authors wish to sincerely thank these people.
The authors also deeply appreciate
D.~Baye,
P.~Capel,
P.~Descouvemont,
M.~Fukuda,
T.~Furumoto,
K.~Hagino,
T.~Kajino,
A.~M.~Moro,
T.~Motobayashi,
G.~V.~Rogachev,
Y.~Sakuragi,
H.~Sakurai,
Y.~Suzuki,
M.~Takechi,
and
K.~Yabana
for valuable suggestions and fruitful discussions.
This work has been supported in part by
the Grants-in-Aid for Scientific Research
from Japan Society for the Promotion of Science.
The computation was carried out using mainly
the computer facilities at
the Research Institute for Information Technology, Kyushu University.

%\appendix
%\section{First Appendix} %Empty argument \section{} yields `Appendix'.
%
%\section{Second Appendix}

%%--------------------------------------------------------------------%%
%%                           References                               %%
%%--------------------------------------------------------------------%%

\end{document}